\documentclass[a4paper,10pt]{article}
\usepackage[english]{babel}

\begin{document}

\newcommand{\bq}{\begin{eqnarray}}
\newcommand{\eq}{\end{eqnarray}}

\newcommand{\gapprox}{\raisebox{-0.5ex}{$\
\stackrel{\textstyle>}{\textstyle\sim}\ $}}
\newcommand{\lsim}{\raisebox{-0.5ex}{$\
\stackrel{\textstyle<}{\textstyle\sim}\ $}}

\newcommand{\baa}{begin{eqnarray}}
\newcommand{\eaa}{end{eqnarray}}

\renewcommand{\theequation}{\arabic{section}.\arabic{equation}}
\renewcommand{\thesubsection}{\arabic{subsection}}

\newcommand{\ct}{\cite}
\newcommand{\lb}{\label}
\newcommand{\cen}{\centerline}
\newcommand{\bc}{\begin{center}}
\newcommand{\ec}{\end{center}}
\newcommand{\bd}{\begin{displaymath}}
\newcommand{\ed}{\end{displaymath}}
\newcommand{\be}{\begin{equation}}
\newcommand{\ee}{\end{equation}}
\newcommand{\ba}{\begin{array}}
\newcommand{\ea}{\end{array}}
\newcommand{\bt}{\begin{tabular}}
\newcommand{\et}{\end{tabular}}
\newcommand{\un}{\underline}
\newcommand{\ov}{\overline}
\newcommand{\bp}{\begin{picture}}
\newcommand{\ep}{\end{picture}}
\newcommand{\bfi}{\begin{figure}}
\newcommand{\efi}{\end{figure}}
\def\la{\mathrel{\mathpalette\fun <}}
\def\ga{\mathrel{\mathpalette\fun >}}
\def\fun#1#2{\lower3.6pt\vbox{\baselineskip0pt\lineskip.9pt
\ialign{$\mathsurround=0pt#1\hfil##\hfil$\crcr#2\crcr\sim\crcr}}}
\newcommand{\rf}{\ref}

\newpage
\thispagestyle{empty}
$\,$
\vskip 3cm
\begin{center}

{\Large Proceedings to the international workshop on}

\vspace{25mm}

\begin{Huge}

{\bf What comes beyond the standard model}

\end{Huge}

\vspace{1cm}

{\Large Bled, Slovenia\\[2mm] 29 June - 9 July 1998}

\vspace{6cm}

{\Large Editors}

\vspace{7mm}

\begin{large}

Norma Manko\v c Bor\v stnik\\

\end{large}

\vspace{2mm}

{\it University of Ljubljana, Jo\v zef Stefan Institute}

\vspace{4mm}

\begin{large}

Holger Bech Nielsen

\end{large}

\vspace{2mm}

{\it University of Copenhagen, Niels Bohr Institute}

\vspace{4mm}

\begin{large}

Colin Froggatt

\end{large}

\vspace{2mm}

{\it University of Glasgow}

\vfill
\centerline{\large DMFA - zalo\v zni\v stvo}
\centerline{\large Ljubljana 1999}
\end{center}

\newpage
\thispagestyle{empty}
$\,$
\vskip 3cm

\begin{large}

\noindent
The international workshop on "What Comes Beyond the Standard
Model"\\[1cm]
was organized by \\[1cm]
{\it  Jo\v zef Stefan Institute},\\[3mm] 
{\it Department of Physics, FMF, University of Ljubljana},\\[3mm] 
{\it National Committee of Physics of Slovenia}

\vspace{1cm}

\noindent and was sponsored by

\vspace{1cm}

\noindent
{\it Ministry of Science and Technology of Slovenia,}\\[3mm]
{\it Ministry of Education and Sport of Slovenia,}\\[3mm]
{\it Department of Physics, FMF, University of Ljubljana,}\\[3mm]
{\it European Physical Society.}\\

\vspace{2cm}

\noindent
Organizing Committee\\

\vspace{5mm}

\noindent
Norma Manko\v c Bor\v stnik\\[3mm]
Holger Bech Nielsen\\[3mm]
Colin Froggatt\\[3mm]

\end{large}

\newpage
\thispagestyle{empty}
\begin{center}

\begin{large}

{\bf CONTENTS}

\end{large}
\end{center}

\vspace{5mm}

\noindent

\begin{itemize}
\item Preface
\item Invited Talks
  \begin{enumerate}
	\item Colin D. Froggatt, {\it The problem of the quark-lepton mass
	spectrum}\dotfill 6
	\item Astri Kleppe, {\it Extending the Standard model by including
	right-handed neutrinos}\dotfill 11
	\item Larisa V. Laperashvili and H.B.Nielsen, \\{\it Multiple point
	principle and phase transition in gauge theories}\dotfill 15
	\item Norma Manko\v c Bor\v stnik, {\it Unification of spins and charges in
	Grassmann space}\dotfill 20
	\item Holger B. Nielsen and Colin D. Froggatt, {\it Masses and mixing
	angles and going beyond the Standard model}\dotfill 29
	\item Berthold Stech, {\it Hints from the Standard model for particle
	masses and mixing}\dotfill 40
	\item Hanns Stremnitzer, {\it Composite models and SUSY}\dotfill 45
  \end{enumerate}  
\item Discussions
  \begin{enumerate}
  	\item Anamarija Bor\v stnik and Norma Manko\v c Bor\v stnik, {\it Are spins
  	and charges unified?
	How can one otherwise understand connection between handedness (spin) and
	weak charge?}\dotfill 52
	\item Colin D. Froggatt and Holger B. Nielsen, {\it Why do we have parity
	violation?}\dotfill 57
	\item Astri Kleppe, {\it The number of families in a scenario with
	right-handed neutrinos}\dotfill 61
	\item Norma Manko\v c Bor\v stnik, Holger B. Nilesen, Colin D. Froggatt,
	{\it Can we make a Majorana field theory starting from the zero mass Weyl
	theory, then adding	a mass term as an interaction?}\dotfill 63
	\item Holger B. Nielsen and Norma Manko\v c Bor\v stnik, {\it Can one
	connect the Dirac-K\" ahler	representation of Dirac spinors and spinor
	representation in Grassmann	space proposed by Manko\v c?}\dotfill 68
	\item Berthold Stech, {\it Comment on the hierarchy problem}\dotfill 73
	\item Hanns Stremnitzer, {\it Parity conservation in broken SO(10)}\dotfill
	76
  \end{enumerate}    
\end{itemize}

\newpage
\thispagestyle{empty}
\begin{large}
\begin{center}
 \bf PREFACE
\end{center}
\end{large}

\vspace{3mm}

\noindent
The workshop {\bf " What comes beyond the standard model? "} was
meant as a real workshop in which participants would spend most of the 
time in discussions, confronting different aproaches and ideas. The
nice town of Bled by the lake of the same name, 
surrounded by beautiful mountains and offering pleasant walks,
was chosen to stimulate the discussions. \\

\noindent
We believe that this really happened. We spent ten fruitful days 
in the house of Josip Plemelj, which belongs to the Physical Society and
Mathematical Society of Slovenia, discussing the open problems in 
high energy physics.\\

\vspace{5mm}

\noindent 
We tried to answer to some of the open questions which the electroweak
Standard Model leaves unanswered. We started the first day of the
workshop with the following list of open questions:

\vspace{5mm}

\begin{itemize}

\item
Are spins and charges unified ? 
How can one otherwise understand the connection between 
spin and weak charge, built into the Standard Model by the following 
requirement: there exist only left handed weak charge doublets
and right handed weak charge singlets, which assures  parity
violation ? 

\item
Why is parity not broken in strong and electromagnetic interactions?

\item
How  is parity conserved  in broken $SO(10)$?

\item
A Majorana particle, if it exists at all, is a rather unusual
particle, since it is a particle and its own antiparticle at the
same time, like a photon, but it is a fermion. 
Can one formulate Majorana quantum field theory so that one sees 
the Dirac sea being filled by Majorana particles?
Can we make a Majorana field theory starting from the zero mass 
Weyl theory, then adding a mass term as an interaction in a
similar way as we can do in the case of a Dirac particle?
What is the Majorana propagator?

\item
How one can understand the hierarchy problem and the scale problem?

\item
Where do the generations of quarks and leptons come from? Do we
have more than three generations?
Is there any approach which would suggest more than three
generations and would not contradict the experimental data?

\item
Where do the Yukawa couplings come from ?

\item
Can we exclude the existence of constituents of quarks and leptons?
Shall we not have problems with confining chiral particles into
chiral clusters of particles?

\item
Is $S(U(2)\times U(3))$ a subgroup of 
(i) $SO(1,13)$,
(ii)$ SU(4)_C \times SU(2)_L \times SU(2)_R $, 
(iii) other unified gauge groups?

\item
Can spin and charges be unified within $SO(1, d-1)$?

\item
Can one connect the Joos representation of Dirac spinors and 
spinor representations in Grassmann space?
\end{itemize}

\vspace{5mm}

We have tried for 10 days to answer these questions. The result of
this effort is collected in the Proceedings. We, of course, only 
succeeded to partly answer some of these questions. 
Everybody tried to answer the questions connected with
her/his own work. Because of that the Proceedings contains a
review of previous work, connected with the above
listed questions. 

\noindent
{\bf Next year we shall meet again at Bled}.

\vspace{5mm}

The organizers are grateful to all the participants for the real 
discussing and working atmosphere. 

\vspace{5mm}

The organizers would like to thank Anamarija Bor\v stnik, Jiannis Pachos, and 
Andreja \v Sarlah for the efficient help in the organization of the workshop 
and in the preparation of the proceedings.

\vspace{5mm}

\begin{flushright}
Norma Manko\v c Borstnik\\
Holger Bech Nielsen\\
Colin Froggatt.
\end{flushright}
\vspace{3mm}

\newpage
\thispagestyle{empty}
\stepcounter{part}
$\,$
\vskip 10cm
\begin{center}
 \begin{Huge}
  \begin{bfseries}
      \sc Invited Talks
  \end{bfseries}
 \end{Huge}
\end{center}
\newpage
\stepcounter{section}
\setcounter{equation}{0}
\section*{\center The Problem of the Quark-Lepton Mass Spectrum }
\centerline{\rm COLIN D. FROGGATT}
\centerline{\it Department of Physics and Astronomy, Glasgow University,} 
\centerline{\it Glasgow G12 8QQ, Scotland, UK}
\vskip 5mm

\subsection{Introduction}

The charged fermion masses and mixing angles arise from the Yukawa 
couplings, which are arbitrary parameters in the Standard Model (SM). 
The masses range over five orders of magnitude, from 1/2 MeV for 
the electron to 175 GeV for the top quark. Also the elements of 
the quark weak coupling matrix, $V_{CKM}$, range from 
$V_{ub} \simeq 0.003$ to $V_{tb} \simeq 1$. This constitutes the 
charged fermion mass and mixing problem. It is only the top 
quark which has a mass of the order of the electroweak scale
$<\phi_{WS}> = 174$ GeV and a Yukawa coupling constant of order 
unity $y_t \simeq 1$. It therefore 
seems likely that the top quark mass will be understood 
dynamically before those of the other fermions.
All of the other Yukawa couplings are suppressed,
suggesting the existence of physics beyond the SM.   Furthermore the 
accumulating evidence for neutrino oscillations provides direct 
evidence for physics beyond the SM, in the form of non-zero neutrino 
masses.

A fermion mass term essentially represents a transition amplitude between 
a left-handed Weyl field $\psi_L$ and a right-handed Weyl field 
$\psi_R$. If $\psi_L$ and $\psi_R$ have different quantum numbers, 
i.e. belong to inequivalent irreducible representations 
of a symmetry group $G$ ($G$ is then called a chiral symmetry), 
the mass term is forbidden in the limit of exact $G$ symmetry and 
they represent two massless Weyl particles. $G$ thus ``protects'' the 
fermion from getting a mass. For example the 
$SU(2)_L \times U(1)$ gauge quantum numbers of the left and right-handed
top quark fields are different and the electroweak gauge 
symmetry protects the top quark from having a mass, 
i.e. the mass term $\overline{t}_Lt_R$ is not gauge invariant. 
It is only after the  
$SU(2)_L \times U(1)$ gauge symmetry is spontaneously broken that the 
top quark gains a mass $m_t=y_t <\phi_{WS}>$, which is consequently 
suppressed relative to the presumed fundamental (GUT, Planck...) 
mass scale M by the symmetry breaking parameter 
$\epsilon = <\phi_{WS}>/M$. The other quarks and leptons have  
masses suppressed relative to $<\phi_{WS}>$ and it is natural to 
assume that they are protected by further approximately 
conserved chiral flavour charges \cite{fn}, as we discuss 
further in section 5.

We first consider dynamical calculations of the top quark and 
the Higgs particle masses, using Infra-Red Quasi-Fixed Points 
in section 2 and the so-called Multiple Point Principle in 
section 3. Mass matrix ans\"{a}tze with texture zeros are considered in 
section 4. 
Finally, the neutrino mass problem is briefly discussed in section 6.

\subsection{Top and Higgs Masses from Infra-red Fixed Point}

The idea that some of the SM mass parameters might be determined as 
infra-red fixed point values of renormalisation 
group equations (RGEs) was first considered \cite{fn} 
some time ago. It was pointed out that the three generation 
fermion mass hierarchy does not naturally arise out of the 
general structure of the RGEs, although it does seem possible 
in special circumstances \cite{abelking}. However it was soon 
\cite{pendleton} realised 
that the top quark mass might correspond to a fixed point value 
or more likely a quasi-fixed point \cite{hill} at the scale 
$\mu = m_t$. 

The SM quasi-fixed point prediction of the top quark mass 
is based on two assumptions: 
(a) the perturbative SM is valid up to 
some high (e.~g.~GUT or Planck) energy scale 
$M \simeq 10^{15} - 10^{19}$ GeV, and 
(b) the top Yukawa coupling 
constant is large at the high scale $g_{t}(M) \gapprox 1$.  
Neglecting the lighter quark masses and mixings, which is 
a good approxmation, the SM one loop RGE for the top quark running
Yukawa coupling $g_t(\mu)$ is:
\begin{equation}
16\pi^2\frac{dg_t}{d\ln\mu} = g_t\left(\frac{9}{2}g_t^2 - 8g_3^2 
- \frac{9}{4}g_2^2 - \frac{17}{12}g_1^2\right)
\end{equation}
Here the $g_i(\mu)$ are the three SM running gauge coupling constants.
The nonlinearity of the RGEs 
then strongly focuses $g_{t}(\mu)$ at the 
electroweak scale to its quasi-fixed point value. 
The RGE for the Higgs self-coupling $\lambda(\mu)$
\begin{equation}
16\pi^2\frac{d\lambda}{d\ln\mu} =12\lambda^2 + 
3\left(4g_t^2 - 3g_2^2 - g_1^2\right)\lambda + 
\frac{9}{4}g_2^4 + \frac{3}{2}g_2^2g_1^2 + \frac{3}{4}g_1^4 - 12g_t^4
\label{rgelam}
\end{equation} 
similarly focuses $\lambda(\mu)$ towards a
quasi-fixed point value, leading to the SM fixed point predictions \cite{hill} 
for the running top quark and Higgs masses:
\begin{equation}
m_{t} \simeq 225\ \mbox{GeV} \quad m_{H} \simeq 250\ \mbox{GeV}
\end{equation}
Unfortunately these predictions are inconsistent with the experimental
running top mass \mbox{$m_{t} \simeq 165 \pm 6$ GeV}. 

The corresponding Minimal Supersymmetric Standard Model (MSSM)
quasi-fixed point prediction for the  
running top quark mass is 
\cite{barger}:
\begin{equation}
m_{t}(m_{t}) \simeq (190\ \mbox{GeV})\sin\beta
\label{mssmfp}
\end{equation} 
which is remarkably close to the experimental value for  
\mbox{$\tan\beta = 2 \pm 0.5$}.
Some of the soft SUSY breaking parameters are also attracted to 
quasi-fixed point values \cite{carena}. For example the trilinear stop 
coupling $A_t(m_t) \rightarrow -0.59 m_{gluino}$. For this low 
$\tan \beta$ fixed point, there is an 
upper limit on the lightest Higgs boson mass:
$m_{h_0} \lsim 100$ GeV. There is also a high $\tan \beta = 60 \pm 5$ 
fixed point solution \cite{fkm}, corresponding to large Yukawa coupling 
constants for the $b$ quark and $\tau$ lepton as well as for 
the $t$ quark, sometimes referred to as the Yukawa Unification scenario.
In this case the lightest Higgs boson mass is $m_{h_0} \simeq 120$ GeV. 
The origin of the large value of $\tan \beta$ is of course a puzzle 
and also SUSY radiative corrections to $m_b$ are then generically large.

\subsection{Top and Higgs Masses from Multiple Point Principle}

According to the Multiple Point Principle (MPP) \cite{glasgowbrioni},
Nature chooses coupling constant values such that a number of 
vacuum states have the same energy density. This principle was 
first used in the Anti-Grand Unification Model 
(AGUT) \cite{bennett,book}, 
as a way of calculating the values of the SM gauge coupling constants. 
In the Euclidean (imaginary time) formulation, the theory has a phase 
transition with the phases corresponding to the degenerate vacua. 
The coupling constants then become dynamical, in much the same way 
as in baby-universe theory, and take on fine-tuned values 
determined by the multiple point. This fine-tuning of the 
coupling constants is similar to that of temperature in a 
microcanonical ensemble, such as a mixture of ice and water 
in a thermally isolated container.   

Here we apply the MPP  
to the pure SM, which we assume valid up to the Planck scale. 
This implies \cite{smtop} that the effective SM
Higgs potential $V_{eff}(|\phi|)$
should have a second minimum
degenerate with the well-known first
minimum at the electroweak scale 
$\langle |\phi_{vac\; 1}| \rangle = 174$ GeV.
Thus we predict that our vacuum is barely stable and we
just lie on the vacuum stability curve in the top quark, Higgs
particle mass ($M_t$, $M_H$) plane.
Furthermore we expect the second minimum to be within an 
order of magnitude or so of the fundamental scale, 
i.e. $\langle |\phi_{vac\; 2}| \rangle \simeq M_{Planck}$.
In this way, we essentially select a particular point on 
the SM vacuum stability curve and hence the MPP condition 
predicts precise values for the pole masses \cite{smtop}:
\begin{equation}
M_{t} = 173 \pm 5\ \mbox{GeV} \quad M_{H} = 135 \pm 9\ \mbox{GeV}
\end{equation} 

\subsection{Mass Matrix Texture and Ans\"{a}tze}

By imposing symmetries and texture zeros on the fermion 
mass matrices, it is possible to obtain testable relations 
between the masses and mixing angles. 
The best known ansatz for the quark mass matrices 
is due to Fritzsch \cite{fritzsch}:
\begin{equation}
M_U =\pmatrix{0                 & C             & 0\cr
                      C                 & 0             & B\cr
                      0                 & B             & A\cr}
\qquad
M_D =\pmatrix{0                 & C^\prime  & 0\cr
                      C^\prime  & 0             & B^\prime\cr
                      0                 & B^\prime  & A^\prime\cr}
\end{equation}
It contains 6 complex parameters A, B, C, $A^\prime$, $B^\prime$ and
$C^\prime$, where it is necessary to {\em assume}:
\begin{equation}
|A| \gg |B| \gg |C|, \qquad |A^\prime| \gg |B^\prime| \gg |C^\prime|
\end{equation}
in order to obtain a good fermion mass hierarchy.
Four of the phases can be rotated away by redefining the phases of
the quark fields, leaving just 8 real parameters (the magnitudes of 
A, B, C, $A^\prime$, $B^\prime$ and $C^\prime$ and two phases 
$\phi_{1}$ and $\phi_{2}$) to reproduce 6 quark masses and 
4 angles parameterising $V_{CKM}$. There are thus two relationships predicted 
by the Fritzsch ansatz:
\begin{equation}
|V_{us}| \simeq 
\left| \sqrt{\frac{m_{d}}{m_{s}}} - 
e^{-i\phi_{1}}\sqrt{\frac{m_{u}}{m_{c}}} \right|,
\qquad \qquad |V_{cb}| \simeq 
\left| \sqrt{\frac{m_{s}}{m_{b}}} - 
e^{-i\phi_{2}}\sqrt{\frac{m_{c}}{m_{t}}} \right|
\label{fritzsch1}
\end{equation} 
The first prediction is a generalised version of the relation 
$\theta_c\simeq\sqrt{\frac{m_d}{m_s}}$ for the Cabibbo angle, 
which originally motivated the two generation Fritzsch ansatz 
and is well satisfied experimentally. However the second relationship
cannot be satisfied with a heavy top quark mass  
\mbox{$m_{t} > 100$ GeV} and 
the original three generation Fritzsch ansatz is excluded by the data.
Consistency with experiment can, for example, be restored by 
introducing a non-zero 22 mass matrix element. A systematic analysis of 
symmetric quark mass matrices with 5 or 6 texture zeros at the 
the SUSY-GUT scale $M_X$ yields five solutions \cite{rrr}. 
An example, in which the 
non-zero elements are expressed in terms of a small parameter 
$\epsilon =\sqrt{\frac{m_c}{m_t}} = 0.058$, is described in the 
Stech's talk \cite{stech}.

The minimal SU(5) SUSY-GUT relation (using a 
Higgs field in the {\bf 5} representation) for the third generation, 
$m_b(M_X) = m_{\tau}(M_X)$, is successful. However it cannot be extended 
to the first two generations as it predicts $m_d/m_s = m_e/m_{\mu}$, 
which fails phenomenologically by an order of magnitude.  
This led Georgi and Jarlskog \cite{georgijarlskog} 
to introduce an ad-hoc coupling 
of the second generation to a Higgs field in the ${\bf \overline{45}}$
representation, giving  
mass matrices with the following texture:
\begin{equation}
M_U =\pmatrix{0                  & C                            & 0\cr
                      C                  & 0                            & B\cr
                      0                  & B                            & A\cr}
\quad
M_D =\pmatrix{0                  & F            & 0\cr
                      F^{\prime} & E                                   & 0\cr
                      0                  & 0                            & D\cr}
\quad
M_E =\pmatrix{0                 & F                             & 0\cr
                      F^{\prime}                & -3E                   
        & 0\cr
                      0                 & 0                             & D\cr}
\label{eq:dhransatz}
\end{equation}
and the successful mass relation $m_d/m_s = 9 m_e/m_{\mu}$. 
 This ansatz has been developed further in the context 
of an SO(10) SUSY-GUT effective operator analysis \cite{anderson}
to give a good fit to all the masses and mixing angles.

\subsection{Mass Hierarchy from Chiral Flavour Charges}

As we pointed out in section 1, a natural resolution to the charged fermion 
mass problem is to postulate the existence of some 
approximately conserved chiral charges beyond the SM. 
These charges, which we assume to be the gauge quantum numbers 
in the fundamental theory beyond the SM, provide selection 
rules forbidding the transitions 
between the various left-handed and right-handed quark-lepton  
states, except for the top quark. In order to generate mass 
terms for the other fermion states, 
we have to introduce new Higgs fields, which break the 
fundamental gauge symmetry group $G$ down to the SM group.
We also need suitable intermediate fermion states to 
mediate the forbidden transitions, which we take to be 
vector-like Dirac fermions with a mass of order the 
fundamental scale $M_F$ of the theory. In this way 
effective SM Yukawa coupling constants are generated, which 
are suppressed by the appropriate product of Higgs field 
vacuum expectation values measured in units of $M_F$.

Consider, for example, an $SMG \times U(1)_f$ model, 
obtained by extending the SM gauge group
$SMG = S(U(3) \times U(2)) \simeq SU(3) \times SU(2) \times U(1)$ 
with a gauged abelian flavour group $U(1)_f$. 
$SMG \times U(1)_f$ is broken to SMG 
by the VEV of a scalar field $\phi_S$ where
$\langle\phi_S\rangle <  M_F$ and $\phi_S$ 
carries  $U(1)_f$ charge $Q_f(\phi_S)$ = 1.  
Suppose further that $Q_f(\phi_{WS})=0$, $Q_f(b_L)=0$ and $Q_f(b_R)=2$.  Then
it is natural to expect the generation of a $b$ mass of order:
\begin{equation}
\left( \frac{\langle\phi_S\rangle }{M_F} \right)^2\langle\phi_{WS}\rangle 
\end{equation}
via the exchange of two $\langle\phi_S\rangle$ tadpoles,
in addition to the usual
$\langle\phi_{WS}\rangle$ tadpole, 
through two appropriately charged vector-like
fermion intermediate states \cite{fn}.
We identify 
$\epsilon_f=\langle\phi_S\rangle/M_F$
as the $U(1)_f$ flavour symmetry breaking parameter. 
In general we expect mass matrix elements of the form:
\begin{equation}
M(i,j) = \gamma_{ij} \epsilon_{f}^{n_{ij}}\langle\phi_{WS}\rangle,
\quad \gamma_{ij} = {\cal O} (1),
\quad n_{ij}= \mid Q_f(\psi_{L_{i}}) - Q_f(\psi_{R_{j}})\mid
\label{eq:mij}
\end{equation}
between the left- and right-handed fermion components. 
So the {\em effective\/}
SM Yukawa couplings of the quarks and leptons to the 
Weinberg-Salam Higgs field 
$y_{ij} = \gamma_{ij}\epsilon_{f}^{n_{ij}}$
can consequently be small even though all 
{\em fundamental\/} Yukawa couplings of
the ``true'' underlying theory are of $\cal O$(1). 
However it appears \cite{bijnens} not possible to explain the 
fermion mass spectrum with an anomaly free set of flavour charges 
in an $SMG \times U(1)_f$ model. It is necessary to introduce 
SMG-singlet fermions with non-zero $U(1)_f$ charge to 
cancel the $U(1)_f^3$ gauge anomaly (as  
in $MSSM \times U(1)_f$ models which also use 
anomaly cancellation via the Green-Schwarz mechanism \cite{ibanezross})
or by extending the SM gauge group further (as in the 
AGUT model \cite{fns}
based on the $SMG^3 \times U(1)_f$ gauge group).

\subsection{Neutrino Mass and Mixing Problem}

Physics beyond the SM 
can generate an effective light neutrino mass term
\begin{equation}
{\cal L}_{\nu-mass} = \sum_{i, j} \psi_{i\alpha}
\psi_{j\beta} \epsilon^{\alpha \beta} (M_{\nu})_{ij}
\end{equation}
in the Lagrangian, where $\psi_{i, j}$ are the Weyl spinors
of flavour $i$ and $j$, and $\alpha, \beta = 1, 2$.
Fermi-Dirac statistics mean that the mass matrix $M_{\nu}$
must be symmetric.
In models with chiral flavour symmetry we typically expect the elements
of the mass matrices to have different orders of magnitude. The charged
lepton matrix is then expected to give only a small contribution
to the lepton mixing. As a result of the symmetry of the neutrino mass
matrix and the hierarchy of the mass matrix elements it is natural
to have an almost degenerate pair of neutrinos, with
nearly maximal mixing\cite{degneut}. This occurs when an off-diagonal
element dominates the mass matrix.

The recent Super-Kamiokande data on the atmospheric neutrino anomaly
strongly suggests large $\nu_{\mu}-\nu_{\tau}$ mixing
with a mass squared difference of order 
$\Delta m^2_{\nu_{\mu} \nu_{\tau}} \sim 10^{-3}$ eV$^2$.
Large $\nu_{\mu} - \nu_{\tau}$ mixing is given by the mass matrix
\begin{equation}
M_{\nu} =
\left(
\begin{array}{ccc}\times & \times & \times \\
\times & \times & A \\
\times & A & \times \end{array}\right )
\label{Mnu1}
\end{equation}
where $\times$ denotes small elements and we have
$\Delta m^2_{23} \ll \Delta m^2_{12} \sim \Delta m^2_{13}$, 
$\sin^2 \theta_{23} \sim 1$
However, this hierarchy in $\Delta m^2$'s is inconsistent with
the small angle (MSW) solution to the solar neutrino problem,
which requires $\Delta m^2_{12} \sim 10^{-5} \ \mbox{eV}^2$.
Also the theoretically attractive solution \cite{fgn} of the 
atmospheric and solar neutrino problems, using maximal 
$\nu_e - \nu_{\mu}$ mixing, seems to be ruled out by the zenith 
angular distribution of the Super-Kamiokande data. 

Hence we need extra structure for the mass matrix such as having
several elements of the same order of magnitude. For example:
\begin{equation}
M_{\nu} =
\left(
\begin{array}{ccc}a & A & B\\
A & \times & \times \\
B & \times & \times \end{array}\right )
\label{Mnu2}
\end{equation}
with $A \sim B \gg a$.
This gives
\begin{equation}
\frac{\Delta m^2_{12}}{\Delta m^2_{23}} \sim \frac{a}{\sqrt{A^2 + B^2}}.
\label{m2rat}
\end{equation}
The mixing is between all three flavours and is given by
the mixing matrix
\begin{equation}
U_{\nu} \sim \left( \begin{array}{ccc}
\frac{1}{\sqrt{2}} & -\frac{1}{\sqrt{2}} & 0\\
\frac{1}{\sqrt{2}} \cos \theta & \frac{1}{\sqrt{2}} \cos \theta &
        -\sin \theta\\
\frac{1}{\sqrt{2}} \sin \theta & \frac{1}{\sqrt{2}} \sin \theta &
        \cos \theta\\
\end{array} \right)
\end{equation}
where $\theta = \tan^{-1} \frac{B}{A}$.
So we have large $\nu_{\mu} - \nu_{\tau}$ mixing with
$\Delta m^2 = \Delta m^2_{23}$, and nearly maximal electron neutrino
mixing with $\Delta m^2 = \Delta m^2_{12}$.
The atmospheric neutrino anomaly requires $\sin^2 2 \theta \gapprox 0.7$ 
or $1/2 \lsim B/A \lsim 2$. The solar neutrino problem is 
explained by vacuum oscillations, although whether it is an 
`energy independent' or a `just-so' solution depends on the 
value of the mass squared difference ratio in eq.~(\ref{m2rat}).

There is also some difficulty in obtaining the required
mass scale for the neutrinos. In models such as the AGUT the neutrino
masses are generated via super-heavy intermediate fermions in
a see-saw type mechanism. This leads to too small neutrino masses:
\begin{equation} m_{\nu} \lsim \frac{\langle{\phi_{WS}} \rangle^2}{M_F}
\sim 10^{-5}\ \mbox{eV},
\end{equation}
for $M_F = M_{Planck}$ (in general $m_{\nu}$ is also supressed 
by the chiral charges). So we need to introduce a new mass scale 
into the theory.
Either some intermediate particles with
mass $M_F \lsim 10^{15}\ \mbox{GeV}$, or an $SU(2)$ triplet
Higgs field $\Delta$ with
$\langle \Delta^0 \rangle \sim 1$ eV is required.
Without further motivation the introduction of such particles
is {\em ad hoc}.


\subsection{Acknowledgement}

Financial support of INTAS grant INTAS-93-3316-ext 
is gratefully acknowledged.

\newpage
\stepcounter{section}
\setcounter{equation}{0}
\section*{\center Extending the Standard Model by Includind Right-Handed
Neutrinos }
\centerline{\rm ASTRI KLEPPE}
\centerline{\it Dep. of Theoretical Physics,
Box 6730, 113 85 Stockholm, Sweden } 
\vskip 5mm
\begin{abstract}
The structure of the neutrino mass matrices is investigated, in a scheme 
where the minimal 
three-family Standard Model is extended by including right-handed neutrinos.
No assumption is made about the presence of a large mass scale, like in 
the see-saw 
scheme.
By demanding that the neutrino mass matrices have a specific form 
with a "Majorana democratic texture", Majorana mass spectra with three 
massless 
(light) neutrinos and either two or three massive neutrinos, are obtained.
\end{abstract}

\subsection{The Standard Model with two right-handed neutrinos}

In the minimal Standard Model it is assumed that the neutrinos have 
no mass, and no right-handed neutrinos are included in the model. 
Although there is still no conclusive experimental evidence 
for massive neutrinos, there is no good physical reason for excluding 
right-handed neutrinos. 
To introduce right-handed neutrinos is actually the simplest way of extending 
the Standard Model. Whereas for example the addition of a fourth 
standard family does not add any new features to the model,
the introduction of right-handed neutrinos adds structures such as massive 
neutrinos, which could be the answer to the solar neutrino deficit and 
the atmospheric neutrino puzzle. Massive neutrinos are also prime candidates 
for hot dark matter. In such a scheme, we may also get lepton mixing 
and CP-violation in the leptonic sector.

In the Standard Model adding right-handed neutrinos results in a generic 
neutrino mass matrix of the form
\begin{equation}\label{dima}
 \cal{M}=\left(\begin{array}{rcl}
                m_{L}          &   m_{D}\\
                m_{D}          &   m_{R}\nonumber
               \end{array}
         \right)
\end{equation}
Unless the Higgs sector is modified, $m_L$ is zero.
If the lepton number is to be conserved, $m_{L}=m_{R}=0$, and the neutrino is a
four-component Dirac spinor endowed with mass by the standard Higgs mechanism
with one Higgs doublet. If however lepton number conservation is not imposed, 
nonvanishing Majorana mass terms from $m_L$ and/or $m_R$ are allowed.

In the case with one left-handed and one right-handed neutrino and
with $m_{L}=0$, $ \cal{M}$ 
corresponds to two nonvanishing mass eigenvalues. With the the assumption 
$m_{D} \ll m_{R}$, one obtains one very light and one very heavy mass 
value, $m_{D}^{2}/m_{R}$ and  $m_{R}$ correspondingly. This is the 
"standard" see-saw mechanism for generating light neutrino masses. 

We investigate an alternative scheme, namely the possibility 
of obtaining very light neutrino masses by including right-handed neutrinos, 
but without making any mass scale assumptions. 

The simplest case, with three left-handed but only one right-handed 
neutrino included, gives rise to two 
very light and two massive neutrino states \cite{cej1}. We however want a 
situation with three light neutrinos.
Therefore, we consider the case with two right-handed neutrinos added to the 
minimal standard
model with three families and one Higgs doublet. 
In the mass basis of the charged lepton sector the most general form of the 
neutrino mass term is 
\begin{equation}\label{gener}
{\cal{L}}_{(\nu-mass)}=-\frac{1}{2}\bar{N} {\cal{M}} N^{C} +h.c.
\end{equation}
where $N$ contains the neutrino fields, and $\cal{M}$ is the neutrino mass 
matrix
\begin{equation}\label{ssam}
 \cal{M}=\left(\begin{array}{rcl}
                {\large\bf{0}}          &   {\large\bf{A}}\\
                  \tilde{\large\bf{A}}  &   {\large\bf{M}}\nonumber
               \end{array}
         \right)
\end{equation}
Here the (Dirac) matrix ${\large\bf{A}}$ comes from the non-diagonal 
interactions of the left-handed and the right-handed neutrinos with the
Higgs doublet, ${\tilde{\large\bf{A}}}$ is the transpose of ${\large\bf{A}}$, 
and the (Majorana) matrix ${\large\bf{M}}$ corresponds to the self-couplings 
of the right-handed neutrinos. As they are singlets they do not need the 
Higgs to acquire mass.

In the case of two right-handed neutrinos,
\begin{eqnarray}
N={\rm{col}}(\nu''_{1L},\nu''_{2L},\nu''_{3L},\nu'^{C}_{1R},\nu'^{C}_{2R})
\nonumber
\end{eqnarray}
and 
\begin{equation}\label{etti}
 \cal{M}=\left
(\begin{array}{rcl}
0&&0{\hspace{6mm}}0{\hspace{7mm}}a_{1}{\hspace{6mm}}b_{1}\nonumber\\
0&&0{\hspace{6mm}}0{\hspace{7mm}}a_{2}{\hspace{6mm}}b_{2}\\
0&&0{\hspace{6mm}}0{\hspace{7mm}}a_{3}{\hspace{6mm}}b_{3}\\
a_{1}&&a_{2}{\hspace{4mm}}a_{3}{\hspace{5mm}}M_{1}{\hspace{4mm}}0    \\
b_{1}&&b_{2}{\hspace{4mm}}b_{3}{\hspace{7mm}}0{\hspace{5mm}}M_{2}\nonumber
\end{array}
        \right)
\end{equation}
\\
We want three zero and two nonvanishing mass eigenvalues.
The characteristic equation reads
\begin{equation}\label{char}
\lambda[\lambda^{4}-\lambda^{3}trM+\lambda^{2}[detM-tr(\tilde{A}A)]+
\lambda tr(\hat{M}\tilde{A}A)+det(\tilde{A}A)] = 0,
\end{equation}
\\
so there is already one "automatically" vanishing mass eigenvalue, 
only two conditions need therefore to be satisfied
in 
order to have three vanishing 
neutrino mass eigenvalues and a nonvanishing Majorana sector (i.e. 
${det M \neq 0}$), 
viz.\\
\begin{equation}\label{flex}
{det(\tilde{A}A)      =0{\hspace{3mm}} {\rm{and}} 
{\hspace{3mm}}tr(\hat{M}\tilde{A}A)=0}
\end{equation}
where
\begin{equation}
 {\hat{\large\bf{M}}}=\left(\begin{array}{rcl}
                M_{2}&0\\
    0{\hspace{2mm}}  &M_{1}\nonumber
                            \end{array}
                      \right)
\end{equation}
This can be expressed as the conditions
\begin{equation}\label{dex}
\bar{a}=x\bar{c},{\hspace{3mm}} \bar{b}=y\bar{c},
{\hspace{3mm}}{\rm{and}}{\hspace{3mm}}M_{1}y^{2}+M_{2}x^{2}=0
\end{equation}
where $x$, $y$ are real numbers, ${\bar{a}=(a_{1},a_{2},a_{3})}$, 
${\bar{b}=(b_{1},b_{2},b_{3})}$ and ${\bar{c}}$ is a unit 3-vector.

The states with vanishing masses at tree level are expected to 
acquire small radiative masses. These radiative corrections are due to the 
two massive states $\lambda_{\pm}$, and generated by one-loop 
diagrams 
with contributions from both Z and the neutral physical Higgs, as well as from
two-W and two-Z exchange diagrams, since in this model there are flavour 
changing neutral currents.

The mass matrix ${\cal{M}}$ is diagonalized by means of 
a unitary 5x5-matrix ${\bf{U}}$, which may be parametrized in terms of four 
angles, ($\psi,{\hspace {2mm}}\phi,{\hspace {2mm}}\theta,
{\hspace {2mm}}\gamma)$,
such that $\bar{c}=(-\cos\psi,\sin\phi\sin\psi,\cos\phi\sin\psi)$, and
\begin{equation}\label{radi}
\frac{x}{y}=\tan\theta,{\hspace{2mm}}{\rm{whereby}}{\hspace{2mm}}
M_{1}=-\tan^{2}\theta M_{2}
\end{equation}
In terms of these angles, the non-vanishing mass eigenvalues and the mixing
matrix are
\begin{eqnarray}\label{toni}
\lambda_{+}&=&M_{1}\frac{(1-\tan^2\theta)}{\tan^2\theta}
 \frac{ \tan^2\gamma}{(1- \tan^2\gamma)}\nonumber\\
\lambda_{-}&=&M_{1}\frac{(1-\tan^2\theta)}{\tan^2\theta}
 \frac{1}{(1- \tan^{2}\gamma)}
\end{eqnarray}
and
\begin{equation}\label{fred}
 {\bf{U}}=\frac{{\bf{\Upsilon}}} {s_{2\gamma}c_{2\theta}}
                 \left(\begin{array}{rcl}
   s_{2\gamma}c_{2\theta}s_{\psi}&&{\hspace{3mm}}s_{2\gamma}c_{2\theta}s_{\phi}
   c_{\psi}{\hspace{12mm}} s_{2\gamma}c_{2\theta}c_{\phi}c_{\psi}
     {\hspace{15mm}}0{\hspace{31mm}}0\nonumber\\
    0{\hspace{4mm}}&&{\hspace{3mm}}s_{2\gamma}c_{2\theta}c_{\phi}{\hspace{12mm}
    }-s_{2\gamma}c_{2\theta}s_{\phi}{\hspace{18mm}}0{\hspace{31mm}}0\\
  -c_{2\gamma}s_{2\theta}c_{\psi}&&{\hspace{2mm}}c_{2\gamma}s_{2\theta}s_{\phi}
  s_{\psi}{\hspace{10mm}}c_{2\gamma}s_{2\theta}c_{\phi}s_{\psi}{\hspace{15mm}}
  c_{\theta}S {\hspace{23mm}}-s_{\theta}S\\
 -c_{\gamma}c_{\psi}S{\hspace{2mm}}&&{\hspace{2mm}}c_{\gamma}s_{\phi}s_{\psi}S
 {\hspace{14mm}} c_{\gamma}c_{\phi}s_{\psi}S
 {\hspace{9mm}}-s_{\theta}c_{\gamma}(c_{2\theta}+c_{2\gamma}){\hspace{5mm}}
 -c_{\theta}c_{\gamma}(c_{2\theta}-c_{2\gamma})\\
-s_{\gamma} c_{\psi}S{\hspace{2mm}}&&{\hspace{2mm}} s_{\gamma}s_{\phi}s_{\psi}S
{\hspace{15mm}} s_{\gamma}c_{\phi}s_{\psi}S {\hspace{14mm}}
s_{\theta}s_{\gamma}(c_{2\theta}-c_{2\gamma}) {\hspace{9mm}}
c_{\theta}s_{\gamma}(c_{2\gamma}+c_{2\theta})\nonumber
                       \end{array}
                 \right),
\end{equation}
\\
correspondingly. Here
$S=\sqrt{c^{2}_{2\theta}-c^{2}_{2\gamma}}$,{\hspace{3mm}}and
${\bf{\Upsilon}}$ is a phase matrix which was
introduced in order to make all mass eigenvalues positive; a reminder 
of the fact that the neutrino and the antineutrino have opposite $CP$ 
parities.
With this mixing matrix, the neutral current term takes the form
\begin{equation}\label{ncur}
{\cal{L}}_{NC}=\frac{g}{2\cos\theta_W}(\bar{\nu}_{1L},\bar{\nu}_{2L},\bar{\nu}_
{
3L},\bar{\nu}_{+L},\bar{\nu}_{-L}) \gamma^{\lambda}{\bf{\Omega}}
\left(\begin{array}{rcl}
                    \nu_{1L}\nonumber\\
                    \nu_{2L}\nonumber\\
                    \nu_{3L}\nonumber\\
                    \nu_{+L}\nonumber\\
                    \nu_{-L}\nonumber\\
                             \end{array}
                      \right)Z_{\lambda}
\end{equation}
where $\bf{\Omega}$ is the matrix
${\bf{\Omega}}={\bf{U}} diag(1,1,1,0,0){\bf{U}}^{\dagger}$, which is but
a unitary transformation of the matrix $diag(1,1,1,0,0)$.
The trace of ${\bf{\Omega}}$ is therefore $tr({\bf{\Omega}})=3$, and from 
${\bf{\Omega}}{\bf{\Omega}}^{\dagger}={\bf{\Omega}}^{2}={\bf{\Omega}}$,
we see that
$tr({\bf{\Omega}}{\bf{\Omega}}^{\dagger})=tr({\bf{\Omega}})=3$.
The neutral current coupling coefficients thus satisfy
\begin{equation}\label{jawa}
\displaystyle\sum_{j,k=1}^{5}|{\bf{\Omega}}_{jk}|^{2}=3,
\end{equation}
where the right-hand side is just the number of left-handed leptonic 
doublets, i.e the number of families.

The invisible width of the Z is determined from studies of Z-production
in $e^+e^-$ collisions \cite{LEP}, by subtracting the measured visible
partial widths, corresponding to Z decays into quarks and leptons, from
the total Z width. 
In our scenario the invisible width of the Z's is
always smaller than predicted by the minimal Standard Model.
The reason for this is that in a model with $n$ left-handed lepton 
doublets
and $k-n$ right-handed neutrinos, the effective number 
$<N_{\nu}>$ of neutrinos is defined by the invisible Z width, i.e.
\begin{eqnarray} 
\Gamma(\rm{Z} \rightarrow \nu's)=\Gamma_{0}<N_{\nu}>=
\Gamma_{0}\sum_{i,j=1}^kX_{ij}|\Omega_{ij}|^{2}
\end{eqnarray}
where $\Gamma_{0}$ is the standard width for a massless neutrino pair
and the $X_{ij}$ are
the phase space and matrix element suppression factors due to the nonvanishing
neutrino masses.
Now, as the $X_{ij}'s$ are bounded by unity, 
$\Gamma(\rm{Z} \rightarrow \bar{\nu}\nu)\leq n\Gamma_{0}$, and in our case,
$n$ = 3.
This means that in a scheme with right-handed neutrinos, no definite conclusion
can be drawn from neutrino-counting at the $Z$-peak.

In neutrinoless double-beta ($\beta \beta)_{0\nu}$ decay, when the neutrino 
masses are very small, we can define
the effective neutrino mass $<m_{\nu}>$ as
\begin{equation}
<m_{\nu}>=|\displaystyle\sum_{j}m_{\nu_{j}}U^{2}_{\nu_{j} e}|,
\end{equation}
In our model, $<m_{\nu}>=0$, and
the current experimental limit \cite{bbd} from the research for 
($\beta \beta)_{0\nu}$ decay is
$|<m_{\nu}>|\stackrel{\large{<}}{\sim} 1-2 eV$. 
Similar results hold for double-muon and double-tau decays.

\subsection{The Standard Model with three right-handed neutrinos}
The case of three right-handed neutrinos added to the Standard Model
is analogous to the case with two right-handed neutrinos. 
Like in the case with two neutrinos, the mass matrix has the 
form ($\ref{ssam})$, but now the matrices $\bf{M}$ and $\bf{A}$ are 3x3. 
In order to get three non-vanishing and three vanishing 
neutrino masses (at tree level), it is necessary that 
$det(\tilde{A}A)=0,{\hspace{2mm}}
tr(M\hat{A}\tilde{\hat{A}})=0{\hspace{2mm}}{\rm{and}}{\hspace{2mm}}
tr({\hat{A}}{\tilde{\hat{A}}})-tr(\hat{M}\tilde{A}A)=0$, where
\begin{eqnarray}
\hat{A}_{ j}=\frac{1}{2}\epsilon_{jkl}\epsilon_{\beta\gamma}
A_{k\beta}A_{l\gamma}\nonumber\\
\hat{M}_{ j}=\frac{1}{2}\epsilon_{jkl}\epsilon_{\beta\gamma}
M_{k\beta}M_{l\gamma},\nonumber
\end{eqnarray}
In analogy with ($\ref{dex}$), this can be obtained by demanding that
\begin{eqnarray}\label{iao.rels}
&&(a_{1},a_{2},a_{3})=x\bar{c},{\hspace{3mm}} (b_{1},b_{2},b_{3})=y\bar{c},
{\hspace{3mm}}(d_{1},d_{2},d_{3})=z\bar{c}\nonumber\\
&&{\rm{and}}{\hspace{3mm}}
M_{2}M_{3}x^{2}+M_{1}M_{3}y^{2}+M_{1}M_{2}z^{2}=0
\end{eqnarray}
where $x,{\hspace{2mm}}y,{\hspace{2mm}}z$ are real non-zero numbers and 
$\bar{c}$ is a unit vector.
${\bf{A}}$ can then be written as 
\begin{equation}\label{uxii}
{\bf{A}}={\bf{C}}{\bf{N}}{\bf{X}}=
\left
(\begin{array}{rcl}
c_{1}&&0{\hspace{7mm}}0\nonumber\\
    0&&c_{2}{\hspace{6mm}}0\nonumber\\
    0&&0{\hspace{7mm}}c_{3}\nonumber
\end{array}
        \right)
\left
(\begin{array}{rcl}
1&&1{\hspace{7mm}}1\nonumber\\
1&&1{\hspace{7mm}}1\\
1&&1{\hspace{7mm}}1\nonumber
\end{array}
        \right)
\left(\begin{array}{rcl}
x&&0{\hspace{7mm}}0\nonumber\\
0&&y{\hspace{7mm}}0\nonumber\\
0&&0{\hspace{7mm}}z\nonumber
\end{array}
        \right)
\end{equation}
\noindent
The mass matrix can now be written in a form that
displays the $Majorana$ $democratic$ $texture$
\begin{equation}\label{iao.majde}
{\cal{M}}= \left
(\begin{array}{rcl}
{\bf{C}}              &&{\bf{0}}\nonumber\\
{\bf{0}}{\hspace{1mm}}&&{\bf{X}}\nonumber
\end{array}
        \right)
\left
(\begin{array}{rcl}
{\bf{0}}&&{\bf{N}}\nonumber\\
{\bf{N}}&&{\bf{m}}\nonumber
\end{array}
        \right)
\left
(\begin{array}{rcl}
              {\bf{C}}&&{\bf{0}}\nonumber\\
{\bf{0}}{\hspace{1mm}}&&{\bf{X}}\nonumber
\end{array}
        \right),
\end{equation}
\\
where ${\bf{0}}$ is a 3x3 matrix where all the matrix elements are zero,
and ${\bf{N}}$, ${\bf{C}}$ and ${\bf{X}}$ are defined by (\ref{uxii}), and
\begin{eqnarray}
{\bf{m}}=
\left
(\begin{array}{rcl}
m_1&&0{\hspace{8mm}}0\nonumber\\
  0&&m_2{\hspace{5mm}}0\nonumber\\
  0&&0{\hspace{6mm}}m_3\nonumber
\end{array}
        \right)
{\hspace{7mm}}
\end{eqnarray}
\\
where $m_1$ = $M_1/x^2$, $m_2$ = $M_2/y^2$, 
$m_3$ = $M_3/z^2$ satisfy
$m_1m_2+m_1m_3+m_2m_3=0$.\\
The neutrino mass matrix (\ref{iao.majde}) is a general Dirac-Majorana 
mass matrix. Its form gives rise to three light and three massive 
neutrinos, in a scheme without any assumption about the presence of a 
large mass scale.
It is tempting to speculate that the democratic texture displayed by this
matrix tells us something about the structure of the mass matrices of the 
charged fermion sector. The ansatz that naturally occurs, is
a charged fermion mass matrix of the form
\begin{equation} 
m= X{\bf{N}}Y+\Lambda
\end{equation} 
where $X$ and $Y$ are (diagonal) 3x3 matrices, and $\Lambda$ is a matrix such 
that $\Lambda_{ij}\ll X_iY_j$.

\newpage
\stepcounter{section}
\setcounter{equation}{0}
\section*{\center Multiple Point Principle and Phase Transition in Gauge
Theories }

\centerline{\rm LARISA V. LAPERASHVILI}
\centerline{\it Institute of Theoretical and Experimental Physics,
B.Cheremushkinskay 25, }
\centerline{\it Moscow 117259, Russia }
\vspace{0.5cm}
\centerline{\rm HOLGER BECH NIELSEN}
\centerline{\it The Niels Bohr Institute, Blegdamsvej 17, DK-2100, Copenhagen,
Denmark}
\vspace{5mm}

Standard model unifying QCD with Glashow--Salam--Weinberg electroweak
theory well describes all experimental results known today. Most
efforts to explain the Standard model are devoted to Grand
unification theory (GUT). The precision of the LEP--data allows to
extrapolate three running constants $\alpha_i(\mu)$ of the Standard
model ( $i=1,2,3$ corresponds to U(1), SU(2), SU(3) groups)
to high energies with small errors and we are able to perform consistency
checks of GUTs.

In the Standard model based on the group
\be
SMG=SU(3)_c\otimes SU(2)_L\otimes U(1)  \lb{lar1}
\ee
the usual definitions of the coupling constants are used:
\be
\alpha_1=\frac{5}{3}\frac{\alpha}{\cos^2\theta_{\overline{MS}}},\quad
\alpha_2=\frac{\alpha}{\sin^2\theta_{\ov{MS}}},\quad
\alpha_3\equiv\alpha_S=\frac{g^2_S}{4\pi},    \lb{lar2}
\ee
where $\alpha$ and $\alpha_s$ are the electromagnetic and strong fine
structure constants, respectively. All of these couplings, as well as the weak
angle, are defined here in the Modified minimal substraction scheme
($\ov{MS}$).
Using experimentally given parameters and the renormalization group
equations, it is possible to extrapolate the experimental values of
three inverse running constants $\alpha^{-1}_i(\mu)$ to the Planck
scale:  $\mu_{Pl}=1.22\cdot10^{19}$GeV.

The comparison of the evolutions of the inverses of the running
coupling constants to high energies in the Minimal Standard model
(MSM) (with one Higgs doublet) and in the Minimal Supersymmetric
Standard model (MSSM) (with two Higgs doublets) gives rise to the existence
of the grand unification point at $\mu_{GUT}\sim10^{16}$GeV only in
the case of MSSM (see Ref.\ct{lar1}).
This observation is true for a whole class of GUT's that break to the
Standard model group in one step, and which predict a "grand desert"
between the weak (low) and the grand unification (high) scales. If grand
desert indeed exists, and the supersymmetry is established at future
colliders then we shall eventually be able to use the coupling constant
unification to probe the new physics near the unification and Planck scales.

Scenarios based on the Anti-grand unification theory ($AGUT$) was
developed in Refs.\ct{lar2}-\ct{lar10} as an alternative to GUT's. The
assumption
of AGUT is: the supersymmetry doesn't exist up to the Planck scale. There
is no new physics (new particles, superpartners) up to this scale. This means
that the renormalization group extrapolation of experimentally determined
couplings to the Planck scale is contingent not encountering new particles.

AGUT suggests that at the Planck scale $\mu_{Pl}$, considered as a fundamental
scale, there exists the more fundamental gauge group $G$, containing $N_{gen}$
copies of the Standard model group $SMG$:
\be
G=SMG_1\otimes SMG_2\otimes \ldots\otimes
SMG_{N_{gen}}\equiv(SMG)^{N_{gen}},    \lb{lar4}
\ee
where the integer $N_{gen}$ designates the number of quark and lepton
generations .

SMG by definition is the following factor group:
\be
SMG = S(U(2)\times U(3)) = \frac{U(1)\times SU(2)\times SU(3)}
     {\{{(2\pi, -1^{2\times 2}, e^{i2\pi/3}1^{3\times 3})}^n|n\in
      Z\}}.  \lb{larSMG}
\ee

If $N_{gen}=3$, then the fundamental gauge group G is:
\be
G=(SMG)^3=SMG_1\otimes SMG_2\otimes SMG_3 ,      \lb{lar5}
\ee
or the generalized G:
\be
   G = {(SMG)}^3\otimes U(1)_f     \lb{larGf}
\ee
which follows from the fitting of fermion masses (see Ref.\ct{lar11}).

The group $G = {(SMG)}^3\otimes U(1)_f$ is a maximal gauge transforming
(nontrivially) the 45 Weyl fermions of the Standard model (which it
extends) without unifying any of the irreducible representations of
the group of the latter.

A base of the AGUT is the Multiple Point Principle (MPP) proposed
several years ago by D.L.Bennett and H.B.Nielsen \ct{lar7}-\ct{lar8}.
Another name for the same principle is the "maximally degenerate vacuum
principle" (MDVP).

According to this Principle, Nature seeks a special point -- the multiple
critical  point (MCP) where the group $G$ undergoes spontaneous
breakdown to the diagonal subgroup:
\be
G\to G_{diag.subgr.}=\left\{g, g, g\parallel g\in SMG\right\} \lb{lar6}
\ee
which is identified with the usual (lowenergy) group SMG.

The idea of the MPP has its origin from the lattice investigations of
gauge theories. In particular, Monte Carlo simulations on lattice
of U(1)--, SU(2)-- and  SU(3)-- gauge theories indicate the
existence of a triple critical point. Using theoretical corrections
to the Monte Carlo results on lattice, it is possible to make slightly
more accurate predictions of AGUT for the Standard model fine-structure
constants.

MPP assumes that SM gauge couplings do not unify and predicts their
values at the Planck scale in terms of critical couplings taken from
the lattice gauge theory:
\be
 \alpha_i(M_{Pl}) = \frac{\alpha_i^{crit}}{N_{gen}}
                  = \frac{\alpha_i^{crit}}{3}
                            \lb{lar7a}
\ee
for $i=2,3$ and
\be
 \alpha_1(M_{Pl}) = \frac{\alpha_1^{crit}}{{\frac 12}N_{gen}(N_{gen}+1)}=
                  = \frac{\alpha_1^{crit}}{6}   \lb{lar7b}
\ee
for U(1).

This means that at the Planck scale the fine structure constants
$\alpha_Y\equiv\frac{3}{5}\alpha_1$, $\alpha_2$ and $\alpha_3,$
as chosen by Nature, are just the ones corresponding to the multiple
critical point (MCP) which is a point where all action parameter (coupling)
values meet in the phase diagram of the regularized  Yang-Mills
$(SMG)^3$ - gauge theory. Nature chooses coupling constant values
such that a number of vacuum states have the same energy density -
degenerate vacua. Then all (or at least maximum) number of phases
convene at the Multiple Critical Point and the different vacua are
degenerate.

The extrapolation of the experimental values of the inverses
$\alpha^{-1}_{Y,2,3}(\mu)$ to the Planck scale $\mu_{Pl}$ by the
renormalization group formulas (under the assumption of a "desert"
in doing the extrapolation with one Higgs doublet) gives us the
following result:
\be
\alpha^{-1}_Y(\mu_{Pl})=55.5;\quad\alpha^{-1}_2(\mu_{Pl})=49.5;\quad\alpha
^{-1}_3(\mu_{Pl})=54.        \lb{lar7}
\ee
Using Monte Carlo results on lattice, AGUT predicts
(Refs.\ct{lar6}-\ct{lar8}):
\vspace{0.2cm} \\
\bc {\bf Table 1}\\
\ec
\bc
\begin{tabular}{|p{1.5cm}|p{4cm}|p{5cm}|}
\hline
Group & AGUT predictions & "Experiment" -- the extrapolation
                                of the SM results to the Planck
                                scale\\

\hline $U(1)$ & $\alpha^{-1}_Y(\mu_{MCP})=55\pm6$
&$\alpha^{-1}_Y(\mu_{Pl})\approx55.5$\\

\hline $SU(2)$ &
$\alpha^{-1}_2(\mu_{MCP})=49.5\pm3$
&$\alpha^{-1}_2(\mu_{Pl})\approx49.5$\\

\hline $SU(3)$&
$\alpha^{-1}_3(\mu_{MCP})=57\pm3$
&$\alpha^{-1}_3(\mu_{Pl})\approx54$\\

\hline
\end{tabular}
\ec
\vspace{0.3cm}
For $U(1)$ - gauge lattice theory the authors of Ref.\ct{lar12} have
investigated
the behaviour of the effective fine structure constant in the vicinity of
the critical point and they have obtained:
\be
\alpha_{crit}\approx{0.2}.     \lb{lar9}
\ee

We gave put forward the calculations of the fine
structure constant in $U(1)$ - gauge theory, suggesting that the
modification of the action form might not change too much the
critical value of the effective coupling constant.

The phase transition between the confinement and "Coulomb" phases
in the regularized U(1)-gauge theory was investigated in Ref.\ct{lar13}.
Instead of the lattice hypercubic regularization it was considered
rather new regularization using Wilson loop (nonlocal) action :
\be
S=\int\limits^{\infty}_0d\log(\frac{R}{a})\beta(R)R^{-4}\sum
\limits_{average}
Re Tr exp\left [i \oint_{C(R)} \hat A_{\mu}(x)dx^{\mu}\right]   \lb{lar10}
\ee
in approximation of circular Wilson loops $C(R)$ of radii $R\ge a$.
Here $(\sum\limits_{average})$ denotes the average
over all positions and orientations of the Wilson loops $C(R)$ in
4-dimensional (Euclidean) space.
It was shown:
\be
\alpha_{crit}\approx{0.204},     \lb{lar11}
\ee
in correspondence with Monte Carlo simulation result (\rf{lar9}) on
the lattice.

The further investigations confirm the "universality" of the critical
coupling constants.

In lattice gauge theories monopoles are artifacts of the regularization.
Let us consider a new assumption: \un{monopoles exist physically}.
With aim to confirm the "universality" of the critical coupling constant
in the regularized U(1)--gauge theory with matter we have investigated
quantum electrodynamics with scalar Higgsed monopoles for a phase
transition (Re.\ct{lar14}).

Considering the Lagrangian which describes the interaction of the
Higgsed scalar monopole field $\Phi (x)$ with dual gauge field
${\hat C}_{\mu} = gC_{\mu}$, we have:
\be
 L(x) = -\frac{1}{4g^2}{({\hat G}_{\mu\nu})}^2 +
{|D_{\mu}\Phi|}^2 - U(\Phi), \lb{lar12}
\ee
where
$$
G_{\mu\nu} = \partial_{\mu}C_{\nu} - \partial_{\nu}C_{\mu},
           \quad\quad D_{\mu} = \partial_{\mu} +i{\hat C}_{\mu}
$$
and
\be
    U(\Phi) = - {\mu}^2 {|\Phi|}^2 + \frac{\lambda}{4}{|\Phi|}^4
                                    \lb{lar15}
\ee
are the dual field strength, covariant derivative and Higgs
potential, respectively.

The complex scalar field $\; \Phi = \frac{1}{\sqrt 2}(\phi + i\chi
)\;$ contains the Higgs and Goldstone boson fields $\phi(x)$ and
$\chi(x)$, respectively.

The free energy $F[\Phi_B]$ may be expressed in terms of the functional
integral (in Euclidean space) over the shifted action:
\be
         S = \int d^4x L^{(E)}(x)    \lb{lar17}
\ee
with a shift:
\be
             \Phi(x)\;->\;\Phi_B + \tilde{\Phi}(x)     \lb{lar18}
\ee
where $\Phi_B$ is a background field. We can obtain the effective
potential $V_{eff}=F[\Phi_B,g^2,\mu^2,\lambda]$ for a constant
background field:$\;\Phi_B = \phi_B = const.$ It was calculated in
the one-loop approximation (see Refs.\ct{lar15}):
{\large $$
             V_{eff} = \frac 12{\mu}^2 {\phi_B}^2 + \frac 14
             \lambda {\phi_B}^4
      +\frac{3g^4}{64{\pi}^2}{\phi_B}^4\log{\frac{{\phi_B}^2}{M^2}}
$$ }
\be {\large
           + \frac{1}{64{\pi}^2}{(\mu^2 + 3\lambda{\phi_B}^2)}^2
             \log{\frac{\mu^2+3\lambda{\phi_B}^2}{M^2}} +
             \frac{1}{64{\pi}^2}{(\mu^2 + \lambda{\phi_B}^2)}^2
             \log{\frac{\mu^2+\lambda{\phi_B}^2}{M^2}}.    \lb{lar19}
}
\ee
This effective potential has several minima, the position of which is
dependent of $g^2,\mu^2$ and $\lambda$.

It is easy to see that the first local minimum occurs at $\phi_B = 0$
and corresponds to so-called "symmetric phase" ("Coulomb phase" in
our description). The phase transition from "Coulomb phase" to the
confinement phase occurs if the second local minimum at
$\phi_B = \phi_0$ is degenerate with the first local minimum
at $\phi = 0$. As a result, we have two equations:
\be
        V_{eff}(\phi_0) = 0 ,    \lb{lar20}
\ee
\be
       {V'}_{eff} = \frac{dV_{eff}}{d\phi_B}|_{\phi_B=\phi_0} = 0.
             \lb{lar21}
\ee
The solution of these equations gives us
the following relation:
\be
   g^4 + \frac{16{\pi}^2}{3}\lambda_{run} +
         \frac{10}{3}{\lambda}^2_{run} = 0,                     \lb{lar22}
\ee
where, according to Eq.(\rf{lar19}):
\be
     \lambda_{run} =
\lambda + \frac{1}{16{\pi}^2}( 3g^4\log{\frac{\phi^2} {M^2}} +
    9\lambda^2\log{\frac{\mu^2 + 3\lambda\phi^2}{M^2}} +
     \lambda^2\log{\frac{\mu^2+\lambda\phi^2}{M^2}}).
                \lb{lar23}
\ee

The third requirement ${ V''}_{eff}\ge 0$,
which means the existence of minimum of $V_{eff}$ at the $\phi = \phi_0$,
leads to the following equation at the border of two phases:
\be
      {V''}_{eff}|_{\phi_B=\phi_0} = 0 .       \lb{lar24}
\ee
The solution of this equation, together with Eq.(\rf{lar22}),
gives us the following results:
$$
       \lambda_{run} = - \frac 45{\pi}^2,
$$
\be
      { g^4}_{crit} = \frac{32}{15}{\pi}^4.
                                   \lb{lar26}
\ee
Using Dirac relations:
\be
           eg = 2\pi ,      \quad\quad
            \alpha_e \alpha_m = \frac 14,   \lb{lar27}
\ee
where $e$ is the electric charge,
$\alpha_e \equiv \alpha = e^2/4\pi$  and $\alpha_m = g^2/4\pi$
is the electric and magnetic fine structure constants, respectively,
we can easily obtain (with help of the result (\rf{lar26})):
\be
       \alpha_{crit}\approx{0.22} .   \lb{lar28}
\ee
This value of the critical fine structure constant is in correspondence
with the results of Monte Carlo simulations on lattice
$(\alpha_{crit}\approx{0.20}$ \ct{lar12}) and the
regularized gauge theory using nonlocal Wilson loop action:
$\alpha_{crit}\approx{0.204}$ \ct{lar13}.

Thus, an idea of the "universality" (regularization independence)
of the critical couplings was confirmed in U(1)-- gauge theories.
Such a (maybe approximate) "universality"  of the critical coupling
constants is needed for the fine structure constant predictions
claimed  from AGUT \ct{lar6}-\ct{lar8}.

We were interested also in a question: is it possible to
include gravity for a phase transition at the Planck scale?
The main idea of our new work \ct{lar16} is to include the gravity,
considering the simplest action for scalar or fermion monopoles which
interact with dual gauge fields in presence of the gravity.
In the case of the fermion monopoles we have the following action:
\be
     S=\int d^4x\sqrt{-\tilde g} [\frac{1}{16\pi G}R +
           \frac{1}{4g^2}{({\hat G}_{\mu\nu})}^2 +
           {\bar \psi}(ig^{\mu\nu}\gamma_{\mu}D_{\nu} - M)\psi ].
                                      \lb{lar29}
\ee
Here $g^{\mu\nu}$ is a metric tensor, $\tilde g$ is its determinant
and $D_{\mu}$ is a covariant derivative, containing also the
connection. $G$ is the gravitational constant and $R$ is the scalar
curvature. But now $g^{\mu\nu}$ plays a role of the field variable.
The variation of the action (\rf{lar29}) over $g^{\mu\nu}, \psi$ and
$C_{\mu}$ gives us three equations of motion:  1) the equation for
$g^{\mu\nu}$; 2) the second one for $\psi$ and 3) the third one for
$C_{\mu}$. Using the first equation, we can consider the second order
formalism, excluding the gravitational field. This procedure leads to
the appearance of the 4-fermion term in the resulting effective
action. Such a term has the coupling constant related with the
gravitational constant $G$ and is responsible for the phenomenon
similar to the formation of Cooper pairs in superconductivity.  Why
is it possible?

Let us assume that monopoles have a large mass $M < M_{Pl}$, but
comparable with the Planck mass $M_{Pl}$. At $\mu \ge M$ the
running (electric) fine structure constant $\alpha(\mu)$ would be
also renormalized by monopole loops and increase rapidly.  Then,
according to the relations (\rf{lar27}), $\alpha_m = \frac{1}{4\alpha}$
decreases rapidly and at some
\be
       \alpha_m^{crit} = {(M/M_{Pl})}^2
          \lb{lar31}
\ee
the repulsion of two monopoles (with the same  magnetic charges $g$)
becomes equal to the gravitational attraction between them. But for
$\alpha_m < \alpha_m^{crit}$ the gravitational attraction will be
larger than electromagnetic repulsion of monopoles and the formation
of scalar bound states (with magnetic charge $2g$) is quite
possible. Their condensate is analogous to the Cooper pairs in
superconductivity and leads to the formation of the vortices
(strings). In such a theory the thikness of "strings" plays role
of the regularization parameter.

Maybe this is a way to construct supersymmetric strings and to
consider the violation of the supersymmetry at the Planck scale.

Financial support of INTAS (grants INTAS-93-3316-ext and
INTAS-RFBR-95-0567) is gratefully acknowledged.

\newpage
\stepcounter{section}
\setcounter{equation}{0}
\section*{\center Unification of Spins and Charges Enabels Unification of All
Interactions }
\centerline{\rm NORMA MANKO\v C BOR\v STNIK}
\centerline{\it Department of Physics, University of
Ljubljana, Jadranska 19 } 
\centerline{\rm and}
\centerline{\it J. Stefan Institute, Jamova 39}
\centerline{\it  Ljubljana , 1 111, Slovenia }
\vskip 5mm

\begin{abstract}
In a space of $d $ Grassmann ( anticommuting ) coordinates 
two types of generators of  Lorentz transformations, one 
of spinorial and the other of vectorial character, define the
representations of the group $SO(1,13)$ and of the subgroups   
$SO(1,3), SU(3), SU(2), U(1)$, for fermions and bosons, respectively,
unifying all the internal degrees of freedom - spins and
Yang-Mills charges.  When 
accordingly all the interactions are unified, Yang-Mills fields
appear as a part of a 
gravitational field. The theory suggests that elementary
particles are either in the fermionic representations with
respect to the groups determining the spin and  the charges,
or they are in the  bosonic representations with respect to the
groups, which determine the spin and the charges. 
It also suggests four rather than three generations of quarks and
leptons and says that the left handed weak charge doublets are
in the same 
multiplet as the right handed weak charge singlets.
\end{abstract}

\subsection{Introduction}

Since Newton, the understanding of the laws of Nature has
developed from the laws leaving (almost) infinitely many
parameters free  (all possible masses as well as forces) to be
determined by the experiment, to the unified quantum theory of
electromagnetic, weak and colour interaction, which only has
around 20 parameters.

What today is accepted as elementary particles and fields are either
 fermionic fields with  
 all the charges   in the
fundamental representations with respect to the groups $ 
U(1), SU(2), SU(3) $ or  bosonic fields with all the  charges  in
the adjoint representations with respect to the same
groups. There exists no 
known fermion (yet) with charges in the adjoint
representations and no known boson yet with charges in the
fundamental representations. The not yet observed Higgs scalar,
however, which 
appears in the Standard model as
a weak charge doublet, seems to be of such a "mixed" type (or it
might be a
constituent particle\footnote[1]{We show that Higgs scalar is
described by one of the bosonic representations of the proposed approach.}).

Not only has the Standard electroweak model free parameters, it
also has several assumptions, which have no theoretical
support (not in the Standard model concept): i) There are three
families of quarks and 
leptons, which are massless and which gain masses through the
Yukawa couplings with the Higgs field. ii) Quarks carry
colour ($SU(3)$) charges. If left 
handed they carry weak ($SU(2)$) and $Y(U(1))$ charges, if right
handed they carry only $Y$ charges. The left handed leptons carry weak
and $Y$ charges, right handed ones carry either only $Y$
charges, or no charge at all. iii) The $Y$ charges are chosen in 
a way to reproduce the electromagnetic charges of physical
particles. iv) Charges of fermions are described by either the
fundamental representations of the groups $SU(3), SU(2)$ and
$U(1)$, or they are singlets with respect to some or all of
these groups, charges of bosons are described by
the adjoint representations of these groups, or they are singlets 
with respect to some or all of these groups. v) There exists a
(complex) scalar field with respect to the group $SO(1,3)$,
which is the colour 
singlet and the weak doublet and carry the $Y$ charge.

The Standard model does not say: i) Why $SO(1,3) \times SU(3)
\times SU(2) \times U(1) $ are the imput symmetries of the
model? ii) Why fermions are left handed weak doublets and right
handed weak singlets? iii) Where do the generations of fermions
come from? Why there exist only three generations? iv) Where do
the Yukawa couplings come from? Where does the Higgs come from?

In the Standard model the $Y$ charges are free parameters of the
model. Embedding the charge groups into $SU(5)$ fixes the $Y$
charge uniquely, but leaves the connection between the
handedness (that is spin) and charges undetermined. {\bf To
connect spins and charges, spins and charges should unify.}

In this talk I am proposing the approach which unifies spins and
charges requiring that elementary fermions are in the spinorial
representations with respect to all gauge groups and
elementary bosons are in the vectorial representations with
respect to all gauge groups. It also suggests that fermions,
which are left handed
weak charge doublets appear together with right handed weak
charge singlets in the same  multiplet, offering a
way of understanding why left handed weak charge doublets and
right handed weak charge singlets  appear in the
Standard electroweak model. The Yukawa couplings appear as a
part of spin connections, which also define all gauge fields.

The space in the approach of mine has $d$ commuting (ordinary) and
$d$ anticommuting ( Grassmann ) coordinates.  
{\bf All the internal degrees of freedom, spins and charges, are
described by the generators of the Lorentz transformations} in 
Grassmann space. All gauge fields - gravitational as well as
Yang-Mills - are defined by supervielbeins.

In Grassmann space there are two types of  generators of 
Lorentz transformations and  translations: one is of spinorial
character and determines properties of fermions, the other is of
vectorial character and determines properties of bosons. Both types
of generators are linear differential operators in Grassmann
space. Their representations can be expressed as monomials of
Grassmann 
coordinates $\theta^a$. If $d\geq 14$ the generators of the
subgroup $ SO(1,3) $ of the group $SO(1,13)$ determine spins of
fields, while generators 
of the subgroups $ SU(3), SU(2), U(1) $ determine their charges.

The Lagrange function describing a particle on a supergeodesics,
leads to the momentum of the particle in Grassmann space which is
proportional to the Grassmann coordinate. This brings  the Clifford
algebra and the spinorial degrees of freedom into the theory. 
The supervielbeins, transforming the geodesics from the freely
falling to the external coordinate system, depend on ordinary and
Grassmann coordinates (the later determine spins and charges
of fields) and carry accordingly the bosonic 
and the fermionic degrees of freedom, if expressed in terms of
monomials of a Grassmann even and odd character, respectively .

The Yang-Mills fields appear as the contribution of gravity
through spin connections and not through vielbeins as in the
Kaluza-Klein theories. Because of that and because 
the generators of the Lorentz transformations in Grassmann
space  rather than  in
ordinary space determine charges of fields, the Planck mass of
charged particles as in Kaluza-Klein theories seems not to
appear in this approach.  
The Yukawa couplings may be explained by having the origin in
spin connections as well. In such a case, however, mass terms
are of the order of a Planck mass. ( More about this approach
can be found in Refs.\cite{man1,man2}.) 
\subsection{ Coordinate Grassmann Space and Linear Operators }

In this section we briefly repeat a few definitions concerning
a d-dimensional Grassmann space,  linear Grassmann space 
spanned over the coordinate space, linear operators defined in
this space and the Lie algebra of generators of the Lorentz
transformations \cite{man2,ber}.

We define a d-dimensional Grassmann space of real anticommuting
coordinates $ \{ \theta ^a \} $, $ a=0,1,2,3,5,6,...,d,$
satisfying the anticommutation relations
$ \theta^a \theta^b + \theta^b \theta^a := \{ \theta^a, \theta^b
\} = 0,$
called the Grassmann algebra \cite{man2,ber}. The metric tensor $ \eta
_{ab}$ $ = diag (1,$ $ -1, -1,$ $ -1,..., -1) $ lowers the
indices of a 
vector $\{ \theta^a \} = \{ \theta^0, \theta^1,..., \theta^d \},
\theta_a = \eta_{ab} \theta^b$. Linear transformation actions on
vectors $ (\alpha \theta^a + \beta  x^a),\;\; $
$ (\alpha \acute{\theta}^a + \beta \acute{x}^a ) = L^a{ }_b
(\alpha \theta^b + \beta x^b ), $
which leave forms
$ ( \alpha \theta^a + \beta x^a ) ( \alpha \theta ^b + \beta
x^b ) \eta_{ab} $
invariant, are called the Lorentz transformations. Here $
(\alpha \theta^a + \beta x^a ) $ is a  vector of d
anticommuting components  and d commuting $ (x^ax^b -
x^bx^a = 0) $ components, and $ \alpha $ and $ \beta$ are two
complex numbers. The requirement that forms $ ( \alpha \theta^a
+ \beta x^a ) ( \alpha \theta ^b + \beta 
x^b ) \eta_{ab} $ are scalars with
respect to the above linear transformations, leads to the
equations 
$ L^a{ }_c L^b{ }_d \eta_{ab} = \eta_{cd}.$

A linear space spanned over a Grassmann coordinate space of d
coordinates has the dimension $ 2^d$. If monomials $
\theta^{\alpha_1} \theta^{\alpha_2}....\theta^{\alpha_n} $
are taken as a set of basic vectors with $\alpha_i \neq
\alpha_j,$ half of the vectors have an even (those with an even n)
and half of the vectors have an odd (those with an odd n)
Grassmann character. Any vector in this space may be represented
as a linear superposition of monomials 
\be
f(\theta) = \alpha_0 + \sum_{i=1}^{d}  \alpha _{a_1a_2 ..a_i}
\theta^{a_1} \theta^{a_2}....\theta^{a_i},\;\; a_k< a_{k+1},
\label{2.1}
\ee

where constants $\alpha_0, \alpha_{a_1a_2..a_i}$ are complex
numbers and the ascending order of coefficients $a_1 < a_2 <...<
a_i$ is assumed.

In  Grassmann space the left derivatives have to be
distinguished  from the right derivatives, due to the 
anticommuting nature of the coordinates \cite{man2,ber}. We
shall make use 
of left derivatives $ {\overrightarrow {{\partial}^{\theta}}}{
}_a:= \frac{\overrightarrow {\partial}}{\partial
\theta^a},\;\;\; {\overrightarrow {{\partial}^{\theta}}}{ }^{a}:=
{\eta ^{ab}} \overrightarrow {{\partial}^{\theta}}{ }_b \; $, on
vectors of the 
linear space of monomials $ f(\theta)$,  defined as follows:  
$ {\overrightarrow{{\partial}^{\theta}}}{ }_a\; \theta^b f(\theta)
= \delta^b{ }_a f(\theta) - \theta^b  
{\overrightarrow{{\partial}^{\theta}}}{ }_a\; f(\theta).$

Here  $ \alpha $ is a constant of either commuting $( \alpha
\theta^a - \theta^a \alpha = 0 )$ or anticommuting $( \alpha
\theta^a + \theta^a \alpha = 0 )$ character, and $n_{a
\partial}$ is defined as follows
$ n_{AB} = \left\{ \begin{array} {ll} +1, & if\; A\; and\; B \;
have\;Grassmann\; odd\; character\\
0, & otherwise \end{array} \right\} $.

We define the following linear operators \cite{man1,man2}
\be
p^{\theta} { }_a := -i {\overrightarrow{{\partial}^{\theta}}}{
}_a , \;\;
 \tilde{a} ^a := i(p^{\theta a} - i \theta^a) ,\;\;
\tilde{\tilde{a}}{}^a := -(p^{\theta a} + i \theta^a). 
\label{2.2}
\ee

According to the inner product defined in what follows,
the operators $  \tilde{a} ^a $ and $ \tilde{\tilde{a}}{}^a $
are either hermitian or antihermitian operators.

We define the generalized commutation relations
(which follow from the corresponding Poisson brackets
\cite{man1,man2}): 
\be
\{ A,B \} := AB - (-1) ^{n_{AB}} BA, 
\label{2.3} 
\ee

fulfilling the equation  
$ \{A,B\}  = (-1)^{n_{AB}+1} \{B,A\},$

We find
\be
\{p^{\theta a}, p^{\theta b} \} = 0 = \{ \theta^{a},
\theta^{b}\},\;\; \{p^{\theta a}, \theta^{b}\} = -i \eta^{ab},\;\;
\{\tilde{a}^{a}, \tilde{a}^{b} \} = 2 \eta^{ab} 
= \{\tilde{\tilde{a}}{ }^{a}, \tilde{\tilde{a}}{ }^{b} \},\;\; \{
\tilde{a}^{a}, \tilde{\tilde{a}}{ }^{b} \} = 0. 
\label{2.4}
\ee

We see that $\theta ^a $ and $ p^{\theta a} $ form a Grassmann
odd Heisenberg algebra, while $ \tilde a^a $ and $
\tilde{\tilde{a}}{ }^a $ form the Clifford algebra.

We  define the projectors
\be P_{\pm} = \frac{1}{2} ( 1 \pm  \sqrt{ (-)^{\tilde
\Upsilon \tilde{\tilde 
\Upsilon}}} \tilde{ \Upsilon} \tilde{\tilde \Upsilon}),\;\;\;\;
(P_{\pm})^2 = 
P_{\pm}, 
\label{2.5} 
\ee

where $\tilde \Upsilon$ and $ \tilde{\tilde \Upsilon}$ are the two
operators  defined  for any dimension d as follows\\
$ \tilde \Upsilon = i^{\alpha} \prod_{a=0,1,2,3,5,..,d} \tilde{
a}{ }^a \sqrt{\eta^{aa}},\;\; \tilde{\tilde \Upsilon} =
i^{\alpha} \prod_{a=0,1,2,3,5,..,d} 
\tilde{ \tilde{a}}{ }^a \sqrt{\eta^{aa}},$
with $\alpha $ equal either to $ d/2 $ or to $ (d-1)/2 $ for
an even and odd dimension $d$ of the space, respectively. 
It can be checked that $( \tilde \Upsilon )^2 = 1 = ( \tilde{
\tilde{\Upsilon}} )^2 $.

The projectors $P_{\pm}$ project out of any monomials of
Eq.(\ref{2.1}) the Grassmann odd and the Grassmann even part of the monomial,
respectively.
We find that for odd d the operators $\tilde \Upsilon $ and
$\tilde{ \tilde \Upsilon}$ coincide ( up to $\pm i$ or $ \pm1$ )
with $\tilde 
\Gamma$ and $\tilde{\tilde \Gamma}$ of Eq.(2.8), respectively.

We define two kinds of operators \cite{man2}. The first ones are
binomials of operators forming the Grassmann odd Heisenberg
algebra 
\be
S^{ab} : = ( \theta^a p^{\theta
b} - \theta ^b p^{\theta a} ).
\label{2.6a}
\ee

The second ones are binomials of operators forming the Clifford
algebra 
\be 
\tilde S ^{ab}: = - \frac{i}{4} [\tilde a ^a , \tilde a ^ b
], \;\; \tilde {\tilde S} { }^{ab}: = - \frac{i}{4} [ \tilde
{\tilde a}{ }^a , \tilde {\tilde a}{ }^b ] , 
\label{2.6b}
\ee

with $ [A, B]:= AB - BA.$

Either $ S^{ab} $ or $ \tilde S ^{ab}$  or $ \tilde{\tilde S}{
}^{ab} $ fulfil the Lie algebra of the Lorentz group $ SO(1,d-1)
$ in the d-dimensional Grassmann space:
$ \{ M^{ab}, M^{cd} \} = -i ( M^{ad} \eta^{bc} + M^{bc}
\eta^{ab} - M^{ac} \eta^{bd} - M^{bd} \eta^{ac} ),$
with $ M^{ab} $ equal either to $ S ^{ab} $ or to $\tilde S
^{ab} $ or to $ \tilde {\tilde S} { }^{ab} $ and $ M^{ab} = -
M^{ba} $.

We see that 
\be
S^{ab} = \tilde S ^{ab} + \tilde {\tilde S}{ }^{ab},\;\;
 \{ \tilde S ^{ab} , \tilde {
\tilde S }{ }^{cd} \} = 0 = \{ \tilde S ^{ab} , \tilde {\tilde
a}{ }^c \} = \{ \tilde a ^a , \tilde { \tilde S }{ }^{bc} \}.
\label{2.7}
\ee

By solving the eigenvalue problem (see below) we find that
operators $ \tilde S ^{ab} $, as well as 
the  operators $ \tilde {\tilde S}{ }^{ab} $, define the
fundamental 
or the spinorial representations of the Lorentz group, while $
S^{ab} = \tilde S ^{ab} + \tilde{\tilde S} { }^{ab} $ 
define the  vectorial
representations of the Lorentz group $ SO(1,d-1) $.

Group elements are in any of the three cases defined by:
$ {\cal U}(\omega) = e^{ \frac{i}{2} \omega_{ab} M^{ab}},$
where $ \omega_{ab} $ are the parameters of the group.

Linear transformations, defined above, can then be written
in terms of group elements as follows
$ \acute{\theta}^a = L^a{ }_b \theta ^b = e^{- \frac{i}{2}
\omega_{cd} S^{cd}} \theta ^a e^{\frac{i}{2} \omega _{cd}
S^{cd}}. $

It can be proved for any d that
$ M^2 $ is the invariant of the Lorentz group 
$ \{ M^2, M^{cd} \} = 0,\;\; M^2 =  \frac{1}{2} M^{ab} M_{ab},$ 
and that for d=2n we can find the additional invariant $ \Gamma $
\be
\Gamma = \frac{i(-2i)^{n}
}{(2n)!} 
\epsilon_{a_1a_2...a_{2n}} M^{a_1a_2}
....M^{a_{2n-1}a_{2n}},\;\;\{ \Gamma, M^{cd} \} = 0, 
\label{2.8}
\ee

where $\epsilon _{a_1a_2...a_{2n}} $ is the totally antisymmetric
tensor with $ 2n $ indices and with $ \epsilon _{ 1 2 3 ...2n }
= 1 $. 
This means that $ M^2  $ 
and $ \Gamma $ are for $ d = 2n$ the two invariants or Casimir
operators of the group $ SO(d) $ (or $ SO(1,d-1). $)
For  $ d = 2n + 1 $ the 
second invariant cannot be defined.
( It can be checked that
$\tilde \Upsilon$ and 
$\tilde{\tilde \Upsilon} $ of Eqs.(\ref{2.11}) are the two invariants
for the spinorial case for any d. For  even d they coincide with 
$\tilde \Gamma $ and $\tilde{\tilde \Gamma}$, respectively,
while for odd d 
the eigenvectors of these two operators are superpositions of
Grassmann odd and Grassmann even monomials.)

While the invariant $ M^2 $ is trivial in the 
case when $ M^{ab} $ has spinorial character, since  
 $ (\tilde S^{ab})^2 = \frac{1}{4} \eta^{aa} \eta^{bb} = 
(\tilde {\tilde S}{ }^{ab})^2 $ and therefore $ M^2 $ is equal
in both cases  to the number $ \frac{1}{2} \tilde
S^{ab} \tilde S _{ab} = \frac{1}{2} \tilde {\tilde S}{ }^{ab}
\tilde {\tilde 
S}{ }_{ab} = d ( d-1 ) \frac{1}{8}$ , it is a nontrivial
differential operator in  Grassmann space if $ M^{ab}$
have  vectorial character $(M^{ab} = S^{ab})$. The invariant of
Eq.(\ref{2.8}) is always a nontrivial operator.

We assume that differentials of Grassmann coordinates $ d\theta^a
$ fulfil the Grassmann anticommuting relations \cite{man2,ber}
$ \{ d\theta^a, d\theta^b \} = 0$
and we introduce a single integral over the whole interval of
$d\theta^a $ 
$ \int d\theta^a = 0, \;\; \int d\theta^a \theta^a = 1, a =
0,1,2,3,5,..,d,$
and the multiple integral over d coordinates
$ \int d^d \theta^0 \theta^1 \theta^2 \theta^3
\theta^4...\theta^d = 1,$ with 
$ d^d \theta: = d\theta^d...d\theta^3 d\theta^2 d\theta^1
d\theta^0 $ in the standard way.

We define \cite{man2,ber} the inner product of two vectors $
<\varphi|\theta> $ and $ <\theta|\chi> $, with $
<\varphi|\theta> = <\theta|\varphi>^* $ as follows:
\be
<\varphi|\chi> = \int d^d\theta ( \omega<\varphi|\theta>)
<\theta| \chi>,
\label{2.9} 
\ee

with the weight function 
$\omega = \prod_{k=0,1,2,3,..,d}
(\frac{\partial}{\partial \theta^k}  + \theta^k ),$ 
which operates on the first function $ <\varphi|\theta> $ only, 
and we define
$ (\alpha^{a_1 a_2...a_k} \theta^{a_1}
\theta^{a_2}...\theta^{a_k})^{+} =
(\theta^{a_k}).....(\theta^{a_2}) 
(\theta^{a_1}) (\alpha^{a_1 a_2...a_k})^{*}.$

According to the above definition of the inner product it
follows that  $\tilde a^a{ }^+ = - \eta^{aa} \tilde a^a $ and
$\tilde{\tilde a}{ }^a{ }^+ = - \eta^{aa} \tilde{\tilde a}{ }^a
$, $(\tilde a^a \tilde a^b){ }^+ = - \eta^{aa} \eta^{bb} 
\tilde a^a \tilde a^b $, and $(\tilde{\tilde a}{ }^a
\tilde {\tilde a}{ }^b){ }^+ = - \eta^{aa} \eta^{bb} 
\tilde {\tilde a}{ }^a \tilde {\tilde a}{ }^b. $  The
generators of the Lorentz 
transformations (Eqs.(\ref{2.6b})) are self adjoint ( if $
a \neq 0 $ and  $ b \neq 0 $ ) or antiself adjoint ( if $ a =
0 $ or $ b = 0 $ )  operators.

Either the volume element $ d^d\theta $ or the weight function $
\omega$ are invariants with respect to the Lorentz
transformations ( both are scalar densities of weight - 1).

According to Eqs.(\ref{2.2}) and (\ref{2.6a}) (\ref{2.6b})we find
$$
S^{ab} = -i ( \theta^a \frac{\partial}{\partial \theta_b}
- \theta^b \frac{\partial}{\partial \theta_a} ),\;\; \tilde a^a
= (\frac{\partial}{\partial \theta_a} + 
\theta^a ),\;\; \tilde{\tilde a}{ }^a = i
(\frac{\partial}{\partial \theta_a} - 
\theta^a ),  $$
\be 
\tilde S ^{ab} = \frac{-i}{2}( \frac{\partial}{\partial
\theta_a} + 
\theta^a ) ( \frac{\partial}{\partial \theta_b} +
\theta^b ) ,\;\;  \tilde {\tilde S}{ } ^{ab} = \frac{i}{2}(
\frac{\partial}{\partial \theta_a} - 
\theta^a ) ( \frac{\partial}{\partial \theta_b} - 
\theta^b ) ,\; if a \neq b. \; 
\label{2.10}
\ee

To find  eigenvectors of any operator $A$, we
solve the eigenvalue problem 
\be
<\theta|\tilde A_i|\tilde{\varphi}> = \tilde{\alpha}_i
<\theta|\tilde{\varphi}> ,\;\;
<\theta|A_i|\varphi> = \alpha_i <\theta|\varphi>,\;\;i = \{1,r\}
,
\label{2.11}
\ee

where $ \tilde{A}_i $ and $ A_i $ stand for $r$ commuting
operators 
of spinorial and vectorial character, respectively.

To solve  equations (\ref{2.11}) we express the operators in the
coordinate representation  and write the eigenvectors
as polynomials of $\theta^a$. We  orthonormalize the vectors
according to the inner product, defined in Eq.(\ref{2.9}), 
$ <{ }^a \tilde{\varphi}_i |{ }^b \tilde{\varphi}_j> =
\delta^{ab} \delta_{ij},\;\;\; <{ }^a {\varphi}_i |{ }^b
{\varphi}_j> = \delta^{ab} \delta_{ij},$
where index $a$ distinguishes between vectors of different
irreducible representations and index j between vectors of the
same irreducible representation. This determines the
orthonormalization condition for spinorial and vectorial
representations, respectively.

\subsection{ Lorentz Groups and Subgroups }

 The algebra of the group $ SO(1,d-1) $ {\it or} $ SO(d) $ 
contains \cite{man1}  $ n $ subalgebras defined by 
operators $ \tau ^{A i}, A = 1,n ; i = 1,n_A $,
where $ n_A $ is the number of elements of each
subalgebra, with the properties 
\be [ \tau ^{Ai} , \tau ^{B j} ] = i \delta ^{AB
} f^{A ijk } \tau ^{A k}, 
\label{3.1}
\ee

if operators $ \tau ^{A i} $  can be expressed as
linear superpositions  of operators $ M^{ab} $ 
\be \tau^{A i} = {\it c} ^{A i} { }_{ab} M^{ab}, \;\;
{\it c} ^{A i}{ }_{ab} = - {\it c} ^{A i}{ }_{ba}, \;\;
A=1,n, \;\;
i=1,n_{A}, \;\;a,b=1,d. 
\label{3.1a}
\ee

Here $ f^{A ijk} $ are structure constants
of the ($ A $) subgroup with $ n_{A} $ operators.
According to the three kinds of operators $ M^{ab} $, two of
spinorial and one of vectorial character, there are  three kinds
of operators $ \tau^{A i} $ defining subalgebras of
spinorial and vectorial character, respectively, those of
spinorial types being expressed with either $ \tilde S^{ab} $ or
$ \tilde{ \tilde S}{ }^{ab} $ and those of vectorial type being
expressed by $ S^{ab} $. All three kinds
of operators are, according to Eq.(\ref{3.1}), 
defined by the same coefficients $ {\it c}^{A i} { }_{ab} $
and the same structure constants $ f^{A i j k } $.
>From Eq.(\ref{3.1}) the following relations among constants ${\it
c}^{A i}{ }_{ab} $ follow:
\be
-4 {\it c}^{A i}{ }_{ab} {\it c}^{B j b}{ }_c -
\delta^{A B} f^{A ijk} {\it c}^{A k}{ }_{ac}
= 0. 
\label{3.1b}
\ee
In the case when the algebra and the chosen subalgebras are
isomorphic, that is if the number of  generators of
subalgebras is equal to $ \frac{d(d-1)}{2} $ , the inverse
matrix $ e^{A iab} $ to the matrix of coefficients $
c^{A i}{ }_{ab} $ exists \cite{man1}
$M^{ab} = \sum_{A i} e^{A iab} \tau^{A i},$
with the properties $ c^{A i}{
}_{ab} e^{B jab} = \delta^{A B} \delta^{ij}, \;\;
c^{A i}{ }_{cd} e^{A iab} = \delta^a{ }_c \delta^b{
}_d - \delta^b{ }_c \delta^a{ }_d $.

When we look for coefficients $ c^{A i}{ }_{ab} $ which
express operators $ \tau ^{A i} $, forming a subalgebra
$ SU(n) $ of an algebra $ SO(2n) $ in terms of $ M^{ab} $, the
procedure is rather simple \cite{geor,man2}. We
find: 
\be
\tau^{A m} = -\frac{i}{2} (\tilde \sigma^{A m})_{jk}
 \{ M^{(2j-1) (2k-1)} +
M^{(2j) (2k)} + i M^{(2j) (2k-1)} - i M^{(2j-1) (2k)}
\}.
\label{3.2}
\ee

Here $(\tilde \sigma^{A m})_{jk}$ are the traceless matrices
which form the algebra of $ SU(n) $.
One can easily prove that operators $ \tau^{A m} $ fulfill the
algebra of the group $ SU(n) $  for any of three
choices for operators $ M^{ab} : S^{ab}, \tilde S^{ab},
\tilde{\tilde S}{ }^{ab}$.

In reference \cite{man2}  coefficients ${\it
c}^{A i}{ }_{ab} $ for 
a few cases  interesting for  particle
physics can be found. Of special interest is the group $SO(1,13)$ with
the subgroups  $ SO(1,3)$ and $SO(10) \supset SU(3)\times SU(2)
\times U(1)$ which enables the unification of spins and charges.
While the coefficients are the
same for all three kinds of operators, the representations depend on the
operators $M^{ab}$. After solving the
eigenvalue problem (Eqs.(2.11)) for the invariants of the
subgroups, the representations can be presented as polynomials
of coordinates $\theta^a, a = 0,1,2,3,5,..,14 $. The operators
of spinorial character define the fermionic representations of
the group and the subgroups, while the operators of vectorial
character define the bosonic representations of the groups and
the subgroups.
  

\subsection{ Lagrange Function for Free Particles in Ordinary and Grassmann
Space and Canonical Quantization}

We  present in this section the Lagrange function for a
particle which lives in a d-dimensional ordinary space
of commuting coordinates and in a d-dimensional Grassmann space
of anticommuting coordinates $ X^a \equiv \{ x^a, \theta^a \} $
and has its geodesics parametrized by an ordinary Grassmann even
parameter ($\tau$) and a Grassmann odd parameter($\xi$).  We
derive the Hamilton function and the corresponding Poisson
brackets and perform 
the canonical quantization, which leads to the Dirac equation with
operators, which are differential operators in 
ordinary and in  Grassmann space \cite{man1,man2}.

$ X^a = X^a(x^a,\theta^a,\tau,\xi)$ are called supercoordinates.
We define the dynamics of a particle by choosing the 
action \cite{man1,ikem}
$ I= \frac{1}{2} \int d \tau d \xi E E^i_A \partial_i X^a E^j_B
\partial_j X^b  \eta_{a b} \eta ^{A B},$
where $ \partial _i : = ( \partial _ \tau , {\overrightarrow
{\partial}} _\xi ), \tau^i = (\tau, \xi) $, while $ E^i _A
$ determines a metric on a two dimensional superspace $ \tau ^i
$ , $ E = det( E^i _A )$ . We choose $ \eta _{A A} = 0,   
\eta_{1 2} = 1 = \eta_{2 1} $, while $ \eta_{a b} $ is the
Minkowski metric with the diagonal elements $
(1,-1,-1,-1,$ $...,-1) $. The action is invariant under the Lorentz
transformations of supercoordinates: $X'{ }^a = L^A{ }_b X^b $.
(See Eq.(\ref{2.3})).
Since a supermatrix $ E^i{ }_A $ transforms as a vector in a
two-dimensional superspace $\tau^i$ under general coordinate
transformations of $\tau^i$, $ E^i{ }_A \tau_i $ is invariant
under such transformations and so is $d^2 \tau E$. 
The action is locally supersymmetric.

Taking into account that either $ x^a $ or $ \theta^a $ depend
on an ordinary time parameter $ \tau $ and that $ \xi^2 = 0 $ ,
the geodesics can be described  as a
polynomial of  $ \xi $ as follows: 
$ X^a = x^a + \varepsilon \xi \theta^a $. We choose $
\varepsilon^2 $ 
to be equal either to $ +i $ or to $ -i $ so that it defines two
possible combinations of supercoordinates. Accordingly we
also choose  the metric  $ E^i { }_A $ : $ E^1{ }_1 = 1, E^1{
}_2 = - \varepsilon M, E^2{ }_1 = \xi, E^2{ }_2 = N -
\varepsilon \xi M $, with $ N $ and $ M $  Grassmann even and
odd parameters, respectively. We write $ \dot{A} =
\frac{d}{d\tau}A $, for any $ A $.

If we integrate the above action over the Grassmann odd
coordinate $d\xi$, the action for a superparticle follows:
\be
\int d\tau ( \frac{1}{N} \dot{x}^a \dot{x}_a + \varepsilon^2
\dot{\theta}^a \theta_a - \frac{2\varepsilon^2 M}{N} \dot{x}^a
\theta_a ). 
\label{4.1}
\ee

Defining the two momenta 
\be
p^{\theta }_a : = \frac{ \overrightarrow{\partial} L}
{ \partial {\dot{\theta}^a}} = \epsilon^2 \theta^a,\;\; p_a : =
\frac{\partial L}{ \partial \dot{x}^a} = \frac{2}{N}( 
\dot{x}_a - M p^{\theta a}), 
\label{4.2}
\ee

the two Euler-Lagrange equations follow:
\be
\frac{dp^a}{d \tau} = 0,\;\;\; \frac{dp^{\theta a}}{d \tau} =
\varepsilon ^2 \frac{M}{2} p^a. 
\label{4.3}
\ee

Variation of the action(4.1) with respect to $M $ and $N$ gives
the two constraints
\be
\chi^1: = p^a a^{\theta}_a = 0, \chi^2 = p^a p_a = 0, \;\;
a^{\theta}_a:= i p^{\theta}_a + \varepsilon^2 \theta_a,
\label{4.4}
\ee

while  
$ \chi^3{ }_a: = - p^{\theta }_a + \epsilon^2 \theta_a = 0 $
(Eq.(\ref{4.2})) is the third type of constraints of the action(4.1).
For $\varepsilon^2 = -i$ we find (Eq.(2.2)), that $
a^{\theta}{ }_a = \tilde{a} 
^a,\;\; \chi^3{ }_a = \tilde{\tilde a}_a = 0. $

We find the generators  of the Lorentz transformations for the
action(\ref{4.1}) to be (See also Eq.(\ref{2.6a}) and (\ref{2.6b}))
\be
M^{ a b} = L^{a b} + S^{a b} \;,\; L^{a b} = x^a p^b - x^b p^a
\;,\; S^{a b} = \theta^a p^{ \theta b} - \theta^b p^{ \theta a}
=  \tilde{S} ^{a b} + \tilde{\tilde{S}}{}^{a b},
\label{4.5}
\ee

which show that parameters of the Lorentz transformations are the
same in both spaces.

We define the Hamilton function:
\be
H:= \dot{x}^a p_a + \dot{\theta}{ }^a p^{\theta}{ }_a - L =
\frac{1}{4} N p^a p_a + \frac{1}{2} M p^a (\tilde a_a +
i \tilde{\tilde a }{ }_a) 
\label{4.6}
\ee

and the corresponding Poisson brackets
\be
\{A,B\}_p=  
\frac{ \partial A}{ \partial x^a} \frac{ \partial B}{ \partial
p_a}  - \frac{ \partial A}{ \partial p_a} \frac{ \partial B}{ \partial
x^a} +  \frac{ \overrightarrow{ \partial A}}{\partial \theta
^a} \frac{ \overrightarrow{ \partial B}}{\partial p^\theta_a} +
  \frac{ \overrightarrow{ \partial A}}{\partial p^\theta_a} 
\frac{ \overrightarrow{ \partial B}}{\partial \theta^a}, 
\label{4.7}
\ee 
which have the properties of the generalized commutators \cite{man2}.

If we take into account the constraint $\chi^3{ }_a = \tilde{\tilde
a}{ }_a = 0\;$ in the Hamilton function (which just means that
instead of H the Hamilton function $ H + \sum_i \alpha^i
\chi^i + \sum_a \alpha^3{ }_a \chi^3{ }^a $ is taken, with parameters $
\alpha^i, i=1,2 $ and $ \alpha^3{ }_a = -\frac{M}{2} p_a,
a=0,1,2,3,5,..,d $ 
chosen on such a way that the Poisson brackets
of the three types of constraints with the new Hamilton function
are equal to zero) and in all dynamical quantities, we find:
\be
H = \frac{1}{4} N p^a p_a + \frac{1}{2} M p^a \tilde a_a,\;\;
\chi^1 = p^a p_a = 0,\;\; \chi^2 = p^a \tilde a_a =
0,
\label{4.4a}
\ee
\be
\dot{p}_a = \{ p_a, H \}_P = 0,\;\; \dot{\tilde a}{ }_a = \{ 
\tilde{a}_a, H \}_P = iM p_a,
\label{4.3a}
\ee

which agrees with the Euler- Lagrange equations (\ref{4.3}).

We further find 
\be
\dot {\chi}^i = \{ H, \chi^i \}_P = 0,\;\;i =1,2,\;\;\; \dot
{\chi}^3{ }_a = \{ H, \chi^3{ }_a \}_P = 0,\;\;a =
0,1,2,3,5,..,d,
\label{4.3b}
\ee

which guarantees that the three
constraints will not change with the time parameter $\tau$ and
that $\dot{\tilde M}{ }^{ab} = 0 $, with $ \tilde M { }^{ab} =
L^{ab} + \tilde{S}^{ab}$,  saying that $ \tilde M{ }^{ab} $
is the constant of motion.

The Dirac brackets, which can be obtained from the Poisson
brackets of Eq.(\ref{4.7}) by adding to these brackets on the right
hand side a term $ - \{A, \tilde{\tilde a}^c \}_P \cdot$ $ ( -
\frac{1}{2i} \eta_{ce} ) \cdot $ $ \{ \tilde{\tilde a}{ }^e, B
\}_P $, give  for the dynamical quantities, 
which are observables, the same results as the Poisson brackets.
This is true also for $ \tilde a^a,$ ( $\{ \tilde
a^a, \tilde a^b \}_D = i\eta^{ab} = \{ 
\tilde a^a, \tilde a^b \}_P $),  which is the
dynamical quantity but not  an observable since its odd
Grassmann character causes  supersymmetric
transformations. We also find that $\{ \tilde a^a, \tilde{\tilde
a}{ }^b \}_D 
= 0 = \{ \tilde a^a, \tilde{\tilde a}{ }^b \}_P $ .
The Dirac brackets give  different results only for the quantities
$ \theta^a $ and $ p^{\theta a} $ and  for $\tilde {\tilde
a}{ }^a $ among themselves: $ \{ \theta^a, p^{\theta b}
\}_P = \eta^{ab}, \{ \theta^a, p^{\theta b}
\}_D = \frac{1}{2} \eta^{ab} $, $ \{ \tilde {\tilde a}{ }^a, \tilde
{\tilde a}{ }^b \}_P = 2i \eta^{ab}, \{ \tilde {\tilde a}{ }^a,
\tilde {\tilde a}{ }^b \}_D = 0 $. According to the above  properties
of the Poisson brackets, I suggest that in the quantization
procedure the Poisson brackets (\ref{4.7}) rather than the Dirac
brackets are used, so that variables $\tilde{\tilde a}^a $,
which are removed from all dynamical quantities, stay as
operators. Then $\tilde a^a $ and 
$\tilde{\tilde a}{ }^a $ are expressible with $\theta^a $ and
$p^{\theta a} $ (Eq.(\ref{2.2}) and the algebra of linear operators
introduced in Sect.2  can be used. We shall
show, that  suggested quantization procedure leads to the
Dirac equation, which is the differential equation in ordinary
and Grassmann space and has all desired properties.

In the proposed quantization procedure$ -i \{ A,B \}_p $ goes to
either a commutator or to an anticommutator, according to the
Poisson brackets (\ref{4.7}). The operators $\theta ^a , p^{\theta a}
$ ( in the coordinate representation they become $ \theta^a 
\longrightarrow \theta^a , \; p^{\theta}_a \longrightarrow i 
\frac{\overrightarrow{\partial }}{\partial \theta^a} $) fulfil
the Grassmann odd Heisenberg algebra, while the operators 
$ \; \tilde{a}^a \; $ and $\; \tilde{\tilde{a}}{}^{a}\; $ fulfill
the Clifford algebra.

The constraints (Eqs.(\ref{4.4})) lead to
the Dirac like and  the Klein-Gordon equations
\be
p^a \tilde{a} _a | \tilde{\Psi} > = 0 \;,\; p^a p_a |
\tilde{\Psi}> = 0 , \; {\rm with} \;  p^a \tilde{a}_a p^b \tilde{a}_b =
p^a p_a . 
\label{4.8}
\ee

Trying to solve the eigenvalue problem $ \tilde{\tilde a}{ }^a 
| \tilde {\Psi} > = 0,\;\; a=(0,1,2,3,5,...,d), $ we find that no
solution of this eigenvalue problem exists, which means that
the third constraint $ \tilde{\tilde a}{ }^a = 0 $ can't be
fulfilled in the operator form (although we take it into account
in the operators for all dynamical variables in order that
operator equations would agree with classical equations). We can
only take it into account 
in the  expectation value form 
\be
< \tilde{\Psi} | \tilde{\tilde a}{ }^a | \tilde{\Psi} > = 0.
\label{4.9}
\ee

Since $ \tilde{\tilde a}{ }^a $ are Grassmann odd operators,
they change monomials (Eq.(\ref{2.1})) of an Grassmann odd character
into monomials of an Grassmann even character and opposite,
which is the supersymmetry transformation.
It means that Eq.(\ref{4.9}) is fulfilled for monomials of either odd
or even Grassmann character and that superpositions of the
Grassmann odd and the Grassmann even monomials are not solutions
for this system.

We can use the projector $P_{\pm}$ of Eq.(\ref{2.5}) to project
out of monomials  either the Grassmann odd or the Grassmann even
part. Since this projector commutes with the Hamilton function $
( H = 
\frac{N}{4} p^a p_a + \frac{1}{2} M\; p^a  \tilde
a_a,\;\; \{ P_{\pm}, H \} = 0 ) $, 
 it means that eigenfunctions of $
H $, which fulfil the eq.(\ref{4.9}), have either an odd or an even
Grassmann character. 
In order that in the second quantization procedure  fields
$ | \tilde{\Psi} > $ would describe fermions, it is meaningful
to accept  in the fermion case Grassmann  odd monomials only.

We further see that although the operators $ \tilde{a}^a $ fulfill
Clifford algebra, they cannot be recognized 
as the Dirac $ \tilde{\gamma}^a $ operator, since  having an
odd Grassmann 
character they transform fermions into bosons, which is
not the case with the Dirac $ \gamma^a $ matrices. We
therefore recognize the generators of the Lorentz
transformations $ -2i \tilde{S}^{b m},\; m = 0,1,2,3 $, with $ b = 5
$ as the Dirac $ \gamma ^m $ operators.
$\tilde{\gamma} ^m = - \tilde{a} ^5 \tilde{a} ^m = -2i\tilde{S}
^{5m} \;,\; m=0,1,2,3. $ 
(For another possible choice of the Dirac $ \gamma ^m $ operators
see the contribution to the discussions entitled "Can one
connect the Dirac-Ka\" ahler representation of Dirac spinors and
spinor representations in Grassmann space, proposed by Manko\v
c?" written by Norma and Holger.)

We choose  the Dirac operators $\tilde{\gamma}^a$ in
the way which in the case that  
$ < \tilde{\psi}|p^{h}|\tilde{\psi} > = 0 
$,  for $ h \in \{ 5,d \} $, enables to recognize the equation 
\be
(\tilde{\gamma}{ }^m p_{m}  )|\tilde{\psi}> = 0 \;,\;
m=0,1,2,3. 
\label{4.10}
\ee

as the Dirac equation for a massless particle.  
Since $-2i\tilde{S}^{5m}$ appear as $ \tilde{\gamma}{ }^m, $ 
$ SO(1,4) $ rather than $ SO(1,3)$ is needed 
to describe the spin degrees of freedom of fermionic fields.
It can be checked that  $\tilde{\gamma}{ }^{m} $ fulfill 
the Clifford algebra $\{\tilde{\gamma}{ }^{m} , \tilde{\gamma}{
}^{n}\}  = \eta{^{mn}} $ , while $ \tilde{S}{ }^{mn} = -\frac{i}{4}
\lbrack \tilde{\gamma}{ }^{m},\tilde{\gamma}{ }^{n}\rbrack_{-},
m \in \{0,3\}$. Accordingly also the group $SO(1,14)$ instead of
$SO(1,13)$ is needed to unify spins and charges. (In the Norma
and Holger contribution the $\tilde{\gamma}^a = i \tilde{a}^a
\tilde{\tilde{a}}{ }^0$ (Eq.30) are suggested as the Dirac
$\gamma^a$ operators, having all the needed properties. In this
case the additional coordinate $\theta^5$ is not needed.)

We presented in  Ref. \cite{man2} four Dirac four spinors ( the
polynomials of $ \theta^a $) which fulfill  Eq.(\ref{2.11}).

For large enough d not only do  generators of  Lorentz 
transformations  in  Grassmann
space define 
the spins  of   fields in the four
dimensional subspace,  they also define the electromagnetic,
the weak and the colour charges. 
This is  true, for example,  for $d = 13 $, since $ SO(1,13) $ has
the subalgebra $ SO(1,3)\times SO(10) $, while $ SO(10) $ has
the subalgebra $ SU(3)\times SU(2)\times U(1) $.
In this case  $\tilde{ \tau}^{A i}$ are  linear
superpositions of operators $ \tilde S ^{ab},\; a,b \in \{5,d\}$ 
fulfilling the algebras as presented in Eqs.(3.1-3.3)
and defining the algebras of $ SU(3), SU(2), U(1),$
while $ SO(1,3) $ remains to define the spin degrees of
freedom in the four dimensional subspace. We find the spinorial
representations of the corresponding Casimir operators as
functions of $ \theta^a $ determining weak charge doublets,
colour charge
triplets and electromagnetic charge singlets \cite{man2}.

\subsection{Particles in Gauge Fields}

The dynamics of a point spinning particle in gauge fields, the
gravitational and the Yang-Mills fields, can be obtained by 
transforming in the Lagrangean vectors from a freely falling to
an external 
coordinate system \cite{wess}.  
To do this, supervielbeins ${\bf
e}^{ia}{ } _{\mu} $ have to be 
introduced, which in our case depend on ordinary and on
Grassmann coordinates, as well as on 
two types of parameters $ \tau^i = ( \tau, \xi ) $. Since there
are two kinds of derivatives $ \partial_i $, there are two
kinds of vielbeins \cite{man1,man2}. The index a 
refers to a freely falling coordinate system ( a Lorentz index),
the index $\mu$ refers to an external coordinate system ( an
Einstein index). Vielbeins with a Lorentz index smaller than
five will determine ordinary gravitational fields. Spin
connections appear in the theory as ( a part of) 
Grassmann odd fields. Those with
a Lorentz index higher than four  define Yang-Mills
fields.

We write the transformation of vectors as follows
$ \partial_i X^a= {\bf e}^{ a} { }_{\mu} \partial_i X^{\mu} \;,\;
\partial_i X^{\mu} = {\bf f}^{ \mu} { }_a \partial_i X^a \;,\;
\partial_i $ $= ( \partial_{\tau} , \partial_{\xi} ).$
>From here it follows that
$ {\bf e}^{ a} { }_{\mu} {\bf f}^{ \mu} { }_b = \delta^a { }_b \;,\;  
{\bf f}^{ \mu} { }_{a} {\bf e}^{ a} { }_{\nu} = \delta^{\mu} {
}_{\nu}.$

Again we make a Taylor expansion of vielbeins with respect to 
$ \xi: \;\; $
$ {\bf e}^{a} { }_{\mu} = e^{ a} { }_{\mu} + \varepsilon \xi
\theta^b e^{ a} { }_{ \mu b} \;,\; {\bf f}^{ \mu} { }_a $ $= f^{
\mu} { }_a - \varepsilon \xi \theta^b
f^{ \mu} { }_{a b}.$

Both expansion coefficients  again depend  on ordinary
and on Grassmann coordinates. Having an even Grassmann character,
$\; e^{a} { }_{\mu}$  will describe the spin 2 part of a
gravitational field. The coefficients $ \varepsilon \theta^{b}
e^{a} { }_{\mu b}$ have an odd Grassmann
character ($\varepsilon =-i $, so that
$\tilde{\tilde a} ^a = 0 $). They define the
spin connections \cite{man1,man2}.

It follows that
$   e^{ a} { }_{\mu} f^{ \mu} { }_b = \delta^a { }_b \;,\;  
f^{ \mu} { }_{a} e^{ a} { }_{\nu} = \delta^{\mu} { }_{\nu}
\;,\; e^{ a} { }_{\mu b} f^{ \mu} { }_c = e^{ a} { }_{\mu}
f^{ \mu} { }_{c b}.$

We find the metric tensor ${\bf g}_{\mu \nu} = {\bf e}^{a}
{ }_{\mu} {\bf e}_{a \nu} ,\;
{\bf g}^{\mu \nu} ={\bf f}^{ \mu} { }_{a} {\bf f}^{ \nu a}$. 

We use the notation $e^{a} { }_{\nu,\mu^{x}} = \frac{\partial}{
\partial x{^\mu}} e^{a} { }_{\nu},\;\;
\overrightarrow{e^{a}} { }_{\nu,\mu^{\theta}} =
\frac{\overrightarrow{\partial}}{\partial \theta^{\mu}}
e^{a} { }_{\nu}$ .
Rewriting the action from Sect.4 in terms of an external
coordinate system, using the Taylor expansion of 
supercoordinates $ X^{\mu}$ and superfields $ {\bf e}^{a} {
}_{\mu}$ and
integrating the action over the Grassmann odd parameter $\xi$,
the action
$$ I=\int d\tau \{ \frac{1}{N} g_{\mu \nu} \dot{x}^\mu
\dot{x}^\nu \; - \; \epsilon^2 \frac{ 2 M}{N} \theta_a e^{ a} {
}_{\mu} \dot{x}^\mu \; + \; \varepsilon^2 \frac{1}{2}(
\dot{\theta}^\mu \theta_a -\theta_a \dot{\theta}^\mu) e^{ a} {
}_{\mu} \; + $$ 
\be
+ \;  \varepsilon^2 \frac{1}{2} (\theta^b \theta_a
-\theta_{a} \theta^b ) e^{ a} { }_{  \mu b} \dot{x}^\mu \} ,
\label{5.1}
\ee

defines the two momenta of the system
$ p_{\mu} = \frac{\partial L}{\partial \dot{x}^\mu} = p_{0 \mu} +
 \frac{1}{2} \tilde{S}^{ab} e_{a \mu b} , \;\;
 p^\theta_\mu = -i \theta_a e^{ a} { }_{\mu} = -i
(\theta_\mu + \overrightarrow{e}^{ a} { }_{\nu , \mu_{\theta}}
e^ { }_{a \alpha} \theta^{\nu} \theta^{\alpha}),$
( $\varepsilon^2 = -i $ ).
Here $ p_{0 \mu} $ are the covariant ( canonical) momenta of a
particle. 
For $ p^{\theta}_{a} = p^{\theta}_{\mu} f^{ \mu} { }_{a}$ it follows
that $ p^{\theta}_{a}$ is proportional to $\theta_{a}$. Then $
\tilde{a}_{a} = i 
(p^{\theta}_{a} - i \theta_{a}), 
$ while $ \tilde{\tilde{a}}_{a}= 0 $.  We may further write
\be
p_{ 0 \mu} = p_{ \mu} - \frac{1}{2} \tilde{S}^{a b} e_{a \mu b}
= p_{ \mu} - \frac{1}{2} \tilde{S}^{a b} \omega_{a b \mu} \;,\;
\omega_{a b \mu}=\frac{1}{2} (e_{a \mu b} - e_{b \mu a}),
\label{5.2}
\ee

which is the usual expression for the covariant momenta in
gauge gravitational fields \cite{wess}.
One can find  the two constraints
\be
p_0^\mu p_{0 \mu} = 0 = p_{0 \mu} f^{ \mu} { }_a \tilde{a}^a .
\label{5.3}
\ee

To see  how  Yang-Mills fields enter into the theory,
the Dirac-like equation (\ref{5.3}) has to be rewritten in terms of
components of fields which determine  
gravitation in the four dimensional subspace and of those 
which determine  gravitation in higher dimensions, assuming
that the coordinates of ordinary space with indices higher than
four  stay compacted to  unmeasurable small dimensions. 
Since  Grassmann space manifests itself through  average
values of observables only, compactification of a part of 
Grassmann space has no meaning.  However, since
parameters of  Lorentz transformations in a freely falling
coordinate system for both spaces have
to be the same, no transformations to the fifth or
higher coordinates may occur 
at measurable energies. Therefore, the four dimensional subspace
of  Grassmann space with the generators defining the Lorentz
group $ SO(1,3)$ is (almost) decomposed from the rest of the 
Grassmann space with the generators forming the (compact) group
$ SO(d-4) $, because of the decomposition of  ordinary
space. This is valid on the classical level only.

We shall assume the case in which  only some components
of fields differ from zero: 
\be
\left( \begin{array}{cc|cc}
   e^{m}{ }_{\alpha}& & & 0\\
   & & &\\ \hline
   & & &\\
   0 & & & e^{h}{ }_{\sigma}
\end{array} \right), \;\;\;\;
\alpha,m \in (0,3),\: \sigma,h \in (5,d),\: i \in (1,2), 
\label{5.4}
\ee

while vielbeins $ e^{m}{ }_{\alpha}, e^{k}{ }_{\sigma}$ depend
on $ \theta^a$ and $x^{\alpha}, \alpha \in \{0,3\}$, only. 
Accordingly we have only $ \omega_{ab \alpha} \neq 0$. We
recognize, as in the freely falling coordinate system,
that Grassmann coordinates with indices 
from $ 0 $ to $ 3 $ determine spins of fields, while Grassmann
coordinates with indices higher or equal to $5$ determine
charges of fields. 
We shall take   $
< p^h > = 0 $,  $ a \ge 5 $.
We find
\be
\tilde{ \gamma}^a f^{\mu}{ }_a p_{0 \mu} = \tilde{ \gamma}
^m f^{ \alpha}{ }_m ( p_{ \alpha} - \frac{1}{2} \tilde{S}^{mn}
 \omega_{mn \alpha} + {\it A}_{\alpha} ) , \;\;
{\rm  with}\;\; {\it A}_{\alpha} = \sum_{A,i} \tilde{ \tau}^{Ai} {\it
A}^{Ai}_{\alpha} 
\label{5.5}
\ee

and
$\sum_{A,i} \tilde{ \tau}^{Ai} {\it A}^{Ai}_{ \alpha} = 
\frac{1}{2}  \tilde{S}^{hk}  \omega_{hk \alpha} ,
 \;\;h,k = 5,6,7,8,..d. $


As we already stated in Sect.3 for $d = 14 $, 
 $ SO(1,13) $ has 
the subalgebras $ SO(1,3)\times SO(10) $, while $ SO(10) $ has
the subalgebras $ SU(3)\times SU(2)\times U(1) $.
%
%

Therefore, in Eq.(\ref{5.5}) the fields $ \omega_{hk \alpha} $
determine all 
the Yang-Mills fields, including electromagnetic ones. The
proposed unification differs from the Kaluza-Klein types of
unification, since Yang-Mills fields are not determined by
nondiagonal terms of vielbeins $ e^{h}{ }_{\alpha} $. Instead
they are determined by spin connections and 
it seems that in the proposed theory there is no difficulties
with the 
Planck mass of the electron 
 unless spin
connections or vielbeins are supposed to generate the  Yukawa
nonzero masses of fermions.

Torsion and  curvature follow from the Poisson brackets
$ \{ p_{0a},  p_{0b} \}_p $, with $ p_{0a} = f^{\mu}{ }_{a} ( p_{\mu} -
\frac{1}{2} \tilde{S}^{cd} \omega_{cd\mu}) $. 
We find       
$ \{ p_{0 a} , p_{0 b} \}_p = -\frac{1}{2} S^{c d} R_{ c d a b}
+ p_{0 c} T^c{ }_{a b} ,$
$ R_{c d a b} = f^{ \mu}{ }_{[a} f^{ \nu}{ }_{b]} 
( \omega_{cd\nu,\mu^{x} } + \omega_{c}{ }^e{ }_{ \mu} \omega_{e d \nu}
+ \overrightarrow{\omega}_{c d \mu , f^{\theta}} \theta^e \omega_e
{ }^f{ }_{ \nu}), $
$ T^c{ }_{a b}  = e^{c}{ }_{ \mu} ( f^{ \nu}{ }_{[b} f^{ \mu}
{ }_{ a]}{ }_{,
\nu} + \omega_{e \nu}{ }^d \theta^e f^{ \nu}{ }_{[b}
\overrightarrow{f^{ \mu}}{ }_{  a]}{ }_{, d^\theta} ),$
 with $ \; A _{[a} B _{b]} $  $ = A_a B_b - A_b B_a. $
For $e^{m}{ }_{\alpha} = \delta^m{ }_{\alpha}$ one easily sees
that Eq.(5.9) manifests 
the Dirac equation for a particle with Yang-Mills charges in
external fields.


\subsection{Concluding Remarks }

In this talk the theory in which space has d ordinary and d
Grassmann coordinates was presented. Two kinds of generators of
Lorentz tranformations in Grassmann space can be defined.
The generators of spinorial character define the spinorial
representations of the Lorentz group, the generators of the
vectorial character define the vectorial  representations of the
Lorentz group. Both kinds of generators are the linear
differential operators in Grassmann space. The Lorentz group $
SO(1,d-1)$ contains for $d=13$ as subgroups $ SO(1,3), SU(3),
SU(2) $ and $ U(1) $. While $ SO(1,3)$ defines spins of fermionic
and bosonic fields, define $ SU(3), SU(2)$ and $ U(1)$ charges
of both fields. Charges of fermionic fields belong to the
spinorial representations, while charges of bosonic fields
belong to the vectorial representations.

When looking for the representations of the operators $ \tilde
{S}^{mn},\; m,n\in{0,3} $ as polynomials of $ \theta^a,\; a \in{0,3}$
and operators $\tilde{\tau}^{Ai} $ as polynomials of $ \theta^h,
\;a \in{5,13}$, we find  representations of the group
$SO(1,13)$ as  outer products of the representations of
subgroups. The Grassmann odd polynomials, which are the Dirac
four spinors, are triplets or singlets with respect to the
colour charge, doublets or singlets with respect to the weak
charge and may have hypercharge \cite{man2} equal to $\pm
\frac{1}{6}, \pm 
\frac {1}{3}, \pm \frac {2}{3}, \pm \frac {1}{2}, \pm 1, 0 $.
When looking for the representations of $ SO(1,13)$, as
Grassmann even polynomials of $ \theta^a $, in terms of
the subgroups $ SO(1,3), SU(3), SU(2), U(1) $, we find scalars
and vectors, which are singlets and octets  
with respect to the colour charge, triplets and
singlets with respect to the weak charge and may have the
hypercharge equal to $ 0 $ or to $\pm 1$. We also find scalars,
which are  weak
charge doublets. These representations
are presented in Ref. \cite{man2}. I shall discuss properties of
representations in discussions sections. I shall show that the
theory may offer the answer to some of  open problems of the
Standard electroweak model: The approach suggests four
rather than three families of quarks and leptons, it predicts
the left handed weak charge doublets
together with right handed weak charge singlets in the same
fermionic multiplet.

We presented the Lagrange function for a particle living on a
supergeodesics, with the momentum in the Grassmann space
proportional to the Grassmann coordinate. In the quantization
procedure the Dirac equation follows, with $\gamma^a $
operators, which have the even Grassmann character and are
differential operators in  Grassmann space with coordinates
$\theta^a,\;a \in{0,3} $. When transforming the Lagrange
function from the freely falling to the external coordinate
system, vielbeins and spin connections describe not only the
gravitational field but also the Yang-Mills fields. Since the
generators of the Lorentz transformations with indices higher
than four determine charges of particles and spin connections
again with indices higher than four describe the Yang-Mills
fields (rather than vielbeins with one index smaller than four
and another greater than three as in the Kaluza-Klein theories),
the problem of the Kaluza-Klein theories, which is the 
 Planck mass of
charged particles, 
seems not to occur.

\subsection{Acknowledgement } This work was supported by Ministry of 
Science and Technology of Slovenia.

\newpage
\stepcounter{section}
\setcounter{equation}{0}
\section*{Masses and Mixing Angles and Going beyond the Standard Model}
\centerline{\rm HOLGER B. NIELSEN}
\centerline{\it Niels Bohr Institute, Blegdamsvej 17,} 
\centerline{\it Copenhagen $\phi$, 
Denmark}
\vskip 5mm
\centerline{\rm COLIN D. FROGGATT}
\centerline{\it Department of Physics and Astronomy, 
University of Glasgow,} 
\centerline{\it Glasgow G12 8QQ, UK}
\vskip 5mm
\begin{abstract}
The idea of following Michel and O'Raifeartaigh of assigning 
meaning to the (gauge) \underline{group} and not only the Lie algebra
for a Yang Mills theory is reviewed. Hints from the group and
the fermion spectrum of the Standard Model is used to suggest the
putting forward of our AGUT-model, which gives a very good fit
of the orders of magnitudes of the quark and lepton masses and
the mixing angles including the CP-breaking phase. But for
neutrino oscillations modifications of the model is needed.
Baryogenesis is not in conflict with the model.
Masses and mixing angles and going beyond the Standard Model
\end{abstract}

\subsection{ Introduction}

For the purpose of finding out what comes beyond the Standard 
Model it is unfortunate that the latter works so exceedingly well
that it actualy describes satisfactorily almost all we know and can
make experiments about: Just extending with even classical Einsteinian
gravity is sufficient to provide well working laws of nature for 
all to day practical purposes. So the true hints for going beyond 
the Standard Model can except for pure theoretical estetical arguments
only come from the structure and parameters - which are not yet understood 
inside the Standard Model - of the Standard Model or from the extremely little
information we have about the physics outside the below 1 TeV range 
where so far Standard Model could potentially work perfectly.
The extremely little knowledge we have for the very short distances 
comes from the baryon number being presumably not conserved:
1) If baryon number assymetry should be cosmologically produced at the weak 
scale and not for instance be due to a $B-L$ assymetry from 
earlier time we would need some new physics, and even if it was an
earlier $B-L$ assymetry that caused the observed baryon number 
there would at some scale at least have to be produced the $B-L$ 
or it would have to truly primordial. 
2) The lack of proton decay gives information that e.g. a naive SU(5) GUT
is not correct.
Finally we really do see nonstandard model physivcs in the neutrino
oscillations.

But apart from these tiny bits of information we mainly have the 
structure and coupling constants and masses in the Standard Model 
from which to seek to guess the model beyond!

We ( Svend Erik Rugh et al. ) estimated that the amount of information in 
these parameters as measured so far and in the Standard model structure 
was just around a couple of hundred bits. It could all be written on one line.

What is now the inspiring information on this line?

In section 2 we stress how part of the information about the quantum numbers 
of the quarks and the leptons ( really their Weyl components)
can be packed into saying what \underline{group} rather than only 
what Lie algebra is to be represented.

In section 3 we look at another hint : the large mass ratios 
of the quark and lepton masses in the various generatios, and
the small mixing angles.

With good will these hints could be taken to point in the 
direction of the AGUT gauge group whhich is our favorite model.
AGUT stands for anti-grand unification and is indeed in 
a way to be explained based on assumptions oppsite to the 
ones leading to the usual SU(5) GUT.

In section 4 we put forward the model, especially the
gauge group AGUT which we characterize as largest group
not unifying the irreducible fermion representations of the
Standard Model.
AGUT stands for Anti Grand Unification which is
the name we give to the gauge group  $SMG^3 \times U(1)_f$, where 
$SMG \equiv SU(3) \times SU(2) \times U(1)$.

The Higgs fields 
responsible for breaking the AGUT gauge group 
$SMG^3 \times U(1)_f$ to the diagonal $SMG$ subgroup, 
identified as the SM gauge group, are considered in 
section 5.

The structure of the resulting fermion mass 
matrices are presented in section 6, together with 
details of a fit to the charged fermion spectrum. In 
sections 7 and 8, we briefly discuss the problems of 
baryogenesis and neutrino oscillations respectively. 
Finally we mention the relation to the MPP principle
- see the contribution by Larisa Laperashvili - in section 9
and the conclusion is in section 10.

\subsection{Gauge \underline{Group}}
Since the Standard Model is a Yang Mills theory the gauge lie algebra 
is an important structural element to specify in order to specify
the model, and this structure is presumably of some significant 
informative strength as far as it seems not so obvious to say why
the theory working at the present stage of the experiments should
just have this gauge algebra: counting all the many cross products
of various Lie algebras it is not so immediately clear why it 
should be the algebra corresponding to $U(1)\times SU(2) \times SU(3)$ 
that should be the God-selected.

Refering to the works by Michel and Oraifaightaigh \cite{MO}
we have long suggested to consider rather than the gauge lie algebra 
- which is of course what specifies the couplings of the Yang Mills
fields to each other - the gauge \underline{group}. A priori the gauge 
\underline{group} 
only is relevant as far as its Lie algebra determines the couplings 
of the Yang Mills particles or fields , the coupling constants being 
proportional to the structure constants of the Lie algebra. 
If there were a truly ontologically existing lattice it would be
a different matter because in that case there would be place for
specifying a \underline{group} and not only the to the group corresponding 
Lie algebra. There is , however, also a phenemenologically accessible
way of asigning a meaning to the gauge \underline{group} and not 
only the algebra:
Different gauge \underline{groups} with the same Lie algebra allow a different 
set of matter field representations. Certain \underline{groups} are thus
not allowed if one requires that the experimentally found matter 
shall be represented under the \underline{group}. 

Now the connection between Lie algebra and Lie \underline{group} is so that 
there are several groups corresponding to one algebra in general but 
always only one algebra to each group. Considering only connected groups 
as is reasonable here there corresponds to each Lie algebra a unique group,
the covering group characterized by being \underline{simply} connected 
- i.e. that any closed curve on it can be continuosly contracted to a point - 
from which which all the other connected groups with the given Lie algebra 
can be obtained by dividing out of the covering group the various 
dicrete invariant subgroups of it. Now it is mathematically so 
that all representations of the Lie algebra are also representations 
of the covering group, but for the other groups with the given Lie algebra 
it is only some of the algebra representations that are also representations
of the group. You can therefore never exclude that the covering group
can be used, whatever the matter field representations may be, while
many of the other groups can easily be excluded whenever some 
matter field representation is known. If one has found a large number
of matter fields as $ \{ ( 2\pi, -1, exp(2\pi/3)1)^n
|n\in {\bf Z})\}$ is the case in the Standard Model then it might 
be almost remarkable if any group other than the covering group
has all these representations. For the Standard Model it can in fact 
rather easily be computed that there is remarkably enough a group
other than the covering one which contains all the representations found 
in nnature so far ! In the light of the relatively ``many'' matter
representations we could then claim that there is a phenomenological 
evidence for that this group is the GROUP of select by nature or
the
Standard Model \underline{group}, which we write by short hand SMG. Indeed the 
the group that in this way deserves to be called the Standard Model Group
is $SMG = S(U(2)\times U(3))= 
({\bf R} \times SU(2) \times SU(3))/\{ ( 2\pi, -1, exp(2\pi/3)1)^n
| n\in {\bf Z})\}$. It may be described as the subset of the cross product 
of $U(2)$ and
$U(3)$ for which the product of the determinat for the U(2) group
element conceived of as a matrix and that of SU(3) is unity.

What this putting forward of a special \underline{group} really means is 
that a regularity in the system of matter field representations 
that occur phenemenologically can be expressed by the \underline{group}
statement.
In the case of the Standard Model \underline{Group} $SMG = S(U(2)\times U(3))$
=$({\bf R} \times SU(2) \times SU(3))/\{ ( 2\pi, -1, exp(2\pi/3)1)^n
|n\in {\bf Z})$ it is actually the regularity required by the 
wellknown rules for electric charge quantization that can be expressed as 
the requirement of the repesentations in nature being representations 
not only of the Lie algebra but really of this \underline{group}. The electric
charge 
quantization rule is: 

For the colorless particles we have the Milikan
charge quantization of all charges being integer when measured in units 
of the elementary charge unit, but for colored particles the charge 
deviate from being integer by $-1/3$ elementary charge for quarks and by 
$+1/3$ for antiquarks.

This rule can be expressed by introducing the concept of
triality $t$, which characterizes the representation of the 
center $ \{ exp(ni2\pi/3) 1^{3\times 3} | n = 0,1,2\}\subset SU(3)$ 
writting under the representation in question and is defined so that
$t=0$ for the trivial representation or for decuplets , octets and so on ,
while it is $t=1$ for triplet ( $\underline{3}$) or anti sixplets 
etc. and $t=-1$ for antitriplet ({$ \underline{\overline{3}}$}) or 
sixtet etc. Then it is written 
\begin{equation}
Q+t/3 = 0 ( \mbox{mod} 1)
\end{equation}
where $Q$ is the electric charge $ Q=y/2 + t_3/2$ ( here $t_3$ is the
third component of the weak isospin
( SU(2)), and y is the weak hypercharge).
We may write this charge quantization rule as
\begin{equation}
y/2 + d/2 + t/3 =0 ( \mbox{mod} 1)
\end{equation}
where we have introduced the duality $d$ which is defined to
be $0$ when the weak isospin is integer and $d=1$ when it is half
integer. It is namely then easily seen that $d/2=t_3/2 (\mbox{ mod} 1)$ 
for all weak isospin representations.

Now the point is that this restriction on the representations 
ensure that the subgroup \\
$\{ ( 2\pi, -1, exp(2\pi/3)1)^n
|n\in {\bf Z})$ is represented trivially and that thus the representations 
allowed  really are  representations of the \underline{group} $SMG =
S(U(2)\times U(3))=$
\\$=
({\bf R} \times SU(2) \times SU(3))/\{ ( 2\pi, -1, exp(2\pi/3)1)^n
|n\in {\bf Z})$.

After having made sense of the \underline{group} it would be natural to
ask if this \underline{group} could somehow give us a hint about what goes on
beyond the Standard Model. Brene and one of us (\cite{BR},\cite{BRsplit}) 
have argued for two indications comming out the \underline{group}-choice 
of nature:

a) The charge quantization rule in the Standard Model is 
in some sense linking the invariant sub Lie algebras more strongly
than in - in a certain way of counting - any other group would do.
To be more specific : There are six different combinations 
of triality and duality - i.e. really of classes of representations of the 
non-abelian part of the gauge Lie algebra - that can be specified by 
providing the abelian charge $y/2$. The logarithm of this number 
of such classes divided by the dimension of the Cartan algebra 
four in the case of the Standard Model group is larger for the 
SMG than for any other group ( except cross products of 
SMG with itself, for which the mentioned ratio must be the same
value). We called this ratio $\chi$.

b) The $SMG$ has rather few automorphisms and can be considered to 
a large extend specified as one of the most ``skew'' groups.

If you would take the point a) to help to guess some group
beyond the Standard Model you could say that we should expect 
also the group behind to have the ratio $chi$ large for the 
group behind. That requirement points in the direction of having
a cross product power of the standard model group, because
such a cross product has just the same $chi$ as the standard model 
\underline{group} itself.a

\subsection{The large mass ratios of leptons and quarks}
What is the origin of the well-known pattern of large ratios 
between the quark and lepton masses and of the small quark 
mixing angles? This is the problem of the hierarchy of 
Yukawa couplings in the Standard Model (SM). 
We suggest \cite{holfn} that the natural resolution to this 
problem is the existence of some approximately conserved chiral 
charges beyond the SM. These charges, which we assume to be 
gauged, provide selection rules forbidding the transitions 
between the various left-handed and right-handed fermion 
states (except for the top quark).

For example, we suppose that there exists some charge (or 
charges) $Q$ for which the quantum number difference between 
left- and right-handed Weyl states is larger for the 
electron than for muon:
\begin{equation}
\left| Q_{eL} - Q_{eR} \right| > \left| 
Q_{\mu L} - Q_{\mu R} \right| 
\end{equation} 
It then follows that the SM Yukawa coupling for the electron 
$g_e$ is suppressed more than that for the muon $g_{\mu}$, 
when $Q$ is taken to be approximately conserved. This is 
what is required if we want to explain the electron-muon 
mass ratio.

 We shall take the point of view 
that, in the fundamental theory beyond the SM, 
the Yukawa couplings allowed by gauge invariance 
are all of order unity and, similarly, 
all the mass terms allowed by gauge invariance are of 
order the fundamental mass scale of the theory---say 
the Planck scale. Then, apart from the matrix element 
responsible for the top quark mass, the quark-lepton 
mass matrix elements are only non-zero due to the 
presence of other Higgs fields having vacuum expectation 
values (VEVs) smaller (typically by one order of magnitude) 
than the fundamental scale. These Higgs fields will, 
of course, be responsible for breaking the fundamental 
gauge group $G$- what ever it may be - down to the SM group. 
In order to generate 
a particular effective SM Yukawa coupling matrix element, 
it is necessary to break the symmetry group $G$ by a 
combination of Higgs fields with the appropriate 
quantum number combination  $\Delta \vec{Q}$. When this 
``$\Delta \vec{Q}$'' is different for two matrix elements 
they will typically deviate by a large factor.
If we want 
to explain the observed spectrum of quarks and leptons in this 
way, it is clear that we need charges which---possibly in a 
complicated way---separate the generations and, at least 
for $t-b$ and $c-s$, also quarks in the same generation. 
Just using the usual simple $SU(5)$ GUT charges does not 
help, because both ($\mu_R$ and $e_R$) and 
($\mu_L$ and $e_L$) have the same $SU(5)$ quantum numbers. 
So we prefer to keep each SM irreducible representation 
in a separate irreducible representation of $G$ and 
introduce extra gauge quantum numbers distinguishing 
the generations, by adding extra cross-product factors to 
the SM gauge group.

What the structure of the quark and lepton spectrum really 
calls for is separation between generations and also between 
at least the c-quark and t-quark from their generations even.
Unification is trictly speaking not called for 
because it is wellknown that the simplest $SU(5)$ unification
can only be made to work by either having complicated Higgs fields
replacing the simple Weinberg salam Higgs field  taken as a fiveplet
- Georgy-Jarlskog model - or even more sophisicated mechanisms
involving other SU(5) breaking than in the minimal SU(5).
The experimental mass ratios predicted by simple SU(5) 
may work for the case of $\tau$ and b-quark adjusted by susy 
or a reasonable scale, but then the $\mu$ to s and the 
e to d cases do not agree with such a simple SU(5) with only
the fiveplet Higgsfield( or two if supersymmetric) replacing
the Weinberg Salam Higgs.

In other words it is called for separation not unification! 

Although one of us (Colin Froggatt) in his contribution 
shows the possibility of extending the standard model with 
just two extra $U(1)$ groups and get a fit of the 
quark and lepton mass order of magnitudes - actually 
that model is the model of the present contribution with
the nonabelian groups amputated away - it is , if one insisists 
on quantum numbers closer to be minimal/small relative to what 
is allowed by the quantization rules ( what is allowed by the 
requirement of representing the \underline{group}), suggested 
to be better to have a larger group extending the SMG. 

\subsection{The ``maximal'' AGUT gauge group}

To limit the search for the gauge group beyond the Standard Model
let us take the point of view that we do not look for the whole
gauge group $G$ say , but only for that factorgroup 
$G'$ = $G/H$ which transform the already known quark and lepton 
weyl fields in a nontrivial way. That is to say we ask for 
the group obtained by dividing out the subgroup $H\subset G$
which leaves the quark and lepton fields unchanged. This
factor group $G'$ can then be identified with its representation 
on the Standard model fermions, i.e. as a subgroup of the
the $U(45)$ group of all possible unitary transformations of the
45 Weyl fields for the Standard model. If one took as $G$ one
of the extensions of SU(5) such as SO(10) or the E-groups 
which are promissing unification groups, the factor group 
$G/H$ would be SU(5) only, the extension parts can be said to
only transform particles that are not in the standard model
( and thus could be pure phantasy a priori). We would like to 
assume that there shall be no gauge or mixed anomalies. So now
we can ask to add some further suggestive properties for
$G'$ that could help us choosing it: 

If we ask the smallest extension of the Standard Model 
unifying as many as possible of the under the standard model 
irreducible representations to irreducible representations 
under $G'$ we get as can be relatively easily seen $SU(5)$
the usual way. That represents all the SO(10) and E-groups,
since we here talked about having divided out the part $H$
that transform the known particles trivially.

But as we above argued there rather from the empirical 
indicators a call for the opposite: separation and a big group!

We have actually calculated that among the subgroups of the $U(45)$
of unitary transformations of the Standard model Weyl fermions 
without anomalies the biggest seperating group is the AGUT-group
which is the gauge group of the model put forward here.    

The AGUT model is based on extending the SM gauge group 
$SMG = S(U(2) \times U(3))$ in a similar way to grand 
unified $SU(5)$, but rather to the non-simple 
$SMG^3 \times U(1)_f$ group. 

The $SMG^3 \times U(1)_f$ group should be understood  
that, near the Planck scale, there are three sets of 
SM-like gauge particles. Each set only couples to its 
own proto-generation [e.g. the proto- $u$, $d$, $e$ and   
$\nu_e$ particles], but not to the other two proto-generations 
[e.g. the proto- $c$, $s$, $\mu$, $\nu_{\mu}$, $t$, $b$, 
$\tau$ and $\nu_{\tau}$ particles]. There is also an extra 
abelian $U(1)_f$ gauge boson, giving altogether 
$3 \times 8 = 24$ gluons, $3 \times 3 = 9$ $W$'s and 
$3 \times 1 + 1 =4$ abelian gauge bosons. The couplings 
of the $SMG_i = S(U(2) \times U(3))_i \approx SU(3)_i 
\times SU(2)_i \times U(1)_i$ group to the $i$'th 
proto-generation are identical to those of the SM 
group. Consequently we have a charge quantisation 
rule,
for each 
of the three proto-generation weak hypercharge 
quantum numbers $y_i$. 

To first approximation---namely in the approximation 
that the the quark mixing angles $V_{us}$, $V_{cb}$, 
$V_{ub}$ are small---we may ignore the prefix $proto-$. 
However we really introduce in our model some 
``proto-fields'' characterized by their couplings 
to the 37 gauge bosons of the $SMG^3 \times U(1)_f$ 
group. The physically observed $u$-quark, $d$-quark 
etc. are then superpositions of the proto-quarks 
(or proto-leptons), with the same named proto-particle 
dominating. Actually there is one deviation from this 
first approximation rule that proto-particles correspond 
to the same named physical particle. In the AGUT fit to 
the quark-lepton mass spectrum, discussed below, 
we find that to first approximation the right-handed components 
of the top and the charm quarks must be permuted:
\begin{equation}
c_{R \ PROTO} \approx t_{R \ PHYSICAL} \qquad 
t_{R \ PROTO} \approx c_{R \ PHYSICAL}
\end{equation}
But for all the other components we have:
\begin{equation}
t_{L \ PROTO} \approx t_{L \ PHYSICAL} \qquad 
b_{R \ PROTO} \approx b_{R \ PHYSICAL}
\end{equation} 
and so on. 

The AGUT group breaks 
down an order of magnitude or so below the Planck 
scale to the SM group as the diagonal subgroup of 
its $SMG^3$ subgroup.

For this breaking we shall fit a relatively complicated 
sytem of Higgses with names $W$, $T$, $\xi$, and $S$.
In order to get neutrino masses fitted we need an even more complicated system.
See the thesis of Mark Gibsosn.

It should however, be said that although at the very hing energies 
just under the Planck energy each generation has its own 
gluons own W's etc. then the breaking makes only one
linear combination of a certain color combination of gluon say 
``survive'' down to the low energy and below the ca 1/10 of the
Planck scale it is only these linear combinations that are 
present and thus the couplings of the gauge particles 
- namely at low energy only these combinations - are the same
for all three generations.

You can also say that the phenomenological gluon say is
a linear combination with amplitude $1/\sqrt{3}$ for
each of the AGUT-gluons of the same color combination.
That then also explains that the coupling constat for the 
phenomenological gluon coules with a strength that is $\sqrt{3}$
times smaller, if as we effectively assume the three AGUT 
$SU(3)$ coulings were equal to each other. Really the 
formula connecting the fine structure constant for the in our model  to
low energy surviving diagonal subgroup 
$\{(U,U,U) | U \in SMG \}\subseteq SMG^3$ is
\begin{equation}
\frac{1}{\alpha_{diag,i}} = \frac{1}{\alpha_{\mbox{1st gen.},i}}
+\frac{1}{\alpha_{\mbox{2nd gen.},i}} + \frac{1}{\alpha_{\mbox{3rd gen.},i}}
\end{equation}.
Here the index $i$ is meant to run over the three groups ina SMG,
namely $i = U(1), SU(2), SU(3)$, so that e.g. $i=3$ means that we talk
abou the gluon couplings ( of the generation in question).   

The gauge coupling constants do not, 
of course, unify, because we not 
built the groups U(1) , SU(2) and SU(3) together,
 but their values have been successfully 
calculated using the so-called multiple point 
principle \cite{hglasgowbrioni}, which is a further 
assumption we put into the model ( see for this (also)
Larisa Laperashvilis contribution to these proceedings). 

At first sight, this $SMG^3 \times U(1)_f$ group with 
its 37 generators seems to be just one among many 
possible SM gauge group extensions. 

However, we shall 
now argue it is not such an arbitrary choice, as it
can be uniquely specified by postulating 4 reasonable 
requirements on the gauge group $G$ beyond the SM.
As a zeroth postulate, of course, we require 
that the gauge group extension must contain the Standard Model
group as a subgroup $G \supseteq SMG$. 
In addition it should obey the
following 4 postulates:
\vspace{2 mm}

\noindent 
The first two are also valid for $SU(5)$ GUT:

\begin{enumerate}
\item $G$ should transform the presently known (left-handed, 
say) Weyl particles into each other, so that
$G \subseteq U(45)$. Here $U(45)$ is the group of all
unitary transformations of the 45 species of Weyl fields (3
generations with 15 in each) in the SM.
\item No anomalies, neither gauge nor mixed. 
We assume that only straightforward anomaly
cancellation takes place and, as in the SM itself, 
do not allow for a Green-Schwarz type anomaly 
cancellation \cite{green-schwarz}.
\vspace{2 mm}

\noindent But the next two are rather just opposite to the properties of the 
$SU(5)$ GUT, thus justifying the name Anti-GUT:

\item The various irreducible representations of Weyl fields
for the SM group remain irreducible under $G$. This is 
the most arbitrary of our assumptions about $G$. It 
is motivated by the observation that combining SM
irreducible representations into larger unified 
representations introduces symmetry relations between 
Yukawa coupling constants, whereas the particle spectrum
doess not exhibit any exact degeneracies (except 
possibly for the case $m_b = m_{\tau}$). In fact 
AGUT only gets the naive $SU(5)$ mass predictions as 
order of magnitude relations: 
$m_b \approx m_{\tau}$, $m_s \approx m_{\mu}$, 
$m_d \approx m_e$.
\item $G$ is the maximal group satisfying the other 3 
postulates. We argued in the previous section that the 
large number of order of magnitude classes of fermion 
mass matrix elements indicates the need for a large 
number of cross product factors in $G$.
\end{enumerate}

With these four postulates a somewhat cumbersome
calculation shows that,
modulo permutations of the various irreducible representations
in the Standard Model
fermion system, we are led to our gauge group
$SMG^3\times U(1)_f$.
Furthermore it shows that the SM group is embedded
as the diagonal subgroup of $SMG^3$, as in our AGUT model.

Several of the anomalies
involving this $U(1)_f$ are in our solution cancelled by assigning
equal and opposite values of the $U(1)_f$ charge to
the analogous particles belonging to second and
third proto-generations, while the
first proto-generation particles have just 
zero charge \cite{davidson}. 

In fact the $U(1)_f$ group does not couple to 
the left-handed particles and the $U(1)_f$ quantum 
numbers can be chosen as follows for the proto-states:
\begin{equation}
Q_f(\tau_R) = Q_f(b_R) = Q_f(c_R) = 1
\end{equation}
\begin{equation}
Q_f(\mu_R) = Q_f(s_R) = Q_f(t_R) = -1
\end{equation}

Thus the quantum numbers of the quarks and leptons 
are uniquely determined in the AGUT model. However 
we do have the freedom of choosing the gauge quantum 
numbers of the Higgs fields responsible for the breaking 
the $SMG^3 \times U(1)_f$ group down to the SM gauge 
group. These quantum numbers are chosen with a view to 
fitting the fermion mass and mixing angle data \cite{smg3m}, 
as discussed in the next section.

\subsection{Symmetry breaking by Higgs fields}

\label{choosinghiggs}

There are obviously many different ways to break down the 
large group $G$ to the much smaller SMG. However, we can 
first greatly simplify the situation by 
assuming that, like the quark and lepton fields, the Higgs 
fields belong to singlet or fundamental representations of 
all non-abelian groups. The non-abelian representations are 
then determined from the $U(1)_i$ weak hypercharge quantum 
numbers, by imposing the charge quantisation rule 
for each of the $SMG_i$ groups.
So now the four abelian charges, which we express in 
the form of a charge vector
\begin{displaymath}
\vec{Q} = \left( \frac{y_1}{2}, \frac{y_2}{2}, 
\frac{y_3}{2}, Q_f \right)
\end{displaymath}
can be used to specify the complete representation of $G$.
The constraint that we must eventually recover the SM 
group as the diagonal subgroup of the $SMG_i$ groups 
is equivalent to the constraint that all the Higgs fields
(except for the Weinberg-Salam Higgs field which of course 
finally breaks the SMG) should have charges $y_i$ satisfying:
\begin{equation}
\label{diagU1}
y=y_1+y_2+y_3=0
\end{equation}
in order that their SM weak hypercharge $y$ be zero.

We wish to choose the charges of the Weinberg-Salam (WS) Higgs
field so that it matches the difference in charges between
the left-handed and right-handed physical top
quarks. This will ensure that the top quark
mass in the SM is not suppressed relative
to the WS Higgs field VEV. However, 
as we remarked in the previous section, it is 
necessary to associate the physical right-handed 
top quark field not with the corresponding third 
proto-generation field $t_R$ but rather with the right-handed 
field $c_R$ of the second proto-generation. Otherwise 
we cannot suppress the bottom quark and tau lepton masses. This is
because, for the proto-fields, the charge differences 
between $t_L$ and $t_R$ are the same as between $b_L$ 
and $b_R$ and also between $\tau_L$ and $\tau_R$. So 
now it is simple to calculate the quantum numbers of 
the WS Higgs field $\phi_{WS}$:
\begin{equation}
\vec{Q}_{\phi_{WS}} = \vec{Q}_{c_R} - \vec{Q}_{t_L}
	= \left( 0,\frac{2}{3},0,1 \right) - 
	\left( 0,0,\frac{1}{6},0 \right)
	= \left( 0,\frac{2}{3},-\frac{1}{6},1 \right)
\label{ws10}
\end{equation}
This means that the WS Higgs field
will in fact be coloured under both $SU(3)_2$ and
$SU(3)_3$. After breaking the symmetry down to the SM 
group, we will be left with the usual WS Higgs field 
of the SM and another scalar which will be an octet of 
$SU(3)$ and a doublet of $SU(2)$.
This should not present any phenomenological problems,
provided this scalar doesn't cause symmetry breaking 
and doesn't have a mass less than the few TeV scale. 
In particular an octet of $SU(3)$ cannot lead to baryon 
decay.
In our model we take it that what in the standard model 
is seen as often very small yukawa-couplings to the 
standard model Higgs field really represent chain Feynmann diagrams
composed of propagators with Planck scale heavy particles 
(fermions) interspaced with the couplings by order of unity 
yukawa couplings to in our model postulated Higgs fields 
with names $ W$,$ T$,$ \xi$, and$ S$ breaking the AGUT to
the Standard Model Group and meaning that the vacuum expectation value is 
active. The smallness of the effective Yukawa coupling in the 
Standard Model is taken to be due to the expectation values of 
$W$, $T$, and $\xi$ relative to the masses occuring in the 
propagators for the Planck scale fermions in the diagrams 
simulated by the effective Yukawa couplings in the Standard Model.

The quantum numbers of our invented Higgs fields $W$, $T$, $\xi$ and $S$
are invented - and it is remarkable that we succeeded so well - so
as to make the order of magnitude for the suppressions of the
mass matrix elements of the various mass matrices fit to the phenmenological 
requirements. 

After the choice of the quantum numbers for the replacement 
of the Weinberg Salam Higgs field in our model ( \ref{ws10})
the quantum numbers further needed to be picked out of vacuum in
order to give say the mass of the b-quark is denoted by $\vec{b}$
and analogously for the other particles, e.g.:

\begin{equation}
\vec{b} = \vec{Q}_{b_L} - \vec{Q}_{b_R} - \vec{Q}_{WS}
\end{equation}

\begin{eqnarray}
\vec{c} & = & \vec{Q}_{c_L} - \vec{Q}_{t_R} + \vec{Q}_{WS} \\
\vec{\mu} & = & \vec{Q}_{\mu_L} - \vec{Q}_{\mu_R} - \vec{Q}_{WS}
\end{eqnarray}
Here we denoted the quantum numbers quarks and leptons
as e.g. $\vec{c_L}$ for the left handed components of the 
proto-charmed quark.
Note that $\vec{c}$ has been defined using the $t_R$ 
proto-field, since we have essentially
swapped the right-handed charm and top quarks. 
Also the charges of the WS Higgs field
are added rather than subtracted for up-type quarks.

Next we attempted to find some Higgs quantum numbers which
if postulated to have ``small'' expectation values compared to
the masses of intermediate particles - i.e. denominators in 
propagators that could go into diagrams and give a fit of the 
orders of magnitudes. We have the proposal:

\begin{equation}
\vec{Q}_W = \frac{1}{3}(2\vec{b}+\vec{\mu}) =
		\left( 0,-\frac{1}{2},\frac{1}{2},-\frac{4}{3} \right)
\end{equation}

\begin{equation}
\vec{Q}_T = \vec{b} - \vec{Q}_W =
\left( 0,-\frac{1}{6},\frac{1}{6},-\frac{2}{3} \right)
\end{equation}

\begin{equation}
\vec{Q}_{\xi} = \vec{Q}_{d_L} - \vec{Q}_{s_L}
	= \left( \frac{1}{6},0,0,0 \right) - 
	\left( 0,\frac{1}{6},0,0 \right)
	= \left( \frac{1}{6},-\frac{1}{6},0,0 \right)
\end{equation}

{}From the well-known Fritzsch relation \cite{Fritschrule} 
$V_{us} \simeq \sqrt{\frac{m_d}{m_s}}$,
it is suggested that the two off-diagonal mass 
matrix elements connecting the
d-quark and the s-quark be equally big.
We achieve this approximately 
in our model by introducing a special Higgs field
$S$, with quantum numbers equal to the
difference between the quantum number
differences for these 2 matrix elements in the 
down quark matrix.
Then we postulate that this Higgs field has
a VEV of order unity in fundamental units,
so that it does not cause any suppression but
rather ensures that the two matrix elements 
get equally suppressed. Henceforth we will 
consider the VEVs of the new Higgs fields as 
measured in units of $M_F$ and so we have:
\begin{equation}
<S> = 1
\end{equation}

\begin{eqnarray}
\vec{Q}_{S} & = & [\vec{Q}_{s_L} - \vec{Q}_{d_R}]
		- [\vec{Q}_{d_L} - \vec{Q}_{s_R}] \nonumber \\
 & = & \left[ \left( 0,\frac{1}{6},0,0 \right) -
		\left( -\frac{1}{3},0,0,0 \right) \right] -
	\left[ \left( \frac{1}{6},0,0,0 \right) -
		\left( 0,-\frac{1}{3},0,-1 \right) \right] \nonumber \\
 & = & \left( \frac{1}{6},-\frac{1}{6},0,-1 \right)
\end{eqnarray}
The existence of a non-suppressing
field $S$ means that we cannot
control phenomenologically when this $S$-field is used.
Thus the quantum numbers of the other
Higgs fields $W$, $T$, $\xi$ and $\phi_{WS}$ 
given above have only been determined modulo those
of the field $S$.

\subsection{Mass matrices, predictions}

We define the mass matrices
by considering the mass terms in the SM to be given by:
\begin{equation}
{\cal L}=Q_LM_uU_R+Q_LM_dD_R+L_LM_lE_R+{\rm h.c.}
\end{equation}
The mass matrices can be expressed in terms of the 
effective SM Yukawa matrices and the WS Higgs VEV by:
\begin{equation}
M_f = Y_f \frac{<\phi_{WS}>}{\sqrt{2}}
\end{equation}
We can now calculate the suppression factors for 
all elements in the Yukawa matrices, by expressing the
charge differences between the left-handed and
right-handed fermions in terms of the
charges of the Higgs fields. They are 
given by products of the small numbers
denoting the VEVs in the fundamental units
of the fields $W$, $T$, $\xi$ and 
the of order unity VEV of $S$. 
In the following matrices we simply write $W$ instead of 
$<W>$ etc. for the VEVs. With the quantum number 
choice given above, the resulting matrix elements 
are---but remember that ``random'' order
unity factors are supposed to multiply all the matrix
elements---for the uct-quarks:
\begin{equation}
Y_U \simeq \left ( \begin{array}{ccc}
	S^{\dagger}W^{\dagger}T^2(\xi^{\dagger})^2 
	& W^{\dagger}T^2\xi & (W^{\dagger})^2T\xi \\
	S^{\dagger}W^{\dagger}T^2(\xi^{\dagger})^3 
	& W^{\dagger}T^2 & (W^{\dagger})^2T \\
	S^{\dagger}(\xi^{\dagger})^3 & 1 & W^{\dagger}T^{\dagger}
			\end{array} \right ) \label{Y_U}
\end{equation}
the dsb-quarks:
\begin{equation}
Y_D \simeq \left ( \begin{array}{ccc}
	SW(T^{\dagger})^2\xi^2 & W(T^{\dagger})^2\xi & T^3\xi \\
	SW(T^{\dagger})^2\xi & W(T^{\dagger})^2 & T^3 \\
	SW^2(T^{\dagger})^4\xi & W^2(T^{\dagger})^4 & WT
			\end{array} \right ) \label{Y_D}
\end{equation}
and the charged leptons:
\begin{equation}
Y_E \simeq \left ( \hspace{-0.2 cm}\begin{array}{ccc}
	SW(T^{\dagger})^2\xi^2 & W(T^{\dagger})^2(\xi^{\dagger})^3 
	& (S^{\dagger})^2WT^4\xi^{\dagger} \\
	SW(T^{\dagger})^2\xi^5 & W(T^{\dagger})^2 &
	(S^{\dagger})^2WT^4\xi^2 \\
	S^3W(T^{\dagger})^5\xi^3 & (W^{\dagger})^2T^4 & WT
			\end{array} \hspace{-0.2 cm}\right ) \label{Y_E}
\end{equation}

We can now set $S = 1$ and fit the nine quark and lepton masses 
and three mixing angles, using 3 parameters: $W$, $T$ 
and $\xi$. That really means we have effectively omitted 
the Higgs field $S$ and replaced the maximal AGUT gauge 
group $SMG^3 \times U(1)_f$ by the reduced AGUT group 
$SMG_{12} \times SMG_3 \times U(1)$, which survives the 
spontaneous breakdown due to $S$.
In order to find the best possible fit we
must use some function which measures how 
good a fit is. Since we are expecting
an order of magnitude fit, this function 
should depend only on the ratios of
the fitted masses to the experimentally 
determined masses. The obvious choice
for such a function is:
\begin{equation}
\chi^2=\sum \left[\ln \left(
\frac{m}{m_{\mbox{\small{exp}}}} \right) \right]^2
\end{equation}
where $m$ are the fitted masses and mixing angles and
$m_{\mbox{\small{exp}}}$ are the
corresponding experimental values. The Yukawa
matrices are calculated at the fundamental scale 
which we take to be the
Planck scale. We use the first order renormalisation 
group equations (RGEs) for
the SM to calculate the matrices at lower scales.

\begin{table}
\caption{Best fit to conventional experimental data. 
All masses are running
masses at 1 GeV except the top quark mass 
which is the pole mass.}
\begin{displaymath}
\begin{array}{ccc}
\hline
 & {\rm Fitted} & {\rm Experimental} \\ \hline
m_u & 3.6 {\rm \; MeV} & 4 {\rm \; MeV} \\
m_d & 7.0 {\rm \; MeV} & 9 {\rm \; MeV} \\
m_e & 0.87 {\rm \; MeV} & 0.5 {\rm \; MeV} \\
m_c & 1.02 {\rm \; GeV} & 1.4 {\rm \; GeV} \\
m_s & 400 {\rm \; MeV} & 200 {\rm \; MeV} \\
m_{\mu} & 88 {\rm \; MeV} & 105 {\rm \; MeV} \\
M_t & 192 {\rm \; GeV} & 180 {\rm \; GeV} \\
m_b & 8.3 {\rm \; GeV} & 6.3 {\rm \; GeV} \\
m_{\tau} & 1.27 {\rm \; GeV} & 1.78 {\rm \; GeV} \\
V_{us} & 0.18 & 0.22 \\
V_{cb} & 0.018 & 0.041 \\
V_{ub} & 0.0039 & 0.0035 \\ \hline
\end{array}
\end{displaymath}
\label{convbestfit}
\end{table}

We cannot simply use the 3 matrices given by
eqs.~(\ref{Y_U})--(\ref{Y_E}) to calculate 
the masses and mixing angles, since
only the order of magnitude of the elements is defined. 
Therefore we calculate 
statistically, by giving each
element a random complex phase and then 
finding the masses and mixing angles.
We repeat this several times and calculate 
the geometrical mean
for each mass and mixing
angle. In fact we also vary the magnitude 
of each element randomly, by
multiplying by a factor chosen to be 
the exponential of a number picked from a
Gaussian distribution with mean value 0 and standard deviation 1.

We then vary the 3 free parameters to 
find the best fit given by the $\chi^2$
function. We get the lowest value of $\chi^2$ for the VEVs:
\begin{eqnarray}
\langle W\rangle & = & 0.179   \label{Wvev} \\
\langle T\rangle & = & 0.071   \label{Tvev} \\
\langle \xi\rangle & = & 0.099 \label{xivev}
\end{eqnarray}
The fitted value of $\langle \xi\rangle$ is approximately 
a factor of two smaller than the
estimate given in eq. above. 
This is mainly because there are
contributions to $V_{us}$ of the same 
order of magnitude from both $Y_U$ and
$Y_D$. The result \cite{smg3m} of the fit is shown 
in table~\ref{convbestfit}. This fit has a
value of:
\begin{equation}
\chi^2=1.87
\label{chisquared}
\end{equation}
This is equivalent to fitting 9 degrees of
freedom (9 masses + 3 mixing angles - 3
Higgs VEVs) to within a factor of 
$\exp(\sqrt{1.87/9}) \simeq 1.58$
of the experimental value. This is 
better than would have been
expected from an order of magnitude 
fit.

We can also fit to different experimental values 
of the 3 light quark
masses by using recent results from lattice QCD, which 
seem to be consistently lower than the conventional
phenomenological values. 
The best fit in this case \cite{smg3m} is 
shown in table~\ref{newbestfit}. 
The values of the Higgs VEVs are:
\begin{eqnarray}
\langle W\rangle & = & 0.123	\\
\langle T\rangle & = & 0.079	\\
\langle \xi\rangle & = & 0.077
\end{eqnarray}
and this fit has a larger value of:
\begin{equation}
\chi^2 = 3.81
\end{equation}
But even this is good for an order of magnitude fit.

\begin{table}
\caption{Best fit using alternative light quark masses 
extracted from lattice QCD. All masses are running
masses at 1 GeV except the top quark mass 
which is the pole mass.}
\begin{displaymath}
\begin{array}{ccc}
\hline
 & {\rm Fitted} & {\rm Experimental} \\ \hline
m_u & 1.9 {\rm \; MeV} & 1.3 {\rm \; MeV} \\
m_d & 3.7 {\rm \; MeV} & 4.2 {\rm \; MeV} \\
m_e & 0.45 {\rm \; MeV} & 0.5 {\rm \; MeV} \\
m_c & 0.53 {\rm \; GeV} & 1.4 {\rm \; GeV} \\
m_s & 327 {\rm \; MeV} & 85 {\rm \; MeV} \\
m_{\mu} & 75 {\rm \; MeV} & 105 {\rm \; MeV} \\
M_t & 192 {\rm \; GeV} & 180 {\rm \; GeV} \\
m_b & 6.4 {\rm \; GeV} & 6.3 {\rm \; GeV} \\
m_{\tau} & 0.98 {\rm \; GeV} & 1.78 {\rm \; GeV} \\
V_{us} & 0.15 & 0.22 \\
V_{cb} & 0.033 & 0.041 \\
V_{ub} & 0.0054 & 0.0035 \\ \hline
\end{array}
\end{displaymath}
\label{newbestfit}
\end{table}

\subsection{ Baryogenesis}

A very important check of our model is if it can be consistent with the 
baryogenesis. In our model we have just the SM interactions up 
to about 
one or two orders of magnitude under the Planck scale.
So we have no way, at the electroweak scale, to produce 
the baryon number in the universe. There is insufficient 
CP violation in the SM. Furthermore, even if created, the baryon number 
would immediately be washed out by sphaleron transitions 
after the electroweak phase transition.  
Our only chance to avoid 
the baryon number being washed out at the electroweak scale is to have 
a non-zero
B-L (i.e. baryon number minus lepton number) produced from the high,
i.e. Planck, scale action of the theory. 
That could then in turn give rise to the baryon number at the 
electroweak scale.
Now in our model the B-L quantum number is broken by an anomaly
involving the $U(1)_f$ gauge group. This part of the 
gauge group in turn  is broken by the Higgs field 
$\xi$ which, in Planck units, is fitted to have an expectation value 
around 1/10. 
The anomaly keeps washing out any net $B-L$ that might appear, 
due to CP-violating forces from the Planck scale physics, until the 
temperature $T$ of the universe has fallen to $\xi = 1/10$.  
The $U(1)_f$ gauge particle then disappears from the thermal soup and thus 
the conservation of B-L sets in. 
The amount of $B-L$ produced at that time should then 
be fixed and would 
essentially make itself felt, at the electroweak scale, 
by giving rise to an amount of 
baryon number of the same order of magnitude. 

The question now is whether we should expect in our model to have a 
suffient amount of time reversal symmetry breaking at the epoch
when the B-L settles down to be conserved, such that the 
amount of B-L relative to say the entropy (essentially
the amount of 3 degree Kelvin background radiation) becomes large enough
to agree with the well-known phenomenological value of the 
order of $10^{-9}$ or $10^{-10}$.
At the time of the order of the Planck scale, when the temparature was also
of the order of the of the Planck temparature, even the CP or 
time reversal 
violations were of order unity (in Planck units). 
So at that time there existed 
particles, say, with order of unity CP-violating decays. However, they 
had also, in our pure dimensional argument approximation, lifetimes
of the order of the Planck scale too. Thus the B-L biased decay products 
would be dumped at time 1 in Planck units, rather 
than at time of $B-L$ conservation setting in. 
In a radiation dominated 
universe, as we shall assume, the temperature will go like 1/a where
a is the radius parameter---the size or scale parameter of the universe.
Now the time goes as the square of this size parameter a.
Thus the time in Planck units is given 
as the temperature to the negative second power 
\begin{equation}
t= \frac{0.3}{\sqrt{ g} \times T^2}
\end{equation}
where \cite{utpal} $g$ is the number of degrees of freedom---counted 
as 1 for bosons but as 7/8 per fermion degrees of freedom---entering 
into the radiation density. In our model $g$ gets a 
contribution of $\frac{7}{8} \times 45 \times 2$ from the fermions and 
$2 \times37$ from the gauge bosons, and in addition 
there is some contribution from the Higgs particles. 
So we take $g$ to be of order 100, in our crude estimate of the 
time t corresponding to the temperature $T = xi = \frac{1}{10}$
in Planck units, when $B-L$ conservation sets in: 
\begin{equation}
t \simeq \frac{0.3}{100^{1/2}} \times (\frac{1}{10})^2 = 3 
\end{equation}
By that time we expect of the order of $\exp{-3}$ 
particles from the Planck era are still present and able 
to dump their CP-violating
decay products. Of course here the uncertainty of an order of magnitude 
would be in the exponent, meaning a suppression 
anywhere between say $\exp{0}$ and $\exp{-30}$ and could 
thus easily be in agreement with the wanted value of order 
$5 \times 10^{-10}$. This result is encouraging, but clearly a more careful 
analysis is required.

\subsection{ Neutrino oscillations, a problem?}
 At first it seems a problem to incorporate the neutrino oscillations 
into our model. The a priori prediction would be that the neutrino masses 
are predicted so small that they could not be seen with presetn accuracy 
because see-saw mass of the order of Planck scale combined with further
suppression leads to too small neutrino masses. However, by
changing the system of Higgses  and getting the neutrino overall 
mass scale a fitted number it has been possible to get a 
satisfactory scheme involving further Higgs fields making shortcuts
in the sense of producing transitions that couls already occur by
other Higgs field combinations. The extension to neutrinos 
is not too attracktive, but tolerable. 

\subsection{ Connection to MPP}
Originally the idea of having $SMG^3$ type model was developped 
in connection with Random dynamics ideas of confusion and long 
time in connection with the idea of requiring many phases to meet 
( multiple point principle MPP) in order to get predictions for the 
fine structure constants. This type of calculations predicted even that there 
be three generations at a time when that was not known experimentally by
fitting the fine structure constants !( see e.g.( \cite{Benthesis}))

\subsection{ Conclusion}
We have looked at some of the hints in the Standard Model 
that may be usefull in going beyond and have put forward our own 
model shown to be to some extend inspired by such features:

We stressed to look for the gauge \underline{group} rather
than just the Lie algebra, which a priori is what is relevant 
for describing a Yang Mills theory.  

We have found surprisingly good fits of masses and mixing angles,
and in a related model even of the finestructure constants, in
a model in which the gauge group at a bit below the Planck scale 
is the maximal one transforming the already known fermions around
and not having anomalies. Although at first it
looked a failure it has turned out that even the baryon number generation
in big bang is not excluded from being in agreement with the present
model. It must , however, then be done by getting first an B-L contribution
made at a time when temeperature was only an order of magnitude under the 
Planck temperature.

To incorporate neutrino oscillations severe but tolerable modifications 
of the model are needed.

\newpage
\stepcounter{section}
\setcounter{equation}{0}

\section*{\center Hints from the Standard Model for Particle Masses and Mixing}
\centerline{\rm BERTHOLD STECH}
\centerline{\it  Institut  f\"ur Theoretische Physik,
Universit\"at Heidelberg,} 
\centerline{\it Philosophenweg 16, D-69120 Heidelberg }
\vskip 5mm
\begin{abstract}
The standard model taken with a momentum
space cut-off may be viewed as an effective low energy
theory. The structure of it and its known parameters can give
us hints for relations between these parameters and for
possible extensions. In the present investigation the Higgs
problem will be discussed, effective potentials, the possible
connection of the Higgs meson with the heavy top quark and the
geometric structure of the quark and lepton mass matrices.
\end{abstract}

\subsection{Introduction}
The standard model is likely to describe the effective
interaction at low energy of an underlying more fundamental theory.
One may speculate that some parameters which emerge at long
distances are insensitive to details of what is going on at much
higher scales. For instance, their values may be given by the fix
points of renormalization group equations and thus being rather
independent of the starting numbers at small distances\footnote{Even
the group structure of the standard model and 4-dimensional space
could possibly be selected through such mechanisms \cite{1}.}
\cite{1}. Or, these
parameters could arise in a bootstrap-type scenario \cite{2}.

In this talk I do not want to discuss specific models of this
type, but will simply look at the measured parameters of the
standard model and at its divergence structure in order to
find hints for possible connections between these parameters.
I will concentrate on the vacuum expectation value of the Higgs 
field, which is defined without reference to external particles 
and their momenta. Thus, its divergence property may be quite 
different from those of ordinary coupling constants, which 
can be renormalized using a momentum subtraction scheme.

By taking the standard model as an effective theory, one should use
a momentum cut-off. The dependence of measurable quantities on
the cut-off reflects the influence of new physics on the low
energy domain. The minimization of this influence provides
suggestions for the sought relations.

\subsection{The vacuum expectation value of the Higgs field and
the invariant Higgs potential}

We write the Higgs part of the Lagrangian in the form
\begin{eqnarray}
{\cal L}_H&=&(D_\mu\Phi)^\dagger D_\mu\Phi+\frac{J}{2}
\Phi^\dagger\Phi-\frac{\lambda}
{2}(\Phi^\dagger\Phi)^2\nonumber\\
&&+\frac{1}{\sqrt2}j\cdot\Phi\quad+\quad{\rm Yukawa\ couplings}\nonumber\\
\Phi&=&\frac{1}{\sqrt2}\left(\begin{array}{c}
\varphi_1+i\varphi_2 \nonumber \\
\varphi_0+i\varphi_3\end{array}\right). \label{bert2.1}
\end{eqnarray}
The quantity $J$ is taken to be $J=J_0+J_1$
with $J_0>0$ describing the Higgs mass parameter responsible
for spontaneous symmetry breaking. $J_1$ can be viewed as an
outside field. It is used for generating a gauge invariant potential 
and will finally be set to zero. The quantity $j$ determines
the field direction of the spontaneous symmetry breaking. It could be
caused by a light quark condensate and may be neglected after
the occurrence of the symmetry breaking. Accordingly, the tree potential
takes the form
\be\label{bert2.2}
V_0=\frac{\lambda}{8}(\sum_i\varphi^2_i)^2-\frac{J}{4}
\sum_i\varphi^2_i-\frac{1}{2}j\varphi_0 ~~.\ee
For $J>0$ the minimum of $V_0$ occurs
for $\varphi_i=\hat\varphi_i(J,j)$
with
\be\label{bert2.3}
\hat\varphi_{1,2,3}=0,\quad \lambda\hat\varphi_0^3-J\hat\varphi_0=j ~~.\ee
We have to select the real solution of the cubic equation. In the
limit $j\to0$, $\lambda\not=0$ one gets
\begin{eqnarray}
\hat\varphi_0(J)&=&\sqrt{\frac{J}{\lambda}},\qquad m^2_H(J)=\frac
{\partial^2V_0}{\partial\varphi^2_0}\big|_{\varphi=\hat\varphi}
=J\nonumber\\
&& m^2_i=\frac{\partial^2V_0}{\partial\varphi^2_i}\big|
_{\varphi=\hat\varphi}=0\qquad~ i=1,2,3~~.\label{bert2.4}
\end{eqnarray}
By replacing as usual $\varphi_0(x)$ by
\be\label{bert2.5}
\varphi_0(x)=\hat\varphi_0(J,j)+H(x)\ee
the interaction part ${\cal L}_{int}'$ of the
shifted Higgs Lagrangian allows to evaluate $<H>$. The
lowest-order expression is
\be\label{bert2.6}
<H>=i\int d^4x<0|T(H(0),{\cal L}_{int}'(x))|0>.\ee
The result \cite{3} obtained from (\ref{bert2.6}) can be used to write
the gauge-invariant vacuum expectation value of the square
of the Higgs field $\sigma(J)=<\sum_i\varphi_i^2>$ for $J>0$
and $j=0$
in the form
\begin{equation}
\label{bert2.7}
\begin{array}{rcl}
\sigma (J) &=&\frac{J}{\lambda}-2<H^2>+2\frac{g^2+{g'}^2}{4\lambda}
<Z_\mu Z^\mu> \\
         & &+4\frac{g^2}{4\lambda}<W^+_\mu W^{-\mu}>-\frac{g^2_t}{\lambda}
<\bar tt>/m_t ~~.
\end{array}
\end{equation}
$g,g'$ are the gauge couplings for the vector bosons
$W$ and $Z$, and $g_t$ denotes the Yukawa coupling for the top
quark. Fermions of lower mass are neglected, but could easily be
added. The ``vacuum leaks'' $<H^2>, <Z_\mu Z^\mu>,...$
could be finite in the full theory with a correspondingly
modified vacuum structure. For the
effective theory with a particle momentum cut-off chosen to
be universal for all propagators (\ref{bert2.7}) leads to
\begin{equation}
\label{bert2.8}
\ba{c}
\sigma(J)=\frac{J}{\lambda}+\frac{\Lambda^2}{8\pi^2}2\\
-\frac{\Lambda^2}{8\pi^2}[3+3\frac{g^2+g^{'2}}{4\lambda}
+6\frac{g^2}{4\lambda}-12\frac{g^2_t}{2\lambda}]\\
+\frac{J}{8\pi^2}
\left\{1+3\left(\frac{g^2+g^{'2}}{4\lambda}\right)^2
+6\left(\frac{g^2}{4\lambda}\right)^2-12\left(\frac{g^2_t}
{2\lambda}\right)^2\right\}\ln\frac{\Lambda^2}{J/\lambda}\\
-\frac{J}{8\pi^2}\left(\ln\lambda+3\left(\frac{g^2+g^{'2}}
{4\lambda}\right)^2\ln\frac{g^2+g^{'2}}{4}\right.\\
\left.+6\left(\frac{g^2}{4\lambda}\right)^2\ln\frac{g^2}{4}-12\left(
\frac{g^2_t}{2\lambda}\right)^2\ln\frac{g^2_t}{2}\right) ~~.
\ea
\end{equation}
The term $\frac{\Lambda^2}{8\pi^2}2$ which represents the free field
part of $<\Sigma_i\varphi^2_i>$ is written separately. 
It also appears in the case $J<0$ where one has $\hat\varphi_0(J)=0$
and no spontaneous symmetry breaking. For $J < 0$ one finds to one 
loop order
\be\label{bert2.9}
\sigma(J) = \frac{\Lambda^2}{8\pi^2}2+\frac{J}{8\pi^2}
\ln\frac{2\Lambda^2}{-J} ~~.
\ee
Thus, in our approximation\footnote{At $J=0$ the one-loop
approximation for $\sigma(J)$ is presumably
insufficient. However, (\ref{bert2.10}) is expected to hold
for the change of $\sigma$ within a larger region around
$J=0$.}, $\sigma(J)$ is discontinuous
at $J=0$, indicating a first-order phase transition with strength
proportional to $\Lambda^2$:
\be\label{bert2.10}
\Delta\sigma=-\frac{\Lambda^2}{8\pi^2}[3+3\frac{g^2+g^{'2}}
{4\lambda}+6\frac{g^2}{4\lambda}-12\frac{g^2_t}{2\lambda}]\ee
If we would keep $j\not=0$, the jump of $\sigma(J)$
would undergo a rapid, but now continuous, change
in the region $-j^{2/3}\leq J\leq j^{2/3}$. The right-hand
side of (\ref{bert2.10}) with its quadratic divergence also shows up
in the conventional (non-gauge invariant) Higgs potential $V(\varphi)$.
Its appearance constitutes an essential part of the hierarchy
problem and necessitates fine-tuned subtractions or the
postulate of a cancellation of the special combinations of
couplings occurring here. The mass relation for which (\ref{bert2.10})
vanishes, is
\be\label{bert2.11}
S_{\Lambda^2}=(3m^2_H+3m^2_Z+6m^2_W-12m^2_t)/<\varphi_0>^2 = 0 ~~.\ee
It is known as the Veltman condition \cite{4}.

Eq. (\ref{bert2.8}) can also be obtained in a more general context.
By defining the ``free energy'' $W(J,j)$ by the logarithm of
the partition function with the action according to (\ref{bert2.1}),
one can obtain $\sigma $ from
\be\label{bert2.12}
\sigma =-4\frac{\partial W(J,j)}{\partial J}~~.\ee
The Legendre transformation of $W(J,j)$ with respect to
$J$ defines the $\sigma$-dependent effective potential \cite{5}
\be\label{bert2.13}
V(\sigma,j,J_0)=W(J(\sigma,j),j)+\frac{\sigma}{4}(J(\sigma,j)-J_0) ~~.\ee
It has an extremum at $\bar\sigma$, for which $J(\bar\sigma,j)=J_0$
where $J_0$ is the mass parameter in the Higgs potential.

Calculating $W(J,j)$ by the saddle point method
up to one-loop order, one gets
\be\label{bert2.14}
\ba{c}
W(J,j)=\frac{\lambda}{8}\hat\varphi^4_0(J,j)-\frac{J}{4}
\hat\varphi_0^2(J,j)-\frac{j}{2}\hat\varphi_0(J,j)\\
+\frac{1}{(4\pi)^2}\frac{1}{2}\sum_pr_p\int^{\Lambda^2}
_0dK^2K^2\ln\left(1+\frac{m^2_p(J,j)}{K^2}
\right) ~~.
\ea
\ee
The sum is over the particles of the standard model with their
$J$- and $j$-dependent masses. $r_p$ is a statistical factor
(3 for the $Z$, 6 for the $W$, -12 for the top, 1 for the Higgs).
For $j=0$ the result is gauge-invariant. To consider the
dependence of $W(J,j=0)$ on $J$ rather than on $j$ has the
additional advantage that for $J>0$ the Goldstone particles remain
massless and thus do not contribute. Furthermore, to one-loop
order, the potential $V(\sigma,J_0)$ remains real for all values
of $\sigma$. The derivative of $W(J)$ with respect to $J$
according to (\ref{bert2.12})
reproduces eq. (\ref{bert2.8}). As long as $\Delta\sigma$ is not very
small and $\Lambda$ of order TeV or larger, $V(J_0,\sigma)$ calculated
from (\ref{bert2.12}--\ref{bert2.14}) changes rapidly in its 
dependence on $J_0$ for
$J_0$ near zero (and small $j$).

For the purpose of renormalization we can add to $W(J,j)$
a polynomial in $J$  up to second order. The linear piece
could be used to cancel the quadratic divergence in (\ref{bert2.8}).
However, it would then reappear in (\ref{bert2.9}). Instead, one should
normalize $\sigma$ such that it is zero in the limit of all 
particle masses going to zero, starting from $J<0$:
$\sigma(J\to 0_-,j\to0)=0$. To achieve this, we have to
subtract the $2\Lambda^2/8\pi^2$ part in (\ref{bert2.8}) and
(\ref{bert2.9}) by replacing $W(J,j)$ in (\ref{bert2.14}) by
$W(J,j)+\frac{\Lambda^2}{32\pi^2}2J$. A further change of
$W(J,j)$ using a subtraction term proportional to $J^2$ can 
remove the logarithmic 
divergence in (\ref{bert2.8}) and (\ref{bert2.13}). This subtraction can 
be interpreted 
as a renormalization of the Higgs coupling constant $\lambda$. 
Again, we do not perform such a complete subtraction since the 
corresponding term would then appear in (\ref{bert2.9}) where it has 
no physical basis. But we can remove the logarithmic divergence 
for regions of $J$ where the gauge bosons and the fermions remain 
massless. Since I am not completely certain about the necessity 
of this subtraction I will consider two cases i) no subtraction 
proportional to $J^2$ and 
-- the more appealing one -- ii) a subtraction such that in (\ref{bert2.9}) 
besides the quadratic divergence also the logarithmic divergence 
is removed. Accordingly, the factor which governs the logarithmic 
divergence of $\sigma$ and $V(\sigma,J_0)$ in the region of 
spontaneous symmetry breaking is ( to one loop order and in terms of masses)
\be\label{bert2.15}
S_{Log\ \Lambda} = (\zeta m^4_H+3 m^4_Z+6m^4_W-12m^4_t)/
<\varphi_0>^4 ~~.
\ee
$\zeta=1$ corresponds to no subtraction, while $\zeta = 0 $ is 
valid when the subtraction according to ii) is performed.  
(I do not consider here the conventional non gauge invariant 
potential $V(J_0,<\varphi_0>)$\cite{6}. It would lead to $\zeta=3/2$).

As a speculation I will now assume a minimum influence
of new physics on the standard model.
In particular, $\Delta\sigma$ of eq. (\ref{bert2.10}) should
be independent of $\Lambda^2$, i.e. the
square bracket in (\ref{bert2.10}) should be proportional to
$1/\Lambda^2$ or zero. If this is the case, the particle
couplings are not independent of each other, but satisfy
-- at least approximately -- the Veltman condition. 
Here one encounters the problem of the scale $(\mu)$ at which the
particle couplings should be taken. In particular, the Yukawa 
coupling of the top quark is sensitive to it. The vacuum expectation
value of the unrenormalized Higgs field is scale-invariant. But to
take advantage of this fact, higher order calculations and a
knowledge of the scale dependence of $\Lambda$ ($\Lambda$ may
be related to high mass states) would be needed. The natural 
scale for the couplings occuring in the loop integrals is 
$\mu \approx \Lambda$. Let us first 
assume that the cancellation of the coupling terms in 
(\ref{bert2.10}) occurs already at the weak scale of $\approx250~ GeV$, 
where the top mass is still big. 
Then, using for the mass of the top $m_t(m_Z)=173$ GeV 
the Higgs mass is predicted to
be $m_H(m_Z)\approx 280~GeV$.

Now we can 
use this value of the Higgs mass to look at the logarithmic 
divergence and calculate
$S_{Log\ \Lambda}$ .
Using $\zeta=1$ and again the scale $\mu \approx250~GeV$ we find
$S_{Log\ \Lambda}=-0.2$~.
We thus have the surprising result that for a Higgs mass
of about 300 GeV the factors responsible for the quadratic
and the logarithmic divergence of the Higgs potential $V(\sigma,J_0)$
are both small. Let us then consider the extreme case
of strictly vanishing factors in front of $\Lambda^2$ and
$\log \Lambda$ for the one-loop Higgs potential $V(\sigma,J_0)$. 
This provides us with two equations which allow a calculation
of $m_t$ and $m_H$ in terms of the gauge couplings for $W$
and $Z$. The result is (for $\mu=250~GeV $ and a correspondingly 
very low cut off value)
\be\label{bert2.16}
m_t(m_Z)= 198~GeV~,\qquad m_H(m_Z)= 320~GeV~~.\ee
The fact that the value of $m_t$ obtained this way is not far away from
the experimental result may be a fortuitous coincidence.
But if not, it indicates a very close connection of Higgs
and top with a Higgs mass not much different from $2m_t\approx
350$ GeV! On the other hand, the prefered value $\zeta = 0 $ gives 
no admissible 
solution at the very low scale considered so far.

Owing to the quadratic form of the two equations which simultaneously 
suppress quadratic and logarithmic divergences, another type of 
solution exists with smaller values for the Higgs mass. For this 
solution the scale relevant for the particles running in the 
loops and thus the value of the cut off $\Lambda$ must be extremly high 
in order to get a large enough value 
for the top mass at the weak scale. We take, therefore, $\mu$ 
equal to the Planck mass, determine $ m_t$ and $m_H$ and apply 
the two loop renormalization group equations to get their values 
at the weak scale. This implies, of course, that physics beyond 
the standard model, which could influence the standard model 
couplings, can occur only near and above the Planck scale. 
The calculation using $\alpha_s(m_Z)=0.12$ and $\zeta=1$ gives
\be\label{bert2.17}
          m_t(m_Z)=169~GeV~,~m_H(m_Z)=140~GeV  \ee
and $\zeta=0$
\be\label{bert2.18}           
          m_t(m_Z)=168~GeV~,~m_H(m_Z)=137~GeV ~~. \ee
Both solution do not differ much since $\lambda$ at the Planck 
scale is found to be small (but not zero). I prefer the solution 
with $\zeta=0$ because of the good behaviour (no divergence) of 
$\sigma$ in the broken as well as in the unbroken phase. 
Furthermore, $S_{Log\Lambda} $ can be expressed in terms of 
the $\beta$-function of $\frac{J^2_0}{8\lambda}$,
i.e. the 
$\beta$-function of the zero order expression of $W(J)$ \cite{6}. 
The numbers obtained in 
(\ref{bert2.17}) and (\ref{bert2.18}) differ little from the result obtained 
by Bennett, Nielsen and Froggatt in the framework of their anti 
grand unification model. This model requires $\lambda(m_{Planck})=0$, 
(not far away from 
$\lambda(m_{Planck})=0.04$ obtained here) and the vanishing 
of $S_{Log\Lambda}$.

We have seen, that the possibility still exists that the particles 
of the standard model have adjusted their couplings such as 
to stabilize their masses. 
If this is so,
the hierarchy problem is no more a problem of protecting the mass 
of the spin zero
Higgs particle compared to the masses of the spin 1/2 particles
which are protected by chiral symmetry.

Clearly, as long as higher-order calculations of $V(\sigma,J_0)$
are not available, the scale dependence of the couplings entering
the expression for $\sigma(J_0)$ brings in large uncertainties.
So far it is only a hope
that the one-loop result is not entirely misleading us.
In higher-loop calculations the quadratic divergence is
not well defined. One needs a lattice regularization or a
regularization by supersymmetry at some high scale. In
the latter case the quadratic divergence
is only an effective one up to the high scale which could 
again be the Planck scale.

\subsection{Masses and Mixings of Quarks and Leptons}
\setcounter{equation}{0}

The possible intimate relation between Higgs and top
discussed in the previous paragraph suggests also a dominant
role of the top for the structure of the quark and lepton mass
matrices. The masses of the lighter particles can then be
expected to be related to the top mass by powers of a small
constant \cite{7,8}. Let us look at the quark and charged lepton masses
at the common scale $m_Z$ in the $\overline{MS}$ scheme (in GeV) \cite{9}
\be\label{bert3.1}
\begin{array}{lll}
m_t=173\pm6& m_b=2.84\pm0.10&
m_\tau=1.78\\
m_c=0.58\pm0.06& m_s=(70\pm14)10^{-3}&
m_\mu=106\times 10^{-3}\\
m_u=(2.0\pm0.5)10^{-3}& m_d=(3.6\pm0.8)10^{-3}&
m_e=0.51\times 10^{-3}\end{array}
\ee
We take as the small parameter $\epsilon\simeq\sqrt{\frac{m_c}{m_t}}
=0.058\pm0.004$. Then, the rule $m_t:m_c:m_n=1:\epsilon^2:\epsilon^4$
may be supposed to hold up to $O(\epsilon)$ corrections.
Similarly, the ratio of down quark masses and the ratio of charged lepton
masses are taken to be simple rational numbers times integer
powers of $\epsilon$. It is then plausible to use a corresponding
geometric structure also for the off-diagonal elements of the mass
matrices. Here, I will not go into details since the paper
containing these suggestions is published \cite{8}. I will
only quote the results: The up-quark matrix can be taken real
and symmetric. For the down-quark and the charged lepton matrices
the simplest possible textures are used for which the 3rd
generation decouples from the first and second. The first and
second generation can mix by a complex entry. Taking this mixing
coefficient of the light generations purely imaginary, one obtains
-- for fixed $\epsilon$ -- a maximal CP-violation. It manifests
itself by making the unitarity triangle to be a right-handed
one with the unitarity angle $\gamma\simeq90^o$. Besides the
top mass and $\epsilon$ there are only two additional parameters:
the beauty to top and the $\tau$ to beauty mass ratios. With
these parameters one gets the following, so far quite successful,
numbers (in GeV):
\be\label{bert3.2}
\begin{array}{lll}
m_t=174& m_b=2.8&
m_\tau=1.77\\
m_c=0.58& m_s=84\times10^{-3}&
m_\mu=103\times 10^{-3}\\
m_u=1.95\times10^{-3}& m_d=4.2\times10^{-3}&
m_e=0.52\times 10^{-3}\end{array}
\ee
\be\label{bert3.3}
|V_{us}|=0.216,\ |V_{cb}|=0.041,\ |V_{ub}|=0.0034,\
|V_{td}|=0.0094,\ee
and
$\alpha=70^0,\ \beta=20^o,\ \gamma=90^o$.

\bigskip
It is a pleasure to thank Norma Mankoc Borstnik and Holger Nielsen
for the organization of the stimulating workshop in pleasant
surroundings. For lively and helpful discussions about the
possible Higgs-top connection I am much indebted to Ulrich
Ellwanger and Christof Wetterich.

\setcounter{page}{45}
\setcounter{section}{6}
\newpage
\stepcounter{section}
\setcounter{equation}{0}
\section*{\center Composite Models and Susy }
\centerline{ HANNS STREMNITZER}
\centerline{\it Institute for Theoretical Physics, University of Vienna} 
\centerline{\it A-1090 Vienna, Austria}
\vskip 5mm
\subsection{Introduction}

The idea that the standard model of strong and electroweak interactions itself
is an effective theory of another, more fundamental interaction and
quarks, leptons and bosons are composites of more fundamental fields is
almost as old as the standard model itself.  Since quarks carry flavor
and color from a purely spectroscopical point of view, one can think of
them as being glued together from entities having flavor only 
(``flavons'') and others having color only (``chromons'').  To get a
similar picture for leptons, Pati and Salam,$^{(1)}$ who invented this
idea as early as 1974, viewed the lepton number as a fourth color.  A
new QCD-like gauge interaction (``metacolor,'' ``hypercolor'') binds the
constituents (preons) together and its residual Van-der-Waals-type
interactions are the interactions of the standard model.

But there are other reasons of taking the existence of strong
fundamental gauge interactions seriously.  They come from the --- up to
now --- unexplored Higgs-sector of the standard model.  Actually, the
only job of the Higgs doublet in the standard model is to provide
symmetry breaking.  This requires a certain form of the Higgs potential,
so the ground state of the system is no longer invariant under
$SU(2)_{L}\times U(1)\,$.  Although there is nothing wrong with this
mechanism, we have not observed such a pattern for elementary boson
fields, but what we have observed is spontaneous symmetry breaking in
QCD, where the global chiral symmetry of the quarks is broken by
condensates $\langle\bar{q}_{L}q_{R}\rangle$ $\neq$ $0$ as effect of the
strong color force.  In fact, technicolor theories$^{(2)}$ have been
invented only for this purpose.

Once familiar with the idea, one will address the whole flavor problem
to composite models.  In fact, any underlying strong interaction theory
must yield the Yukawa couplings, the condensate values and therefore the
masses of fermions as a result rather than as fundamental parameters.
It is therefore a challenging task for any model builder to provide at
least a gross understanding of the mass hierarchies, as well as to give
an argument how many families we should expect.

Some composite models$^{(3,4)}$ also address problems usually associated
with Grand Unified Theories, like the existence and incorporation of
additional baryon-number (or even B-L)-violating interactions.  Here the
experimental 
evidence  is the existence of the baryon asymmetry in the
universe, a measurement to be taken seriously by any particle physicist.
This, however, means having the compositeness scale very high $\Lambda>
10^{11}$ GeV, which usually makes it difficult but not impossible$^{(5)}$ to
make low energy predictions.

\subsection{The Size of Composite Quarks}

If a particle is composite, its most important parameter is its size.
The inverse size of a particle shows up as a scale parameter in all
couplings with dimension, like magnetic moments.  The easiest way to
define the size of  a fermion$^{(3)}$ is by the strong
four-fermi-interaction
\be
{\cal L}_{ST} = {k^{2}\over M^{2}}\cdot\bar{\psi}_{\alpha}
\psi_{\alpha} \cdot \bar{\psi}_{\alpha}\psi_{\alpha}\; .
\ee

For a strong interaction, we expect ${k^{2}\over 4\pi}\sim 0(1)$ and
define
\be
\Lambda_{\alpha}^{2} = {4\pi\over k^{2}}\cdot M^{2}\; .
\ee
For instance, from low-energy proton-proton scattering, we find for the
nucleon
\be
\Lambda_{p}\sim m_{\rho} \sim 700\; {\rm MeV}\; .
\ee

In a preonic theory, all the inverse sizes of the composites should be
related to a fundamental scale $\Lambda\,$.  This scale is usually
defined by the value of the condensates
\be
\langle\bar{f}_{L}^a f_{R}^b \rangle \simeq A^{ab} \Lambda^{3}
\ee
where the flavor matrix $A^{ab}$ is assumed to be $O(1)$ unless there 
exists a peculiar suppression mechanism, as we will discuss below.

If we assume that the composite fermion $\psi_{\alpha}$ acquires its
mass through dynamical breaking of chiral symmetry, and if in addition
the fermion $f$ is a constituent of $\psi\,$, we find the
relation$^{(6)}$
\be
\Lambda_{\alpha} = {1\over r_{\alpha}} = k\left({\Lambda^{3}\over
m_{\alpha}}\right)^{1\over 2}\; .
\ee
This relation is well known for QCD, where $\Lambda = \Lambda_{QCD}$
$\simeq$ 250 MeV, $m_{\alpha}$ $\equiv$ $m_{p}$ $\simeq$ 1 GeV, $k\sim
g_{QCD}$ $=$ $g_{\rho NN}$ $=$ 5.5.  We find $\Lambda_{p}\sim$ 700 MeV
in agreement with equ.(3).

Now let us impose equ. (5) on a preonic-type theory, yielding
composite quarks.  In this case the condensate in (4) is the one
relevant for $SU(2)_{L}\times U(1)$-breaking, $\psi_{\alpha}$ is a
quark, and $f$ is one of its constituents.  Since, however, the
condensates as well as the masses form a matrix in flavor space, we can
only give a statement about the heaviest eigenvalue.  This yields,
\be
\left(\Lambda_{q}\right)_{H} = k\left[ \Lambda_{HC}^{3} /
m(q_{H})\right]^{1/2} \; .
\ee
Assuming $m(q_{H}) = m_{top} = $ 175 GeV, $k^{2} \simeq 10\,$, 
$\Lambda_{HC}\simeq (\sqrt{2}G_{F})^{-1/2}$ $\simeq$ 150 GeV from 
electroweak symmetry breaking, we
arrive at $\left(\Lambda_{a}\right)_{HC}$ $<$ 1,6 TeV.  We would like
to stress the point that this result is quite general, as long as the
{\it same} constituents are used making up the quark as well as the
condensate and as long as there is no special suppression mechanism. \\

\subsection{Constraints on Preon Dynamics}

If we take the more radical point of view and assume {\it all} particles
are composite, even leptons, gluons, and photons, the compositeness scale
is presumably very high, $\Lambda = \Lambda_{M} \simeq
10^{11}$ GeV. Such a theory must have a spectrum of massless composites, 
and therefore also
a residual chiral symmetry to protect fermion masses and a residual
gauge symmetry to prevent vector meson masses.  Let us outline a few
general theorems which are important in such a case.

Assume we start with a preonic theory (QPD) based on fundamental
fermions (with or without fundamental bosons), which has an invariance
group $G_{M}({\rm local})$ $\times$ $G_{fc}$ (global or local).  At a
certain scale $\Lambda_{M}$ the local $G_{M}$ interactions become strong
and form bound states (composites), which are $G_{M}$ singlets.  In
addition, part of $G_{fc}$ is spontaneously broken by condensates
$\langle \bar{f}f\rangle_{o}\neq 0\,$, yielding an effective low energy
symmetry group $G_{LE}\subset G_{fc}$ for the composites.  This
effective symmetry group is local, if the composite vector mesons
carrying the quantum numbers of $G_{LE}$ are born massless at
$\Lambda_{M}\,$, otherwise it is global.  Can $G_{LE}$ contain the
familiar $SU(3)_{C}\times SU(2)_{L}\times U(1)_{Y}$ gauge group? 

Before going to certain details, we should remark the following:
Certainly any effective theory among composite objects is
nonrenormalizable,  because one can write down arbitrarily high
dimensional operators in terms of the fundamental fields.  However, in
terms of composite fields, $d$-dimensional operators are suppressed by a
scale factor $\Lambda_{M}^{-(d-4)}$ $(d>4)\,$, and their contribution is
of order $\left( m / \Lambda_{M}\right)^{d-4}\,$, where $m$ is the mass
of the composites.  Therefore, if we restrict ourselves to the mass zero
composites ({\it i.e.} $m/\Lambda_{M} << 1$), the remaining
low-dimensional operators have to be renormalizable and therefore are
assumed to be a gauge theory (with certain Yukawa couplings).

One has further to distinguish whether $G_{M}$ is vectorlike or
chiral. In the latter case, the preons have to be in a complex
representation $R$ alone without being accompanied by another set of
preons belonging to $\bar{R}\,$. In the first case, one has the appealing 
feature that the underlying vectorlike gauge theory is responsible for the 
formation of chiral composites and parity violation. This is a challenging
task, in particular since for such QCD-like theories a number of constraints 
and no-go-theorems have been worked out:

\newcounter{00}
\begin{list}
{(\roman{00})}{\usecounter{00}\labelwidth=1cm\labelsep=.4cm}
\item The Weingarten,$^{(7)}$ Nussinov,$^{(8)}$ and Witten$^{(9)}$
mass inequalities require, among other things, that the lightest
composite fermion be heavier than the light composite boson:
\be
m_{F} >  (N/2) m_{B}
\ee
(assuming $G_{M} = SU(N))\,$;
\item The Weinberg-Witten theorem$^{(10)}$ shows that in a vectorlike
theory with a Lorentz-covariant conserved current (energy-momentum
tensor), no composite particle with spin $j > 1/2$ $(j> 1)$ with mass
zero and corresponding charge unequal zero can be formed.
\item The Vafa-Witten theorem$^{(11)}$ shows that vectorlike {\it
global} symmetries $U_{L+R}(N)$ are not broken by condensates,
and massless bound states do not form from massive constituents.
\item Contrary to this, Weingarten$^{(7)}$has shown , using lattice
field theory calculations, that global chiral symmetries $U_{L}(N)
\times U_{R}(N)$ are broken at the composite level. 
\item Furthermore, even in chiral QPD, one has to fulfill 't Hooft's
anomaly-match\-ing conditions$^{(12)}$ which state that the
Adler-Bell-Jackiw anomalies$^{(13)}$ associated with background gauge
fields of $G_{fc}$ have to be the same at the preon as at the composite
level.  (It should be noted that $G_{M}\,$, as a gauge theory, has to be
anomaly-free anyway in order to be consistent.)  This puts certain
severe restrictions on the preon representation of $G_{M}$ in chiral
QPD. 
\end{list}

We see that the mass inequalities$^{(7-9)}$ and the Vafa-Witten
constraint prohibit the appearance of a residual $SU(2)_{L}$ protecting
the masses of composite quarks and leptons (without  having lighter
composite bosons) in a vectorlike QPD.  Fortunately, these constraints
can be avoided in a {\it supersymmetric} QPD for the following reasons:
\newcounter{000}
\begin{list}
{(\roman{000})}{\usecounter{000}\labelwidth=1cm\labelsep=.4cm}
\item Both the Weingarten inequality and the Vafa-Witten theorem use the
positivity of the fermion determinant in QCD-type theories in their
proof.  However, in supersymmetric QCD-theories, there exist
Yukawa-interactions between gauginos, matter fermions and spin-zero
matter bosons which destroy the positivity.$^{(14)}$
\item The proof of the breakdown of chiral symmetry in lattice QCD also
ignores the presence of Yukawa interactions.
\item In a local supersymmetric Yang-Mills theory there is no
Lorentz-covariant conserved current and energy-momentum tensor.
\item The anomaly matching conditions are easily fulfilled in a rather
economical way, if quarks and leptons are constructed as fermion-boson
composites (as is the case in supersymmetric theories).$^{(15)}$
\end{list}

\subsection{SUSY-Quantumpreondynamics: A specific model}

The necessity for supersymmetry at the preon level opens two new
aspects.  First, one is led to the intriguing possibility that the
supersymmetric QPD itself is the result of a supergravity or superstring
theory at a higher scale, presumably the Planck scale.  Secondly, one
has to answer how the supersymmetry is broken.  The most promising way
is to assume the breakdown of local supersymmetry through gaugino
condensates$^{(16)}$ formed by the $G_{M}$-gauge force itself at
$\Lambda_{M}\,$.  This in turn implies the appearance of a vacuum
expectation value for the auxiliary fields and therefore a breakdown of
global supersymmetry.  There exists however, certain restrictions on
global SUSY-breaking, depending on the representation and mass of your
fundamental preon fields.$^{(17)}$  For such cases, one finds that the
flavor matrix $A^{ab}$ in front of the scale parameter $\Lambda^3$
has to be damped by powers of $\Lambda_{M} /
M_{P\ell}\,$. In the following, we give an example of a SUSY-composite 
model of this type.

   The model under consideration$^{18,19}$ is based on a set of
masseless chiral superfields, each belonging to the
fundamental representation of the metacolor gauge symmetry
$SU(N)$. The superfields carry also flavor-color quantum numbers
according to the gauge group $SU(4)_c \times SU(2)_L\times SU(2)_R$:
\be
\Phi^{a,\sigma}_{\pm} = (\varphi,\psi,F)^{a,\sigma}_{L,R}
\ee
where $\sigma$ denotes the metacolor index, $a$ denotes flavor-color
indices up/down and red, yellow, blue and lilac for the lepton color.

It is assumed that the metacolor force becomes strong and
confining at a scale $\Lambda_M \simeq 10^{11}$ GeV, with the
following effects:
\begin{itemize}
\item Three light chiral families$^{(19)}$ of composite quarks and
leptons $\left( q_{L,R}^{i}\right)_{i=1,2,3}$  and two 
 vector-like families $Q_{L,R}$ and $Q^{\prime}_{L,R}$, coupling
 vectorially to $W_{L}$'s and $W_{R}$'s, are formed:
\begin{eqnarray}
q_L = \Phi^a_+ (\Phi^r_-)^* \simeq (2,1,4), \nonumber \\
q_R = \Phi^a_- (\Phi^r_+)^* \simeq (1,2,4), \nonumber \\
Q_{L,R} = \Pi_{\pm} (\Phi^a_+ (\Phi^r_+)^*) \simeq (2,1,4), \\
Q'_{L,R} = \Pi_{\pm} (\Phi^a_- (\Phi^r_-)^*) \simeq (1,2,4) 
\nonumber
\end{eqnarray}

where $\Pi_{\pm}$ are the projection operators onto left/right 
chiral superfields;
\item Supersymmetry-breaking condensates are formed; they include
 the metagaugino condensate
$\langle\vec{\lambda}\cdot\vec{\lambda}\rangle$ and the matter
fermion-condensates $\langle\bar{\psi}^{a}\psi^{a}\rangle\,$. 
  Noting that, within the class of models under consideration, the 
index theorem prohibits a dynamical breaking of supersymmetry
in the absence of gravity$^{(20)}$, so
the formation of these condensates must need the collaboration between
the metacolor force and gravity.  As a result, each of these condensates is
expected to be damped by one power of $\left( \Lambda_{M} /
M_{P\ell}\right) \simeq 10^{-8}$ relative to $\Lambda_{M}\,$ $^{(17)}$:
\begin{eqnarray}
\langle\vec{\lambda}\cdot\vec{\lambda}\rangle & = &
\kappa_{\lambda}\Lambda_{M}^{3}\left(\Lambda_{M} /
M_{P\ell}\right) , \nonumber \\*
\langle\bar{\psi}^{a}\psi^{a}\rangle & = &
\kappa_{\psi_{a}}\Lambda_{M}^{3}\left(\Lambda_{M} / M_{P\ell}\right)
\end{eqnarray}
Here, the indices $a$ are running over color and flavor quantum numbers.
 The condensates $\langle\bar{\psi}^{a}\psi^{a}\rangle\,$,
 break not only SUSY but also the electroweak
symmetry $SU(2)_{L}$ $\times$ $U(1)_{Y}\,$ therefore giving mass
to the electroweak gauge bosons. The coefficients $\kappa_{\lambda}$ 
and $\kappa_{\psi_{a}}\,$, apriori, are expected
to be of order unity within a factor of ten (say), although
$\kappa_{\lambda}$ is expected to be bigger than $\kappa_{\psi}$'s, 
typically by factors of 3 to 10, because the $\psi$'s are in the fundamental
and the $\lambda$'s are in the adjoint representation of the metacolor
group.
\item Furthermore, supersymmetry-preserving condensates, which
however break the gauge group $SU(4)_c \times SU(2)_L\times SU(2)_R$
to the low-energy gauge group $SU(3)_c \times SU(2)_L \times U(1)_Y$
are assumed to form as well. They provide a large superheavy
Majorana mass to the right-handed neutrinos and may play an
interesting role in the discussion of inflationary models $^{(21)}$.
\end{itemize}

Now, the vector-families $Q_{L,R}$ and
$Q_{L,R}^{\prime}$ acquire relatively heavy masses through the
metagaugino condensate $\langle\vec{\lambda}\cdot\vec{\lambda}\rangle$
of order $\kappa_{\lambda}\Lambda_{M}\left(\Lambda_{M} / M_{P\ell}\right)$
$\sim$ 1~TeV which are independent of flavor and color.  But
the chiral families $q_{L,R}^{i}$ acquire masses primarily through their
mixings with the vector-like families $Q_{L,R}$ and $Q_{L,R}^{\prime}$
which are induced by $\langle\bar{\psi}^{a}\psi^{a}\rangle\,$.  This is
because the direct mass-terms  cannot be induced through
two-body condensates. Thus, ignoring $QCD$
corrections and higher order condensates for a moment, 
the Dirac-mass matrices of all four types -
i.e., up, down, charged lepton, and neutrino - have the form:
\begin{eqnarray}
M^{(o)} =
\bordermatrix{& q_{L}^{i} & Q_{L} & Q_{L}^{\prime}\cr
\overline{q_{R}^{i}} & O & X\kappa_{f} & Y\kappa_{c}\cr
\overline{Q_{R}} & Y^{\prime\dagger}\kappa_{c} & \kappa_{\lambda} & O\cr
\overline{Q_{R}^{\prime}} & X^{\prime\dagger}\kappa_{f} & O &
\kappa_{\lambda}\cr}~~~. 
\end{eqnarray}
Here, the index $i$ runs over three families, $f,c$ denotes
flavor- or color-type condensate, and
the quantitites  $X\,$, $Y\,$, $X^{\prime}$ and
$Y^{\prime}$ are column matrices in the family-space and have
their origin in the detailled vertex structure of the
corresponding preonic diagrams. They are expected to be numbers 
of order $\simeq 1 $. As a
result, the Dirac mass-matrices of all four types have a natural
see-saw structure. \\

Naturally, the key ingredient in this prediction is the seesaw mechanism,
provided by the existence of the two set of vector--like quarks.
The prediction of the masses, coupling constants and decay rates
of these quarks and their verification by experiments is 
therefore a crucial test of the model itself. 
For this reason, detailled predictions for the discovery 
of the new heavy vectorlike Quarks have been worked out.
Let us shortly repeat the results:
\begin{itemize}
\item We expect four heavy quarks $U_1,U_2,D_1,D_2$ as well as four 
leptons $E_1,E_2,N_1,N_2$ whose electroweak properties are given 
by $Q, Q'$ including their mixing;
\item Fitting the six parameters of the mass matrix and allowing for 
about 10\% corrections due to electroweak final state interactions, 
one gets
\begin{eqnarray}
U_1 \simeq 1814~ GeV ~~~~~N_1 \simeq 765~GeV \nonumber \\
D_1 \simeq 1787 ~GeV ~~~~~E_1 \simeq 763~GeV \nonumber \\
U_2 \simeq 1504 ~GeV ~~~~~N_2 \simeq 581~GeV  \\
D_2 \simeq 1465 ~GeV ~~~~~E_2 \simeq 667~GeV \nonumber
\end{eqnarray}
\end{itemize}

In summary, two vector--like families, not more not less, with one 
coupling vectorially to $W_L$'s and the other to $W_R$'s (before 
mass--mixing), with masses of order $\simeq 1TeV$, constitute a 
{\it hall--mark} and a crucial prediction of the SUSY preon
model $^{(18,19)}$ under consideration. There does not seem to be any 
other model including superstring--inspired models of elementary 
quarks and leptons which have a good reason to predict two such complete 
vector--like families with masses in the TeV range.\\

\subsection{Acknowledgements}

I would like to thank Prof. M. Borstnik and the organizers of the
workshop in Bled for the hospitality, the interesting discussions and
the beautiful meeting. \\

\newpage
\thispagestyle{empty}
$\,$
\newpage
\thispagestyle{empty}
\stepcounter{part}
$\,$
\vskip 10cm
\begin{center}
 \begin{Huge}
  \begin{bfseries}
      \sc Discussions
  \end{bfseries}
 \end{Huge}
\end{center}
\newpage
\setcounter{section}{0}
\stepcounter{section}
\setcounter{equation}{0}
\section*{\center Are spins and charges unified? How can one otherwise
understand the connection between the handedness (the spin) and the weak
charge? } 

\centerline{\large  Anamarija\footnote[1]{Anamarija Bor\v
stnik, Dept. of Phys., University of Ljubljana, Jadranska 19,
1000 Ljubljana} and Norma}

\vspace{5mm}

\noindent
This question seems to Norma the essential open question
of the Standard electroweak model. The answer to this question
should show the way  beyond the Standard model.

\subsection{Introduction}


\noindent
In Norma's talk the  algebras and
subalgebras of   two kinds of  generators  of the
Lorentz transformations, defining in  $14  $-dimensional Grassmann
space\footnote[2]{In the talk of Norma one additional
dimension to 14 dimensions was proposed, needed to properly
define $\gamma ^a$ matrices, in the contribution to 
the discussions written by Norma and Holger, another definition of
the $\gamma^a$ matrices was proposed, which does not need the
additional dimension. The algebra of the group $SO(1,13)$ does
not depend on the additional dimension.} the group $ SO(1,13) $,
both the linear differential operators, 
were presented and some of their representations were discussed.
According to two kinds of generators 
defined in the linear vector space, spanned over  Grassmann
coordinate space, there are also two kinds of representations: 
we call them spinorial and vectorial representations,
respectively. We choose Grassmann odd polynomials to describe 
spinorial and Grassmann even polynomials to describe 
vectorial kind of vectors.

\noindent
Since the group $ SO(1,d-1) $ contains for $ d=14 (+1) $ as subgroups
the groups $ 
SO(1,3)$, needed to describe spins of fermions and bosons, as
well as $U(1), SU(2) $ and $ SU(3) $, needed to describe the
Yang-Mills charges of fermions and bosons, {\bf the spin and the
Yang - Mills charges of either fermions or  bosons are in the
presented approach unified}. Since spins and charges are described
by the representations of the generators of the Lorentz
transformations of either fermionic or of bosonic character, it
means that {\bf fermionic  states must belong  to the
spinorial representations with respect to the groups,
describing charges,  while
bosonic states must belong  to the vectorial representations
with respect to the groups, describing charges}. 

\noindent
Among  representations of the proposed approach are 
 the ones, needed to 
describe the quarks, the leptons and
the gauge bosons, which appear in the Standard electroweak
model\cite{norma,ana}. 
We find left handed spinors, $SU(3)$ triplets and $SU(2)$
doublets with $U(1)$ charge  $ \frac{1}{6}$ and
right handed spinors, $SU(3)$ triplets and $SU(2)$ singlets 
with $U(1)$ charge $ \frac{2}{3}$ and $ - \frac{1}{3}$, which
describe quarks. We find left handed spinors, $SU(3)$ singlets
and $SU(2)$ doublets with $U(1)$ charge $- \frac{1}{2}$
and right handed spinors, $SU(3)$ singlets and $SU(2)$ singlets
with $U(1)$ charge $0$ or $- 1$, which describe leptons. We find
also the corresponding representations for anti quarks and anti 
leptons. 
We find the four vectors, $SU(3)$ triplets
and $SU(2)$ singlets with $U(1)$ charge $0$, describing gluons
and  $SU(3)$ singlets and $SU(2)$ triplets with
$U(1)$ charge $0$, describing massless weak bosons and $SU(3)$
singlets and $SU(2)$ singlets 
with $U(1)$ charge $0$, describing  a $U(1)$ gauge 
field.
One can
find\cite{ana} for vectorial case besides octets and
singlets of $SU(3)$ also 
triplets and besides triplets and singlets of $SU(2)$ also
doublets. These representations have an odd Grassmann
character in $ SO(10)$ and the correspondingly even or odd
Grassmann character in the rest of space.
Accordingly the Higgs's boson of this model\footnote{Higgs could
appear also as a 
constituent field!} appears 
as a scalar, which is  a $SU(3)$ singlet and
$SU(2)$ doublet,  with an odd
Grassmann character in $SO(1,3) $ and $SU(2) $ part of the
Grassmann space.
Since  the four dimensional subspace of the Grassmann space,
above which the group $SO(1,3)$ is defined, has $2^4 = 16$ basic
functions, {\bf the approach predicts four} rather than three
{\bf families of quarks and leptons}\footnote[4]{ See also the
contribution to the discussions written by Norma and Holger},
provided that this symmetry manifests already on the level of
quarks and leptons.

\noindent
However, the supersymmetric partners of  the gauge
bosons, required by the supersymmetric extension of the
Standard Electroweak Model, can in the proposed theory exist only
as constituent particles.
\noindent
In this contribution  we  demonstrate
that Manko\v c approach, defining all the internal degrees of
freedom in an unique way in Grassmann space and accordingly
unifying spins and charges, offers 
a possible explanation why the postulate of the Standard model
that only left handed fermions carry the weak charge occurs, or
equivalently why the
weak interaction breaks left-right symmetry. 

\subsection{ Subgroups of SO(1,7), SO(1,9) and SO(1,13) and 
representations} 

\noindent
We first define the operators, which in Grassmann space define
vectorial and spinorial representations. We start with
\begin{equation}
p^{\theta a}: = i \overrightarrow{\partial}{ }^{\theta}_{a},\;\;
\tilde{a}^a: = i(p^{\theta a} - i
\theta^a),\;\;\tilde{\tilde{a}}{ }^a: = -(p^{\theta a} +i
\theta^a). 
\end{equation}
We find
\begin{equation}
\{ p^{\theta a}, p^{\theta b} \} = 0 = \{ \theta^a, \theta^b
\},\;\; \{ p^{\theta a}, \theta^b \} = i \eta^{ab},
\end{equation}
\begin{equation}
\{\tilde{a}^a, \tilde{a}^b \} = 2 \eta^{ab} =
\{\tilde{\tilde{a}}{ }^a,
\tilde{\tilde{a}}{ }^b \},\;\;\{\tilde{a}^a,
\tilde{\tilde{a}}{ }^b \} = 0,\;\; \tilde{\gamma}^a: =
i\tilde{\tilde{a}}{ }^0 \tilde{a}^{a}.  
\end{equation}
Then we define two kinds of binomials. The first kind is made of
operators forming the Heisenberg odd algebra
\begin{equation}
S^{ab}: = ( \theta^a p^{\theta b} - \theta^b p^{\theta a} ), 
\end{equation}
the second kind of operators forming the Clifford algebra
\begin{equation}
\tilde{S}^{ab} = \frac{i}{4} [\tilde{a}^a, \tilde{a}^b] =
\frac{i}{4} [\tilde{\gamma}^a, \tilde{\gamma}^b], \;\;{\rm
with}\; [A,B] = AB-BA.
\end{equation}
Either $S^{ab}$ or $\tilde{S}^{ab}$ fulfill the Lorentz algebra.

\noindent
We shall write $M^{ab}$ for either $S^{ab}$ or
$\tilde{S}^{ab}$. We shall use the space of coordinates
$\theta^m$, $m \in \{0,1,2,3\}$ to describe spin degrees of
freedom and $h \in \{5,6,..,d\}$ to describe charges.

\subsubsection{ $SO(1,3)$}

\noindent
We define the representations of the group $SO(1,3)$ through 
two $SU(2)$ subgroups in the standard way
\begin{equation}
N^{\pm}_i: = \frac{1}{2} (\frac{1}{2}\epsilon_{ijk} M^{ik} \pm
iM^{0i}),\;\;[N^{\pm}_i, N^{\pm}_j] = i \epsilon_{ijk}
N^{\pm}_k,\;\; [N^{\pm}_i, N^{\mp}_j] = 0.
\end{equation}
We shall present in Table I the representations of only the
spinorial type of generators. 
Taking into account the expressions for the operators
$\tilde{S}^{mn}$ one easily finds the eigenvectors of the
Casimirs $\sum_i(\tilde{N}^{\pm}_i)^2$ and $\tilde{N}^{\pm}_3$ as the
polynomials of $\theta^m$.

\begin{center}
\begin{tabular}{|cc|c|ccc|c|}
\hline 
a & i &$<\theta|\varphi^a_i>$& $\tilde{S}_3$& $ \tilde{K}_{3}$& 
$\tilde{\Gamma}$& family \\ 
\hline
1 & 1 &$\frac{1}{2}(\tilde{a}^1 - i \tilde{a}^2)(\tilde{a}^0 -
\tilde{a}^3)$ &$-\frac{1}{2}$ &$ \frac{i}{2}$&-1&\\
1 & 2 &$-\frac{1}{2}(1 + i\tilde{a}^1 \tilde{a}^2)(1-
\tilde{a}^0 \tilde{a}^3) $ &$\frac{1}{2} $&$ -\frac{i}{2}$&-1&\\
&&&&&&I\\
2 & 1 &$\frac{1}{2}(\tilde{a}^1 - i \tilde{a}^2)(\tilde{a}^0 +
\tilde{a}^3)$ &$-\frac{1}{2}$ &$ -\frac{i}{2}$&1&\\
2 & 2 &$-\frac{1}{2}(1 + i\tilde{a}^1 \tilde{a}^2)(1+
\tilde{a}^0 \tilde{a}^3)$  &$\frac{1}{2} $&$ \frac{i}{2}$&1&\\
\hline
3 & 1 &$\frac{1}{2}(\tilde{a}^1 - i \tilde{a}^2)(1- \tilde{a}^0 
\tilde{a}^3)$ &$-\frac{1}{2}$ &$ -\frac{i}{2}$&1&\\
3 & 2 &$-\frac{1}{2}(1 + i\tilde{a}^1 \tilde{a}^2)(\tilde{a}^0 -
 \tilde{a}^3) $ &$\frac{1}{2} $&$ \frac{i}{2}$&1&\\
&&&&&&II\\
4 & 1 &$\frac{1}{2}(\tilde{a}^1 - i \tilde{a}^2)(1 + \tilde{a}^0 
\tilde{a}^3)$ &$-\frac{1}{2}$ &$ \frac{i}{2}$&-1&\\
4 & 2 &$-\frac{1}{2}(1 + i\tilde{a}^1 \tilde{a}^2)(\tilde{a}^0 +
\tilde{a}^3)$  &$\frac{1}{2} $&$ -\frac{i}{2}$&-1&\\
\hline
5 & 1 &$\frac{1}{2}(1 -i \tilde{a}^1 \tilde{a}^2)(\tilde{a}^0 -
\tilde{a}^3)$ &$-\frac{1}{2}$ &$ \frac{i}{2}$&-1&\\
5 & 2 &$-\frac{1}{2}(\tilde{a}^1 +i \tilde{a}^2)(1-
\tilde{a}^0 \tilde{a}^3) $ &$\frac{1}{2} $&$ -\frac{i}{2}$&-1&\\
&&&&&&III\\
6 & 1 &$\frac{1}{2}(1- i\tilde{a}^1 \tilde{a}^2)(\tilde{a}^0 +
\tilde{a}^3)$ &$-\frac{1}{2}$ &$ -\frac{i}{2}$&1&\\
6 & 2 &$-\frac{1}{2}(\tilde{a}^1 +i \tilde{a}^2)(1+
\tilde{a}^0 \tilde{a}^3)$  &$\frac{1}{2} $&$ \frac{i}{2}$&1&\\
\hline
7 & 1 &$\frac{1}{2}(! -i\tilde{a}^1 \tilde{a}^2)(1- \tilde{a}^0 
\tilde{a}^3)$ &$-\frac{1}{2}$ &$ -\frac{i}{2}$&1&\\
7 & 2 &$-\frac{1}{2}(\tilde{a}^1 +i\tilde{a}^2)(\tilde{a}^0 -
 \tilde{a}^3) $ &$\frac{1}{2} $&$ \frac{i}{2}$&1&\\
&&&&&&IV\\
8 & 1 &$\frac{1}{2}(1-i\tilde{a}^1 \tilde{a}^2)(1 + \tilde{a}^0 
\tilde{a}^3)$ &$-\frac{1}{2}$ &$ \frac{i}{2}$&-1&\\
8 & 2 &$-\frac{1}{2}(\tilde{a}^1 +i \tilde{a}^2)(\tilde{a}^0 +
\tilde{a}^3)$  &$\frac{1}{2} $&$ -\frac{i}{2}$&-1&\\
\hline
\end{tabular}
\end{center}

\noindent
Table I: The polynomials of $\theta^m$, representing the four
times two Dirac bispinors, are written. For each state the
eigenvalues of $\tilde{S}_3: =\tilde{S}^{1,2}, \tilde{K}_3,
\Gamma: = i \tilde{a}^0 \tilde{a}^1 \tilde{a}^2 \tilde{a}^3$ are
written. The Roman numerals tell the possible family number.
We use the relation $\tilde{a}^a |0> = \theta^a$.

\noindent
One can check that $\tilde{\gamma}^0$ transforms the two
bispinors of each family into one another. One also finds
that the operator with an odd Grassmann character $\tilde{a}^m$
transforms the first family into the second and the third into
the fourth, while $\tilde{\tilde{a}}^0$ transforms the first
family into the third and the second into the fourth. One also
finds that complex conjugation transforms the first family into
the fourth and the second into the third.

\subsubsection{ $SO(4)$}

\noindent
We use the Grassmann coordinates $\theta^5, \theta^6, \theta^7$
and $\theta^8$ to describe the weak and $U(1)$ charge.

\noindent
We define the generators of the two groups
\begin{equation}
\tau^{11}: = \frac{1}{2} ( M^{58} - M^{67}),\;\; \tau^{12}: =
\frac{1}{2} ( M^{57} + M^{68}),\;\; \tau^{13}: = \frac{1}{2} (
M^{56} - M^{78}), 
\end{equation}
while
\begin{equation}
\tau^{2w}: = \frac{1}{2} ( M^{56} + M^{78}).
\end{equation}
One finds
\begin{equation}
[\tau^{A i}, \tau{B j}] = \delta^{A B} \delta^{A 1} i
\epsilon_{ijk} \tau^{A k},
\end{equation}
with $ A \in(1,2),\;\; i\in (1,..,n_A),\;\;
n_A = 3\;\; {\rm for}\;\; SU(2)\;\; {\rm and}\;\; 1\;\; {\rm
for} \;\; U(1).$
We present in Table II the eigenstates of the Casimirs
$\sum_i (\tau^{1i})^2$ and$ \tau^{2 w}$ and of $\tau^{13}$ for
spinorial degrees of freedom.

\subsubsection{ $SO(1,7)$}

\noindent
We assume that the representations of the group $SO(1,7)$ are
the direct product of representations of the subgroups $SO(1,3)
$ and $ SO(4)$.
Taking into account the results of the last two subsections,
one finds that the operator $\tilde{S}^{mh}, m\in(0,1,2,3),
h\in(5,6,7,8)$ transforms the left handed $SU(2)$ doublets into
right handed $SU(2)$ singlets. Accordingly one finds in a
$SO(1,7)$ multiplet  left handed doublets and right handed
singlets. One can, of course, also find multiplets with left
handed singlets and right handed doublets, since the left-right
symmetry is not broken (yet).

\vspace{1mm}

\noindent
Table II: The eigenstates of the $\tau^{13}, \tau^{2w}$ are
presented. We find two doublets and four singlets of an even
Grassmann character and two doublets and four singlets of an odd
Grassmann character. One sees that complex conjugation
transforms one doublet of either odd or even Grassmann character
into another of the same Grassmann character changing the sign
of  the
value of $\tau^{13}$, while it transforms one singlet into
another singlet
 of the same Grassmann character and of the opposite value of $
\tau^{2w}$.  One can check that $\tilde{a}^h, \;\;
h \in (5,6,7,8)$, transforms the doublets of an even Grassmann
character into 
singlets of an odd Grassmann character.

\begin{center}
\begin{tabular}{|cc|c|cc|c|}
\hline 
a & i &$<\theta|\varphi^a{ }_i>$& $\tilde{\tau}^{13}$& $ \tilde{\tau}^{2w}$& 
Grass. character\\ 
\hline
1 & 1 &$\frac{1}{2}(1-i\tilde{a}^5 \tilde{a}^6)(1+ i\tilde{a}^7
\tilde{a}^8)$ &$-\frac{1}{2}$ &$ 0$&\\
1 & 2 &$-\frac{1}{2}(\tilde{a}^5 + i\tilde{a}^6)(
\tilde{a}^7 - i \tilde{a}^8) $ &$\frac{1}{2} $&$ 0$&\\
&&&&&\\
2 & 1 &$\frac{1}{2}(1+i\tilde{a}^5 \tilde{a}^6)(1- i\tilde{a}^7
\tilde{a}^8)$ &$\frac{1}{2}$ &$ 0$&\\
2 & 2 &$-\frac{1}{2}(\tilde{a}^5 - i\tilde{a}^6)(
\tilde{a}^7 + i \tilde{a}^8) $ &$-\frac{1}{2} $&$ 0$&\\
&&&&&even\\
3 & 1 &$\frac{1}{2}(1+i\tilde{a}^5 \tilde{a}^6)(1+ i\tilde{a}^7
\tilde{a}^8)$ &0&$\frac{1}{2}$ &\\
4 & 1 &$\frac{1}{2}(\tilde{a}^5 + i\tilde{a}^6)(
\tilde{a}^7 + i \tilde{a}^8) $ &0&$\frac{1}{2} $&\\
5 & 1 &$\frac{1}{2}(1-i\tilde{a}^5 \tilde{a}^6)(1- i\tilde{a}^7
\tilde{a}^8)$ &0&$-\frac{1}{2}$ &\\
6 & 1 &$\frac{1}{2}(\tilde{a}^5 - i\tilde{a}^6)(
\tilde{a}^7 - i \tilde{a}^8) $ &0&$-\frac{1}{2} $&\\
\hline
\hline
7 & 1 &$\frac{1}{2}(1+i\tilde{a}^5 \tilde{a}^6)(\tilde{a}^7-i
\tilde{a}^8)$ &$\frac{1}{2}$ &$ 0$&\\
7 & 2 &$-\frac{1}{2}(\tilde{a}^5 - i\tilde{a}^6)(1+
\tilde{a}^7 \tilde{a}^8) $ &$-\frac{1}{2} $&$ 0$&\\
&&&&&\\
8 & 1 &$\frac{1}{2}(1-i\tilde{a}^5 \tilde{a}^6)(\tilde{a}^7+i
\tilde{a}^8)$ &$-\frac{1}{2}$ &$ 0$&\\
8 & 2 &$-\frac{1}{2}(\tilde{a}^5 + i\tilde{a}^6)(1-i
\tilde{a}^7 \tilde{a}^8) $ &$\frac{1}{2} $&$ 0$&\\
&&&&&odd\\
9 & 1 &$\frac{1}{2}(1-i\tilde{a}^5 \tilde{a}^6)(\tilde{a}^7-i
\tilde{a}^8)$ &0&$-\frac{1}{2}$ &\\
10 & 1 &$\frac{1}{2}(\tilde{a}^5 + i\tilde{a}^6)(1+
\tilde{a}^7 \tilde{a}^8) $ &0&$\frac{1}{2} $&\\
11 & 1 &$\frac{1}{2}(1+i\tilde{a}^5 \tilde{a}^6)(\tilde{a}^7+i
\tilde{a}^8)$ &0&$\frac{1}{2}$ &\\
12 & 1 &$\frac{1}{2}(\tilde{a}^5 - i\tilde{a}^6)(1-
\tilde{a}^7 \tilde{a}^8) $ &0&$-\frac{1}{2} $&\\
\hline
\end{tabular}
\end{center}

\subsubsection{$SO(6)$}

\noindent
We use the Grassmann coordinates $\theta^9, \theta^{10}, \theta^{11},
\theta^{12}, \theta^{13}
$
and $\theta^{14}$ to describe the colour  and $U(1)$ charge.

\noindent
We define the generators of the two groups as follows
$$
\tau^{3\;1}: = \frac{1}{2} ( M^{9\;12} - M^{10\;11}),\;\; \tau^{3\;2}: =
\frac{1}{2} ( M^{9\;11} + M^{10\;12}),\;\; \tau^{3\;3}: = \frac{1}{2} (
M^{9\;10} - M^{11\;12}), $$
$$\tau^{3\;4}: = \frac{1}{2} ( M^{9\;14} - M^{10\;13}),\;\; \tau^{3\;5}: =
\frac{1}{2} ( M^{9\;13} + M^{10\;14}),\;\; \tau^{3\;6}: = \frac{1}{2} (
M^{11\;14} - M^{12\;13}),\\
$$
\begin{equation}
  \tau^{3\;7}: =
\frac{1}{2} ( M^{11\;13} + M^{12\;14}),\;\; \tau^{3\;8}: =
\frac{1}{2\sqrt{3}} ( M^{9\;10} +
M^{11\;12} - 2 M^{13\;14}), 
\end{equation}
while
\begin{equation}
\tau^{2c}: = -\frac{1}{3}  ( M^{9\;10} +
M^{11\;12} + M^{13\;14}).
\end{equation}
One finds
\begin{equation}
[\tau^{A i}, \tau^{B j}] = \delta^{A B}\;\; \delta^{A 3}\;\; i
f^A_{ijk} \;\; \tau^{A k},
\end{equation}
with $ A \in(1,2),\;\; i \in (1,..,n_A),\;\;
n_A = 8 \;\; {\rm for}\;\; SU(3)\;\; {\rm and}\;\; 1 \;\;{\rm
for} U(1).$ 
The constants $f^A_{ijk} $ are the structure constants of the
group $SU(3)$. 

\noindent
We can again find the eigenstates of $\sum_i (\tau^{3i})^2$,
$\tau^{3\;3}$, $\tau^{3\;8}$ and $\tau^{2\;c}$ and present them
as polynomials of $\theta^k,\;\; k \in (9,10,11,12,13,14)$. We find
triplets, anti triplets, singlets and anti singlets\cite{norma,ana}.
The operator $\tilde{a}^k,\;\; k \;\in\;\;(9,10,11,12,13,14)$ transforms
Grassmann even triplets into Grassmann odd anti triplets and singlets.

\subsubsection{$SO(1,9)$}

\noindent
We assume the representations of the group $SO(1,9)$ to be the
direct product of the representations of the group $SO(1,3)$ and
the group $SO(6)$. Accordingly one finds that left handed
$SU(3)$ triplets and right handed $SU(3)$ anti triplets and
singlets appear in the same $SO(1,9)$ multiplet. We also have
right handed triplets and left handed anti triplets and
anti singlets in the another multiplet. One has to find out
what would the break of the left-right symmetry bring.

\subsubsection{ $SO(1,13)$}

\noindent
We may assume that the representations of the group $ SO(1,13)$
are the direct product of the representations of the groups
$SO(1,3)$, $SU(4)$ and 
$SU(6)$ and that the $U(1)$ charge is the sum of the $\tau^{2w}$
and $\tau^{2c}$ charges:$Y: = \tau^{2w} + \tau^{2w}$. 
 
\noindent
In Table III we present the quantum numbers of fermions, which
appear in the proposed approach
as members of {\bf one multiplet of the group $SO(1,13)$, } if
 members
of the same multiplet are found by the application of the
operators $\tilde{S}^{ab}, a,b \in \{0,1,.,8\}$
on any starting representation.
One finds the four bispinors (rather than one), which differ among
themselves with respect to discrete symmetries of the Lorentz
subgroup $ SO(1,3) $ and which may represent four (rather than
three) families of
quarks and leptons.

\begin{center}
\begin{tabular}{|c|cccc|rrrc|rrrc|}
\hline
  &\multicolumn{4}{c|}{} & 
   \multicolumn{4}{c|}{SU(2) doublets} & 
   \multicolumn{4}{c|}{SU(2) singlets}\\ 
family & I & II& III& IV & $\tilde{\tau}^{13}$& $
\tilde{\tau}^{2w}+\tilde{\tau}^{2c} $& 
$\tilde Q$& $\tilde{\Gamma}^{(4)}$ & $\tilde{\tau}^{13}$&
$\tilde{\tau}^{2w}+ \tilde{\tau}^{2c} $&
$\tilde Q$& $\tilde{\Gamma}^{(4)}$ \\ \hline
SU(3) triplets & & &&&&&&&&&& \\
$\tilde{\tau^{3\;3}}$  =  ( $ \frac{1}{2},$  $ -\frac{1}{2},$
  0  ) & 
$u_1$& $u_2$ & $u_3$ &  $u_4$ & 
1/2 &1/6 & 2/3 & $\pm 1$ & 0 &2/3 & 2/3 & $\mp 1$\\
$\tilde{\tau}^{3\;8} $  =  ( $\frac{1}{2 \sqrt{3}},$
$\frac{1}{2\sqrt{3}},$  $-\frac{1}{\sqrt{3}}$  ) & 
$d_1$ & $d_2$ & $d_3$ & $d_4$ & 
-1/2 &  1/6 &  -1/3  &  $\pm 1$ & 0 &  -1/3 &  -1/3 & $\mp 1$\\ \hline 
SU(3) singlets & & &&&&&&&&&& \\
$\tau^{3\; 3} =  0$ & $\nu_1$ & $\nu_2$ & $\nu_3$ & $\nu_4$ & 
1/2 &  -1/2 & 0 & $\pm 1$ & 0 & 0 & 0 & $\mp 1$\\
$\tau^{3\; 8} = 0$ & 
$e_1$ & $e_2$ & $e_3$ & $e_4$ & 
-1/2 & -1/2 & 1 & $\pm 1$ & 0 & -1 & -1 & $\mp 1$\\ \hline
\end{tabular}
\end{center}

\vspace{3mm}

\noindent
Table III: Quantum numbers of fermions for  chiral multiplets
of the group $SO(1,13 )$, which  represent four families of
 quarks and leptons. Quantum numbers of the corresponding
antifermions have the opposite signs. We introduce $ \tilde{Q} =
\tilde{\tau}^{13} + \tilde{\tau}^{2w} + \tilde{\tau}^{2c}. $ To
the column of $SU(2) 
$ doublets with $ \tilde{\Gamma} = \pm 1$, the column of $SU(2)$
singlets with $ \tilde{\Gamma} = \mp 1 $, correspond.

\subsection{Conclusions and what have we learned}

\noindent
We have presented the approach, which unifying spins and
charges, connects lefthanded $SU(2)$ doublets and 
right handed $SU(2)$ singlets into the same multiplet,
manifesting that handedness and weak charges of fermions are
connected in a way, needed in the Standard model. The
operators $\tilde{S}^{mh}, m \in (0,1,2,3), h \in (5,6..)$,
which cause such transformations, appear in the equations of
motions of the approach (See Eq.(5.2) of Norma's talk) in terms,
which behave  like Yukawa couplings.

\noindent
The generators of the Lorentz transformations in Grasssmann
space, defining the Yang - Mills charges, commute with the
generators of the Lorentz transformations in the four
dimensional subspace in accordance with the Coleman - Mandula
theorem \cite{col} as well as with it extension for the
supersymmetric case \cite{haag} as long as the group $SO(1,13)$
manifests as the product of the subgroups $ SO(1,3) \times SU(3)
\times SU(2) \times U(1)$, which only is true for low enough energies.


\newpage
\stepcounter{section}
\setcounter{equation}{0}
\section*{\center Why do we have parity violation?}
\centerline{\large  Holger and Colin}

\vspace{5mm}

\subsection{Introduction}

Why do we have parity violation, or why is the weak charge dependent 
on handedness?

The short answer to this question is that 
we need at least some of the charges to be different for 
the observed right-handed and left-handed fermion 
states---i.e. handedness dependent or chiral---for the 
purpose of mass protection.
That is to say: in the philosophy that the particles we ``see''---those  
we can afford to produce and measure---are very 
light (essentially massless) from the supposed
fundamental scale point of view and, thus, they need a mechanism for being 
exceptionally light so that we have a chance to ``see' them. This
mechanism should suggestively be that any pair of right and 
left (Weyl) components should have at least one 
gauge quantum number different between them, so that any 
mass term is forbidden by gauge invariance.

Really we should rather ask why is parity conserved in the 
electromagnetic and strong interactions. Our philosophy would be that
a priori there is no reason why these symmetries should be 
there at all, and it is the presence of the 
symmetries (rather than their breaking) 
that needs an explanation. This is the philosophy
of what we call random dynamics, which really means: 
all that is not forbidden occurs. It is very natural since really 
to know a symmetry exists is much more informative 
than to know it not to be there.
So a priori one should rather say that, if there is no reason for them, 
we should not expect symmetries to be present. 

In the case of the question of 
whether the electroweak charges on the Weyl components of the quark and 
lepton fields should be the same for the two handednesses, right and left, 
we can say that, since the Weyl fields transform
under Lorentz transformations without mixing into 
each other 
(i.e.~they transform into themselves only), 
we should consider each 
Weyl field as essentially corresponding to a completely separate 
particle. As separate particles we expect them to a priori have 
completely different charges.
You might of course object that when the particle has a mass, so that we 
are talking about a Dirac particle, there is a connection between  
the left and the right Weyl component fields.  
However in the Standard Model
it is well-known that the masses come about as an effect of the Higgs 
field vacuum expectation value.    So, before the 
effect of the Higgs mechanism,
the fermions are massless and there is no association of the various Weyl
field with each other a priori.

Thus the question that really deserves and needs an answer is rather 
why there is parity conservation for the strong and 
elctromagnetic interactions, 
in the sense that the electromagnetic and colour charges are the same 
on the right and the left components of the same Dirac particle.
In addition the fact that the right-handed components are singlets,
while the left-handed components are doublets, under the weak 
SU(2) gauge group needs an explanation. 
So, assuming the existence of the Standard Model gauge group, 
we now derive the Standard Model fermion representations on 
the basis of a few simple assumptions. 

\subsection{Starting assumptions}
\subsubsection{The assumptions to derive Standard Model representations of 
fermions \label{SMassum}}

\begin{itemize}
\item[(a)] As the starting point for the derivation of the Standard Model 
representations, we
shall assume 
the gauge \underline{group} and not only the gauge Lie algebra 
of the Standard Model to be $S(U(2)\times(U(3))$.
\item[(b)] Further we shall make the assumption that the 
representations---realised by the 
Weyl fermions---of this group are ``small''. More specifically 
we assume
that the weak hypercharge charge $y/2$ is 
at most unity numerically, and that only the trivial and the lowest 
fundamental (defining) representations of the nonabelian groups SU(2) and
SU(3) are used.
\item[(c)] Further  we assume mass 
protection, i.e.~we say that all particles 
for which a mass could be made, without the Higgs field being used, 
would be so heavy that we should not count them as observable particles.
\item[(d)] In our argument we shall also use 
the requirement that there shall 
be no gauge nor mixed anomalies. This is needed 
since otherwise there would be a breaking of the gauge symmetry.
\end{itemize}
These assumptions are of course known to be true in the Standard Model.
Indeed they are rather suggestive regularities of the Standard Model, 
if one is looking for inspiration to go beyond the Standard Model.
You could say that it might not be so difficult to find some rather 
general argumentation for why representations should be ``small''
in some way---not exactly how small perhaps.

\subsubsection{Slightly reduced assumptions for parity in strong and 
electromagnetic interactions}

>From the 
assumptions stated in the foregoing subsection we can indeed 
derive the fermion representations of the whole Standard Model 
and, thus, also the fact that there is parity conservation in 
electromagnetic and strong interactions. 
However, if we replace the requirement $|y/2|\leq 1 $ by the 
slightly modified assumption 
that the electric charge $Q=y/2+I_{W3}$  (where $I_{W3}$
is the third component of the weak isospin)  
has numerical value less than 
or equal to unity for all the Weyl fermion representations, 
the mass protection assumption is not needed 
for this parity derivation. The point of course is 
that the mass protection 
is performed by the weak interaction, and 
the electromagnetic and colour quantum numbers do not provide any 
mass protection themselves--they cannot with parity symmetry.

In other words, for the illustrational derivation of 
parity conservation in strong and electromagnetic interactions 
alone, we assume:

\begin{itemize}
 \item[(a)] Either the gauge \underline{group} $U(3)$ for strong and
electromagnetic interactions, or the total 
gauge group $S(U(2)\otimes U(3))$ as above.
 \item[(b)] The ``small'' representations in the form $|Q|\leq 1$ and
 $\underline{a}
\leq \underline{3}$
 \item[(c)] and then of course that there should still be no anomalies.
\end{itemize} 

\subsection{Derivation of parity for QCD and electromagnetism}

 The program of our proof of parity for strong and electromagnetic 
interactions consists in showing that
the Weyl fields must have quantum number combinations  
that will be paired into 
Dirac fields, so that parity in the electromagnetic and strong interactions
gets preserved. 

What we have to show is that there are 
always equally many Weyl field species
with a given electric and colour charge combination and the opposite.
In this way we could then say that at least the possibility is there for
combining these Weyl fields into Dirac fields, so that the 
electric and colour charges on the right and the left Weyl components 
become the same. We should of course have in mind that, in 
four dimensions, one can consider the right handed Weyl components 
as represented by their CP-conjugates 
so to speak, meaning a corresponding
set of left-handed fields with the opposite charges.
So we actually 
need only discuss the
left-handed components, just letting them represent the right-handed
ones too as antiparticles. 

Now for anomaly calculations it is easily seen that 
left-handed Weyl fields in conjugate 
representations give just equal and 
opposite contributions to the various 
anomalies. Thus we can only hope to say from anomaly considerations 
something about the number of species in one representation minus
the number in the conjugate one. We should therefore introduce names 
for these differences:

We let the symbol $ N_{(y/2, \underline{a}, I_W )}$ 
denote the number of 
left-handed Weyl species with the weak hypercharge y/2, the colour 
representation $\underline{a}$ and the weak isospin $I_W$ minus 
the number of species with the opposite (conjugate)  quantum numbers.
But, in the present section, we 
ignore the weak isospin and use  $N_{(Q , \underline{a})}$ to mean 
the difference of the number of Weyl-species with electric charge 
Q and color representation $\underline{a}$ and the number of Weyl-species
with the conjugate quantum numbers. 

The requirement of the smallness of the representations
means that  $ N_{(Q, \underline{a})}$ is zero unless 
\begin{equation}
|Q|\leq 1 
\label{e1}
\end{equation}
\begin{equation}
|\underline{a}|\leq |\underline{3}|
\label{e2}
\end{equation}
Obviously by our definition $N_{(Q =0,\underline{1})} = n - n =0$.

What we have to show to get parity conservation for these 
interactions is that
\begin{equation}
N_{(Q,\underline{a})}=0.
\end{equation}
for all the quantum number combinations $(Q, \underline{a})$.

The requirements of small representations 
and of the gauge \underline{group} being 
$U(3)$ leaves only the three differences of species numbers
$ N_{(Q=1,\underline{1})}$, $ N_{(Q=2/3 ,\underline{3})}$, 
$ N_{(Q=-1/3,\underline{3})}$ non-zero. 

Now the anomalies in four dimensions come from triangle diagrams
with external gauge fields for the gauge anomalies and with two gravitons 
and one gauge particle assigned in the case of the mixed anomaly.
In order to get rid of the anomalies, so as to avoid 
breaking the gauge symmetry say, 
we must require that the relevant triangle diagrams
have cancellations between the contributions coming from the different 
Weyl field species, the latter circling around the triangle. 
The only mixed anomaly diagram, not already vanishing for other reasons,
is a triangle with Weyl particles circling around it having 
two gravitons attached and the photon 
at the third vertex. The cancellation required to get rid of this 
the mixed anomaly becomes
\begin{equation}
N_{(Q=1,\underline{1})}+\frac{2}{3} \times 3N_{(Q=2/3,\underline{3})}
+(-\frac{1}{3}) \times 3N_{(Q=-1/3,\underline{3})} =0
\end{equation}
To ensure no gauge anomaly there are three triangle diagrams that must
have a cancellation: one with three external gluons which gives
\begin{equation}
N_{(Q=2/3,\underline{3})}
+N_{(Q=-1/3,\underline{3})} =0,
\end{equation} 
one with one photon and two gluons attached, which gives
\begin{equation}
2/3 \times N_{(Q=2/3,\underline{3})}
+(-1/3) \times N_{(Q=-1/3,\underline{3})} =0
\end{equation}
and finally one with three photons attached which gives
\begin{equation}
N_{(Q=1,\underline{1})}+\left(\frac{2}{3}\right)^3 
\times 3N_{(Q=2/3,\underline{3})}
+\left(-\frac{1}{3}\right)^3 \times 3N_{(Q=-1/3,\underline{3})} =0
\end{equation}

We have here got four linear equations for three unknowns,
so it is no wonder that they lead to all the differences being 
zero. That then means to every Weyl representation there is  
the possibility of finding just one with the opposite (conjugate) 
representation. This vanishing of the differences is sufficient 
to give parity conservation, 
provided possible mass mechanisms do not violate the gauge symmetries for 
colour and electromagnetism. It means that one may directly construct 
a parity operator, by diagonalizing a perhaps present  U(3) gauge invariant
mass mechanism and letting it map the right to the corresponding left
mass eigenstate and opposite.

\subsection{Deriving all standard model fermion representations}

Using a very similar technique, but now within all the four assumptions 
stated in the subsection \ref{SMassum}, we can show the fermion 
representations to be those of the Standard Model with some as 
yet not determined number of generations.

For this purpose the assumption about small representations can be taken 
to mean 
that  $ N_{(y/2, \underline{a}, I_W )}$ is zero unless 
\begin{equation}
|y/2|\leq 1 
\label{e11}
\end{equation}
\begin{equation}
|\underline{a}|\leq |\underline{3}|
\label{e22}
\end{equation}
\begin{equation}
|I_W|\leq 1/2.
\label{e33}
\end{equation}
Really it means that we assume zero species for the representations 
not fullfilling this and thus of course the same for the differences 
$N_{(y/2, \underline{a}, I_W)}$. 
The requirement of the gauge \underline{group} being 
$S(U(2)\times U(3))$ means that the species numbers are zero unless 
the congruence 
\begin{equation}
y/2 + t/3 + d/2 = 0 \qquad (\mbox{mod 1})
\end{equation}
is fullfilled, where t is triality and d is duality.

Obviously by our definition $N_{(y/2 =0,\underline{1}, I_W=0)} =n - n=0$.

The small representation and the gauge \underline{group} requirements 
now allow six $N_{(y/2,\underline{a}, I_W )}$'s to be nonzero a priori,
namely one for each of the allowed numerical values of y/2 which runs 
from 1/6, in steps of 1/6, to 1.  

As above we use the cancellation criteria for the anomalies, meaning the 
cancellation of triangle diagrams: This time the mixed anomaly cancellation
diagram has two gravitons and one weak hypercharge coupling and it gives
\begin{eqnarray} 
\frac{1}{6} \times 6N_{(y/2 =1/6,\underline{3},I_W=1/2)}+
\frac{1}{3} \times 3N_{(y/2 = 1/3,\underline{\bar{3}},
I_W=0)} & & \nonumber \\
+\frac{1}{2} \times 2N_{(y/2 = 1/2,\underline{1},I_W=1/2)}
+ \frac{2}{3} \times 3N_{(y/2= 2/3,\underline{3},I_W=0)}
& & \nonumber \\
+ \frac{5}{6} \times 6N_{(y/2 =5/6,
\underline{\bar{3}},I_W=1/2)}+N_{(y/2=1,\underline{1},I_W=0)} & = & 0 
\end{eqnarray}
The no gauge anomaly triangle diagrams consist of one with three external
gluons, while the one with three external W's 
is trivially zero and does not count, then there are two diagrams with 
respectively two gluons and two W's and one weak hypercharge coupling, and
finally there is one diagram with all three attached gauge particles being
the abelian one (coupling to weak hypercharge). The conditions become:
\begin{eqnarray}
2N_{(y/2 =1/6,\underline{3},I_W=1/2)}
-N_{(y/2 = 1/3,\underline{\bar{3}},
I_W=0)} & & \nonumber \\
+ N_{(y/2= 2/3,\underline{3},I_W=0)}-2N_{(y/2 =5/6,
\underline{\bar{3}},I_W=1/2)} & =& 0
\end{eqnarray}
\begin{eqnarray}
\frac{1}{6} \times 2N_{(y/2 =1/6,\underline{3},I_W=1/2)}+
\frac{1}{3} \times N_{(y/2 = 1/3,\underline{\bar{3}},
I_W=0)} & & \nonumber \\
+ \frac{2}{3} \times N_{(y/2= 2/3,\underline{3},I_W=0)}
+\frac{5}{6} \times 2N_{(y/2 =5/6,
\underline{\bar{3}},I_W=1/2)} & = & 0
\end{eqnarray}
\begin{eqnarray}
\frac{1}{6} \times 3N_{(y/2 =1/6,\underline{3},I_W=1/2)}
+\frac{1}{2} \times N_{(y/2 = 1/2,\underline{1},I_W=1/2)} 
& & \nonumber \\
+ \frac{5}{6} \times 3N_{(y/2 =5/6,
\underline{\bar{3}},I_W=1/2)} & = & 0
\end{eqnarray}
\begin{eqnarray}
\left(\frac{1}{6}\right)^3 \times 6N_{(y/2 =1/6,\underline{3},I_W=1/2)}+
\left(\frac{1}{3}\right)^3 \times 3N_{(y/2 = 1/3,\underline{\bar{3}},
I_W=0)} & & \nonumber \\
+\left(\frac{1}{2}\right)^3 \times 2N_{(y/2 = 1/2,\underline{1},I_W=1/2)}+
\left(\frac{2}{3}\right)^3 \times 3N_{(y/2= 2/3,\underline{3},I_W=0)}
& & \nonumber \\
+\left(\frac{5}{6}\right)^3 \times 6N_{(y/2 =5/6,
\underline{\bar{3}},I_W=1/2)}+N_{(y/2=1,\underline{1},I_W=0)} 
& = & 0
\end{eqnarray}
Here we have got 5 equations linear in the $N$'s of which there were 
6. Thus it is not surprising that there is, up to the unavoidable 
scaling by a common factor of all the unknowns---the N's---just
one solution. This must, however, be that of the Standard Model
since the latter satisfies the anomaly cancellation conditions. 
The scaling factor corresponds to the 
generation number we could say. 
So far we have only shown that the N's
are as in the Standard Model. We need now to use the assumption
about mass protection to deduce that we cannot have both 
representations---i.e.~a representation and its 
conjugate---associated with a given N 
present. That implies first that the cases of N's that are zero 
imply that there will be no Weyl fermions at all associated with those
quantum numbers---there will be no vector fermions. Also for the cases
of nonzero N's only one of the two associated representations will
exist, depending on the sign of the N in question.
With this conclusion we almost truly derived the Standard Model
fermion representations; there are however still two ambiguities:
1) the generation number can be any integer, 2) we could have the 
opposite signs for the N's which would correspond to a model
that is, so to speak, a parity reflected version of the Standard Model.

Since we have now derived the whole representation system for the
fermions in the Standard Model, we did not really need the exercise 
of deriving parity for the electromagnetic and colour interactions 
separately; we got it all at once after all, assuming though---as is 
needed---the Higgs mechanism. 

\subsection{Conclusion}
We have shown that, from the four requirements above, 
it is possible to argue for the whole system of the Weyl fermion 
representations of the Standard Model. 
So if one can just argue for these assumptions in some model beyond the
Standard Model one will have the fermion system for free. 

Concerning the question of whether the charges depend on handedness,
we saw that for the colour and electric charges no such dependence is 
allowed, by just using the smallness of electric charge and 
colour representation plus the no anomaly conditions.
Concerning the question of why there is a dependence---namely for the
weak charges---we saw that it was the mass protection requirement that 
enforced it. In fact each of the differences N had to be 
a difference between zero and another number, because the mass protection
would not allow two sets of Weyl fields counted as left-handed having
opposite (conjugate) quantum numbers. They would namely combine
to get a huge mass according to 
the philosophy of mass protection.
Thus indeed the charges must, in one way or another, be different
for the right and left handnesses.

This 
really means that we take the point of view that the fundamental 
scale, or the next level in fundamentality, 
has so huge a characteristic energy
or mass scale that all the particles we know must in first approximation 
be arranged to be massless, i.e.~they must be mass protected. 

\subsection{Acknowledgement}

Financial support of INTAS grant INTAS-93-3316-ext 
is gratefully acknowledged.



\newpage
\stepcounter{section}
\setcounter{equation}{0}
\section*{\center The Number of Families in a Scenario with Right-Handed
Neutrinos}
\centerline{\large  Astri\footnote{A. Kleppe}}

\vspace{5mm}

\noindent In the Standard Model, the number of light neutrino 
flavours, $<N_{\nu}>$, is defined by the invisible Z-width, i.e. 
\begin{eqnarray}
\Gamma_{inv}(\rm{Z} \rightarrow \nu's)=\Gamma_{0}<N_{\nu}>
\end{eqnarray}
where $\Gamma_{0}$ is the standard width for a massless neutrino pair,
$\Gamma_0={G_FM_Z^3/ 12\pi\sqrt{2}}$, and $<N_{\nu}>$ equals the number
of left-handed lepton doublets, i.e. the number of families.

In a scenario where (two or more) right-handed neutrinos are included in 
the Standard Model, we can however draw no definite conclusion 
about the number of families from the invisible Z-width.

In order to see this, consider the neutral current term (in the physical
basis)
in an extended Standard Model scenario, with two right-handed neutrinos
and $n$ families,
\begin{equation}\label{ncur}
{\cal{L}}_{NC}=\frac{g}{2\cos\theta_W}\bar{N} \gamma^{\mu}{\bf{\Omega}} 
N
Z_{\mu}
\end{equation}
Here $N$ contains the neutrino fields,
\begin{eqnarray}
N=
\left(\begin{array}{rcl}
                    \nu_{1L}\nonumber\\
                    \nu_{2L}\nonumber\\
                      .\nonumber\\
                      .\nonumber\\
                      .\nonumber\\
                     \nu_{nL}\nonumber\\
                      0\\
                      0\nonumber
                             \end{array}
                      \right)
\end{eqnarray}
and $\bf{\Omega}$ is the $(n+2)$x$(n+2)$ matrix 
\begin{eqnarray}
{\bf{\Omega}}= \bf{U}\label{uuiu}
\left(\begin{array}{rcl}
1&&{\hspace{1mm}}0{\hspace{8mm}}0{\hspace{8mm}}0{\hspace{8mm}}0{\hspace{8mm}}0
{\hspace{8mm}}0\nonumber\\
0&&{\hspace{1mm}}1{\hspace{8mm}}0{\hspace{8mm}}0{\hspace{8mm}}0{\hspace{8mm}}0
{\hspace{8mm}}0\nonumber\\
.&&{\hspace{1mm}}.{\hspace{8mm}}.{\hspace{8mm}}.{\hspace{8mm}}.{\hspace{8mm}}.
{\hspace{8mm}}.\nonumber\\
.&&{\hspace{1mm}}.{\hspace{8mm}}.{\hspace{8mm}}.{\hspace{8mm}}.{\hspace{8mm}}.
{\hspace{8mm}}.\nonumber\\
0&&{\hspace{1mm}}0{\hspace{8mm}}0{\hspace{8mm}}0{\hspace{8mm}}1{\hspace{8mm}}0
{\hspace{8mm}}0\nonumber\\
0&&{\hspace{1mm}}0{\hspace{8mm}}0{\hspace{8mm}}0{\hspace{8mm}}0{\hspace{8mm}}0
{\hspace{8mm}}0\nonumber\\
0&&{\hspace{1mm}}0{\hspace{8mm}}0{\hspace{8mm}}0{\hspace{8mm}}0{\hspace{8mm}}0
{\hspace{8mm}}0
\end{array}
        \right) {\bf{U}}^{\dagger}
\end{eqnarray}
where $U$ is the $(n+2)$x$(n+2)$ neutrino mixing matrix.\\
The matrix ${\bf{\Omega}}$ is just a unitary transformation of the
matrix 
$diag(1,1,...,1,0,0)$. The trace of ${\bf{\Omega}}$ is thus 
$tr({\bf{\Omega}})=n$, and since
${\bf{\Omega}}{\bf{\Omega}}^{\dagger}=
{\bf{\Omega}}^{2}={\bf{\Omega}}$,\\
${\hspace{45mm}} \displaystyle\sum_{j,k=1}^{n}|{\bf{\Omega}}_{jk}|^{2}=
tr({\bf{\Omega}}{\bf{\Omega}}^{\dagger})=tr({\bf{\Omega}})=n$.\\
\\
The neutral current coupling coefficients thus satisfy
\begin{equation}\label{jawa}
\displaystyle\sum_{j,k=1}^{n}|{\bf{\Omega}}_{jk}|^{2}=n,
\end{equation}
where the right-hand side is just the number of left-handed leptonic 
doublets, i.e the number of families.

>From equation ($\ref{ncur}$) we have that
\begin{eqnarray}
\Gamma(\rm{Z} \rightarrow
\nu's)=\Gamma_{0}\sum_{i,j=1}^nX_{ij}|\Omega_{ij}|^{2},
\end{eqnarray}
where the $X_{ij}$ are
the phase space and matrix element suppression factors due to the
nonvanishing
neutrino masses, bounded by unity. In a scenario with right-handed
neutrinos
present, the invisible width therefore satifies the inequality
\begin{eqnarray}
\Gamma(\rm{Z} \rightarrow \nu's)\leq n\Gamma_{0},
\end{eqnarray}
and no decisive conclusion can thus be drawn about the number of
families. 

Does this mean that provided there are right-handed neutrinos,
data are compatible with more than three families?

What does our model tell about three versus four families?
In our specific scenario, the generic neutrino mass matrix is of
the form
\begin{equation}\label{ssam}
 \cal{M}=\left(\begin{array}{rcl}
                {\large\bf{0}}          &   {\large\bf{A}}\\
                  \tilde{\large\bf{A}}  &   {\large\bf{M}}\nonumber
               \end{array}
         \right),
\end{equation}
and in the case with three families and two right-handed neutrinos, the
matrix ${\large\bf{A}}$ is a 2x3 matrix, and the matrix
${\tilde{\large\bf{A}}}$ is its transpose, i.e.
\begin{equation}\label{etti}
 \cal{M}=\left
(\begin{array}{rcl}
0&&0{\hspace{6mm}}0{\hspace{7mm}}a_{1}{\hspace{6mm}}b_{1}\nonumber\\
0&&0{\hspace{6mm}}0{\hspace{7mm}}a_{2}{\hspace{6mm}}b_{2}\\
0&&0{\hspace{6mm}}0{\hspace{7mm}}a_{3}{\hspace{6mm}}b_{3}\\
a_{1}&&a_{2}{\hspace{4mm}}a_{3}{\hspace{5mm}}M_{1}{\hspace{4mm}}0    \\
b_{1}&&b_{2}{\hspace{4mm}}b_{3}{\hspace{7mm}}0{\hspace{5mm}}M_{2}\nonumber
\end{array}
        \right)
\end{equation}

In this scheme, with the introduction of certain constraints, the
mass spectrum consists of three massless and two massive eigenstates,
and the invisible Z-width takes the form
\begin{eqnarray}
&&\Gamma(\rm{Z} \rightarrow \nu+\bar{\nu})=\Gamma_{0}[1+1+
\frac{1}{s^{4}_{2\gamma}c^{4}_{2\theta}}(c^{4}_{2\gamma}s^{4}_{2\theta}+
2S^{2}c^{2}_{2\gamma}s^{2}_{2\theta}(
c^{2}_{\gamma}X_{34}+s^{2}_{\gamma}X_{35})+\nonumber\\
&&+S^{4}(c^{4}_{\gamma}X_{44}+s^{4}_{\gamma}X_{55}+
2c^{2}_{\gamma}s^{2}_{\gamma}X_{45}))]
\end{eqnarray}
where $S=\sqrt{c^{2}_{2\theta}-c^{2}_{2\gamma}}$ and $c_{2\gamma}$ 
= $\cos 2\gamma$, etc.
The angles $\gamma$ and $\theta$ are defined by $M_1/M_2 = -
\tan^2\theta$
and $\lambda_{+}/\lambda_{-} = \tan^2\gamma$, where $\lambda_{\pm}$ are
the 
non-vanishing neutrino masses, and constrained by $\tan\gamma \neq 0$,
$\tan\gamma \neq 1$, 
$\tan\theta \neq 0$, $\tan\theta \neq 1$, and $\tan\gamma \neq
\tan\theta$.
Since the factors $X_{ij}$ are bounded by unity, 
$\Gamma(\rm{Z} \rightarrow \nu+\bar{\nu}) \leq 3\Gamma_{0}$, in
agreement 
with data.

A scheme with four families can be modeled on the case with three
families by
letting ${\large\bf{A}}$ be a 2x4 matrix, whereby the mass matrix
becomes
\begin{equation}\label{etty}
 \cal{M}=\left
(\begin{array}{rcl}
0&&0{\hspace{6mm}}0{\hspace{7mm}}0{\hspace{7mm}}a_{1}{\hspace{6mm}}b_{1}
\nonumber\\
0&&0{\hspace{6mm}}0{\hspace{7mm}}0{\hspace{7mm}}a_{2}{\hspace{6mm}}b_{2}
\nonumber\\
0&&0{\hspace{6mm}}0{\hspace{7mm}}0{\hspace{7mm}}a_{3}{\hspace{6mm}}b_{3}
\nonumber\\
0&&0{\hspace{6mm}}0{\hspace{7mm}}0{\hspace{7mm}}a_{4}{\hspace{6mm}}b_{4}\\
a_{1}&&a_{2}{\hspace{4mm}}a_{3}{\hspace{5mm}}a_{4}{\hspace{6mm}}M_{1}
{\hspace{4mm}}0\\
b_{1}&&b_{2}{\hspace{5mm}}b_{3}{\hspace{5mm}}b_{4}{\hspace{7mm}}0{\hspace{5mm}}
M
_{2}\nonumber
\end{array}
        \right)
\end{equation}
With a mass spectrum with four zero and two non-vanishing neutrino
masses,
the corresponding invisible Z-width is
\begin{eqnarray}
&&\Gamma(\rm{Z} \rightarrow \nu+\bar{\nu})=\Gamma_{0}[1+1+1+
\frac{1}{s^{4}_{2\gamma}c^{4}_{2\theta}}(c^{4}_{2\gamma}s^{4}_{2\theta}+
2S^{2}c^{2}_{2\gamma}s^{2}_{2\theta}(
c^{2}_{\gamma}X_{45}+s^{2}_{\gamma}X_{46})+\nonumber\\
&&+S^{4}(c^{4}_{\gamma}X_{55}+s^{4}_{\gamma}X_{66}+
2c^{2}_{\gamma}s^{2}_{\gamma}X_{56}))],
\end{eqnarray}
where the definitions and constraints are the same as above.
The suppression factors are given by
\begin{equation}
X_{ij}=\frac{\sqrt{\lambda (M^{2}_{\rm{Z}}, m^{2}_{i}, 
m^{2}_{j}})}{M^{2}_{\rm{Z}}A_{ij}}
\end{equation}
where $\lambda$ is given by
$\lambda(a,b,c)=a^{2}+b^{2}+c^{2}-2(ab+bc+ac)$, and
$A_{ij}$ include the mass dependence of the matrix elements.

For small enough $\tan 2\theta/\tan 2\gamma$, as well as 
$S$ and $X_{ij}$'s, $\Gamma(\rm{Z} \rightarrow
\nu+\bar{\nu})/\Gamma_{0}$ 
is compatible with the experimental value of the number of neutrinos,
$<N_{\nu}>_{exp} =  2.991\pm 0.016$ 
(LEP Electroweak Working Group, CERN/PPE/95-172).
The actual values of the angles and the other parameters of course
depend on
the (unknown) neutrino masses and mixings.

So in this model with right-handed neutrinos,
with suitable neutrino masses and mixing angles, four 
families are in principle not excluded by data.

\newpage
\stepcounter{section}
\setcounter{equation}{0}
\section*{\center Can we make a Majorana field theory starting from the zero
mass Weyl theory, then adding a mass term as an interaction?}
\centerline{\large Norma, Holger and Colin}
\vskip 5mm

The answer to this question is: yes we can. One can proceed 
similarly to the case of the Dirac massive field theory. In both
cases one can start from the zero mass Weyl theory and then 
add a mass 
term as an interacting term of massless particles with a
constant ( external )
field. In both cases the interaction gives rise to a field
theory for a free
massive fermion field. We shall present the procedure for the
creation of a mass term in the case of the Dirac and the
Majorana field and we shall look for a massive field as
a superposition of massless fields. The Majorana particle is an
unusual particle, since it is his own antiparticle.

\subsection{The zero mass Weyl field theory}


Let as start with massless particles, described by the Weyl
massless fields. We pay 
attention on momentum $(p^a = (p^0, \vec{p}))$ and
spin of fields. Charges of fields will   
not be pointed out. It means that in our considerations 
we shall  not distinguish between neutrinos, 
electrons and quarks. The Weyl equation for massless fields

\be
2 \vec{S}.
\vec{p} =   \Gamma p^0,\label{maj1.1}
\ee

\noindent
determines four states.
Here $S^{i} = \frac{1}{2}\varepsilon_{ijk} S^{jk},\;\; S^{ij} =
\frac{i}{4} \;[\gamma^i, \gamma^j],\;\;(i\in\{1,2,3\}) $, which
are the generators of the Lorentz transformations,
determine  the spin of states and $
\gamma^a, \;\;a \in\{0,1,2,3\} $  are the Dirac operators. 
Operator $\Gamma = \frac{-i}{3!} \varepsilon_{abcd}S^{ab}S_{cd}$ 
is one of the two  Casimir operators  of the
Lorentz group $SO(1,3)$ acting in the internal space of spins
only and defines the handedness of states. Two eigenstates of Eq.(1.1)
have left handedness ($< \Gamma > = r,\; r =
-1$), the other two have right handedness\footnote[1] { We shall make
use of the symbol $\Gamma$ for the operator and r (ro\v cnost in
slovenian language means handedness) for the corresponding 
eigenvalue. The symbol $h$ will be used 
for both, for the helicity operator and for his eigenvalue.}
($r = 1$). The left 
handed solutions have either left helicity

\be
h = \frac{2 \vec{S}.
\vec{p}}{ |\vec{p^0}|}, \label{maj1.2}
\ee

\noindent
( $h = -1$) and positive energy ($ p^0 = |p^0| $) or right
helicity ($h = 1$) and negative energy ($ p^0 = -|p^0|$). The
right handed solutions have either  right helicity ($ h = 1$)
and positive energy or left helicity ($h = -1$) and negative
energy. We shall denote the positive energy solutions by a
symbol $u$ and the negative energy solutions by a symbol $v$.
To determine the positive energy solution completely is enough to
tell the momentum $\vec{p}$, ( $p^0 =
|\vec{p}|$) and 
either handedness or helicity: $ u_{\vec{p}, L}
\equiv u_{\vec{p}, h = -1},$  $ u_{\vec{p}, R}
\equiv u_{\vec{p}, h = 1}.$ Equivalently it follows
: $ v_{\vec{p}, L}
\equiv v_{\vec{p}, h = 1},$  $ v_{\vec{p}, R}
\equiv v_{\vec{p}, h = -1}.$
We shall point out either helicity (h) or handedness (r),
depending on what will be more convenient. 
\\

After quantizing the field the creation operators are defined,
creating the 
negative energy particles: $ d_{\vec{p}, L}^{(0)+}
\equiv d^{(0)+}_{\vec{p}, h = 1},$   $
d^{(0)+}_{\vec{p}, R} 
\equiv d^{(0)+}_{\vec{p}, h = -1},$ and the
positive energy particles: $ b^{(0)+}_{\vec{p}, L}
\equiv b^{(0)+}_{\vec{p}, h = -1},$   $
b^{(0)+}_{\vec{p}, R} 
\equiv b^{(0)+}_{\vec{p}, h = 1}.$ 
The field can then according to ref.\cite{bjor} be written as:

\be
\psi(x) = \sum_{r = \pm1} \sum_{\vec{p}, p^{02} 
= {\vec{p}}^2} \frac {1}{\sqrt{(2 \pi)^3}} ( b^{(0)}{
}_{\vec{p}, r} u_{\vec{p}, r} e^{ -ipx} +
d^{(0)}{ }_{\vec{p}, r} v_{\vec{p}, r}
e^{ ipx} ). \label{maj1.3}
\ee 

\noindent
To simplify the discussions we discretize the momentum and replace
the integral  with the sum. 
Then  the energy operator $ H^{(0)} = \int
d^3 \vec{x} \psi^+ p_0 \psi $ can be written as

\be H^{(0)} = \sum_{r = \pm1} \sum_{\vec{p}, p^{02} =
\vec{p}^2} 
|p^0|\; ( b^{(0)+}_{\vec{p}, r}
b^{(0)}_{\vec{p}, r} - d^{(0)+}{
}_{\vec{p}, r} 
d^{(0)}_{\vec{p}, r} ).\label{maj1.4}
\ee

\noindent
If the "totally empty" vacuum state is denoted by $|0>$, then
the vacuum 
state occupied by massless particles up to $ \vec{p}
= 0$ is ( due to discrete values of momenta ) equal to 

\be
|\phi_{(0)}> = \prod_{\vec{p}, r}
d^{(0)+}_{\vec{p}, r} |0>. \label{maj1.5}
\ee 

\noindent
The energy of such a vacuum is
accordingly $<\phi_{0}| H^{(0)}|\phi_{0} > = \sum_{\vec{p}, r} 
E_{\vec{p}, r}^{(0)}, $ with $ E_{\vec{p}, r}^{(0)} =
-|\vec{p}|, $ which is of course infinite.
Accordingly the particle state of momentum $\vec{\acute{p}},
\acute{p}^0 = |\vec{\acute{p}}|, $ and handedness
$r$, with the energy, which is for $\acute{p}^0$ larger than the
energy of the vacuum, can be written as $ b^{(0)+}_{\vec{\acute{p}}}
|\phi_0>.$


\subsection{ Charge conjugation}


The symmetry operation of charge conjugation is associated with
the interchange of particles and antiparticles.
Introducing the charge conjugation operator $C$, with the
properties $ C^2 = C, C^+ = C, C \gamma^{a*} C^{-1} = -\gamma^a $ ,
where $(^+)$ 
stays for hermitian conjugation and $(^*)$ for complex
conjugation, one finds the charge conjugated 
field $ \psi(x)^C $ to the field $\psi(x)$ as $ \psi(x)^C = C
\psi(x). $ One accordingly finds for the charge conjugating
operator ${ \cal C }$, which affects creation and anihilation
operators 

\be
\begin{array}{cl}
{\cal C}\;\; b^{(0)+}_{\vec{p}, h = -1} \;\;{\cal C}^{-1}
= -d^{(0)}_{-\vec{p}, h = 1}, & {\cal C}\;\;
b^{(0)+}_{\vec{p}, h = 1} \;\; {\cal C}^{-1} 
= d^{(0)}_{-\vec{p}, h = -1}, \\
 {\cal C} \;\; d^{(0)+}_{\vec{p}, h = 1} \;\; {\cal C}^{-1}
= -b^{(0)}_{-\vec{p}, h = -1}, & {\cal C}\;\; 
d^{(0)+}_{\vec{p}, h = -1} \;\; {\cal C}^{-1} 
= b^{(0)}_{-\vec{p}, h = 1}.\\
\end{array} \label{maj2.1}
\ee

\noindent
According to Sect. 2.  the left handed column concerns the charge
conjugation of left handed particles while the right handed column
concerns the charge conjugation of right handed particles. 
One easily finds that the hamiltonian $ H^{(0)}$ is invariant
under the charge conjugation operation. The charge conjugation
operation on the vacuum state $|\phi
_{(0)}> = 
\prod_{\vec{p}, r = \pm 1}  d^{(0)+}_
{\vec{p}, r}  |0>$  
should let it be invariant, since we want it as the physical
vacuum to be charge conjugation invariant. To achieve that we
are, however, then forced to let the totally empty vacuum $|0>$
transform under charge conjugation as
\be
{ \cal C}|0> = \prod_{\vec{p},r}(
b^{\dagger}_{\vec{p},r}d^{\dagger}_{\vec{p},r}|0> \label{maj2.2}
\ee
The charge conjugated operator $
b^{(0)+}_{\vec{p}, L}$ ( which generates
on a vacuum $|\phi_{(0)}>$ a one particle positive energy state of
left helicity )
annihilates in the vacuum state
$|\phi_{(0)}>$ a  negative energy particle state of opposite
momentum and helicity and therefore generates a hole, which
manifests as an antiparticle. Handedness stays unchanged.


\subsection{ The massive Dirac field theory}


We shall first treat the case of the massive Dirac field, for
which the procedure is  simpler
than in the massive Majorana case and from which we can learn
the procedure. The
mass term $ \int d^3 \vec{x} m_{D} \bar{\psi} \psi = \int d^3
\vec{x} \; m_{D} \; ( \bar{\psi_{L}} \psi_{R} + \bar{\psi_{R}}
\psi_{L} )$ can be written, if using the 
expression for $\psi$ from Eq.(\ref{maj1.3}),  as follows

\be H^{(1D)} = m_{D} \; \int d^3 \vec{x} \; \bar{\psi} \psi
= \sum_{h = \pm1} 
\sum_{\vec{p}}  
H^{(1D)}_{\vec{p}, h},\;\;\; 
H^{(1D)}_{\vec{p}, h} = m_D \;(\; b^{(0)+}_
{\vec{p}, h} d^{(0)}_{\vec{p}, h}
+ d^{(0)+}_
{\vec{p}, h} b^{(0)}_{\vec{p}, h}\;).
\label{maj3.1}
\ee 

\noindent
If we define

$$ N_{\vec{p}, h} = h^{+}_{\vec{p}, h} + h^{-}_{\vec{p}, h},$$

\be
h^{+}_{\vec{p}, h} = b^{(0)+}_
{\vec{p}, h} b^{(0)}_{\vec{p}, h}, \;\; h^{-}_{\vec{p}, h} = 
d^{(0)+}_{\vec{p}, h} d^{(0)}_
{\vec{p}, h}, \label{maj3.2}
\ee

\noindent
one easily finds that $\;  [N_{\vec{p}, h},
H^{(0)}_{\vec{\acute{p}}, \acute{h}}] = 0 = 
[N_{\vec{p}, h},
H^{(1D)}_{\vec{\acute{p}}, \acute{h}}]$. We see that the
interaction term $ H^{(1D)}_{\vec{p}, h}$ does not mix massless
states of different helicity. The appropriate basic states,
which are eigenstates of the operator for number of particles of
definite helicity $ N_{\vec{p}, h}$ are accordingly defined
either with $b^{(0)+}_ 
{\vec{p}, h = 1}$ and $d^{(0)+}_{\vec{p}, h = 1}$ with $h = 1$
( but of right and left handedness, respectively ) or with $
b^{(0)+}_ 
{\vec{p}, h = -1}$ and $ d^{(0)+}_{\vec{p}, h = -1} $ with $ h =
-1$ (but of left and right handedness, respectively). The first
two basic states have $ < N_{\vec{p}, h = 1} > = 1  $ and 
$ < N_{\vec{p}, h = -1} > = 0 $, while the second two basic
states have   $ < N_{\vec{p}, h = 1} > = 0  $ and 
$ < N_{\vec{p}, h = -1} > = 1 $. 
Diagonalizing $ H^{(D)}_{\vec{p}, h}  =  H^{(0)}_{\vec{p},
h} + H^{(1D)}_{\vec{p}, h}$ within the two basic states of
definite helicity ( but not handedness ), one finds 
that

\be
\begin{array}{cl}
b^{+}_{\vec{p}, h} = \alpha_{\vec{p}} \; b^{(0)+}_{\vec{p},
h} + \beta_{\vec{p}} \; d^{(0)+}_
{\vec{p}, h}, & p^0 = \;\;\;|p^0|, \\
d^{+}_{\vec{p}, h} = \alpha_{\vec{p}} \; d^{(0)+}_{\vec{p},
h} - \beta_{\vec{p}} \; b^{(0)+}_
{\vec{p}, h}, & p^0 = -|p^0|, \\
\alpha_{\vec{p}} = \sqrt{\frac{1}{2}(1 +
\frac{|\vec{p}|}{|p^0|})}, & \beta_{\vec{p}} = \sqrt{\frac{1}{2}(1
- \frac{|\vec{p}|}{|p^0|})}, \\  \;\;|p^0| =
\sqrt{\vec{p}^2 + 
m_{D}{ }^2 }. &   
\end{array} \label{maj3.3}
\ee

\noindent The operator $ b^{+}_{\vec{p}, h} $ creates a massive
positive energy one
particle state  ($ p^0 = |p^0| $)  and $
d^{+}_{\vec{p}, h} $ creates a massive negative energy one
particle state  
($ p^0 = - |p^0|$), both states have momentum $\vec{p}$ and
helicity $h$. Both 
are eigenstates of the hamiltonian for a massive  
Dirac field 
 
\be
H^{(D)} = \sum_{{\vec{p}, h}}\;  H^{(D)}_{\vec{p}, h} = 
\sum_{{\vec{p}, h}} \; ( H^{(0)}_{\vec{p}, h} +
H^{(1D)}_{\vec{p}, h} ) = \sum_{{\vec{p}, h}} \;
|p^0|  ( b^{+}_{\vec{p}, h}
b_{\vec{p}, h} - d^{+}_{\vec{p}, h} d_{\vec{p}, h} ) 
 \label{maj3.4}
\ee 

\noindent
of momentum $\vec{p}$ and  helicity $h$. 
The Dirac sea of massive particles is now

\be
|\phi_{(D)}> = \prod_{\vec{p}, h = \pm1}
d^{+}_{\vec{p}, h} |0> =  \pi_{\alpha} \; e^{- \sum_{\vec{p}}  
\frac{\beta_{\vec{p}}}{\alpha_{\vec{p}}} \; b^{(0)+}_{\vec{p}, h}
d^{(0)}_{\vec{p}, h} } \; |\phi_{(0)} >, \; \pi_{\alpha} =
\prod_{\vec{\acute{p}}} \alpha_{\vec{\acute{p}p}}. \label{maj3.5}
\ee  

\noindent
All states up to $p^0 = -m_D$ are occupied and due to Eq.(\ref{maj2.2})
it follows that ${\cal C}
|\phi_{(D)}> = |\phi_{(D)}>. $
 The interaction term
$H^{(1D)}$ causes the superposition of positive and negative
energy massless states (Eq.(\ref{maj3.3})). One sees that the vacuum state of
massive particles can be understood as a coherent state of
particle and antiparticle pairs on the massless vacuum state.
The energy of the vacuum state of massive Dirac particles
$ <\phi_{(D)}| H^{(D)}|\phi_{(D)} > = \sum_{\vec{p}, h} \; E_{\vec{p},
h}^{(D)}, $ with $ 
E_{\vec{p}, h}^{(D)} = 
-\sqrt{\vec{p}^2 + m_D{ }^2 } $,  which is  infinite.\\

\noindent
According to  Eq.(\ref{maj3.3}), the creation and annihilation operators 
for massive fields go in the limit when $m_D
\longrightarrow 0 $ to the creation and annihilation operators
for the massless case. 
\\

A one particle state of energy $|p^0| =
\sqrt{ \vec{p}^2 + m_D^2 } $
can be written as $ b^{+}_{\vec{p}, h} \; |\phi_{(D)}>,$
with $ b^{+}_{\vec{p}, h}$ defined in Eq.(\ref{maj3.3}). Also this state
becomes in the limit $m_{D} = 0$ a massless Weyl one particle
state of positive energy $|\vec{p}|$ above the Sea of massless
particles.  \\

One easily finds that $H^{(1D)}$ is invariant under charge
conjugation and so is therefore also $H^{(D)}$. Taking into
account  Eqs.(\ref{maj2.1}) it follows

\be
\begin{array}{cl}
{\cal C} \; b^{+}_{\vec{p}, h = -1} \; {\cal C}^{-1}
= \;-d^{}_{-\vec{p}, h = 1}, & {\cal C} \;\;
b^{+}_{\vec{p}, h = 1} \;\; {\cal C}^{-1} \;
= d^{}_{-\vec{p}, h = -1}, \\
 {\cal C}\; d^{+}_{\vec{p}, h = 1} \;\;\; {\cal C}^{-1}
= -b^{}_{-\vec{p}, h = -1}, & {\cal C} \;
d^{+}_{\vec{p}, h = -1} \; {\cal C}^{-1} 
= b^{}_{-\vec{p}, h = 1}.\\
\end{array}
\label{maj3.6}
\ee

\noindent
In the limit $m_D \longrightarrow 0 $ Eqs. (\ref{maj3.6}) coincide with
Eqs. (\ref{maj2.1}).
The charge conjugation transforms the particle of a momentum
$\vec{p}$ and helicity $h$ into the hole in the Dirac
sea of the  momentum $-\vec{p}$ and helicity $-h$.


\subsection{ The Majorana massive field theory}


The Majorana mass term with only left handed fields $ \; m_{ML}
\; \int d^3  \vec{x} \; (\bar{\psi}_L + \bar{\psi}_L{ }^C)
(\psi_{L} + \psi_{L}^C) $
can be written, if using the expression for $\psi$ from
Eq.(1.3), with the summation going over the left handed fields
only and if taking into account the definition of charge
conjugation from Sect. 1.,
as follows

$$ H^{(1M)}_{L} = m_{ML} \; \int d^3 \vec{x} \; ( \bar{\psi}_L +
\bar{\psi}_L{ }^C ) (\psi_L + \psi_L{ }^C)
= \sum_{(\vec{p})^+}  
H^{(1M)}_{\vec{p}, L}, $$

\be
H^{(1M)}_{\vec{p}, L} = m_{ML} \; ( b^{(0)+}_
{\vec{p}, h = -1} b^{(0)+}_{-\vec{p}, h = -1} +
b^{(0)}_{-\vec{p}, h = -1} b^{(0)}_{\vec{p}, h = -1} + 
d^{(0)+}_
{\vec{p}, h = 1} d^{(0)+}_{-\vec{p}, h = 1} +  d^{(0)}_
{-\vec{p}, h = 1} d^{(0)}_{\vec{p}, h = 1} ).
\label{maj4.1}
\ee

\noindent
The symbol $\sum_{(\vec{p})^+} $ means that the sum runs over
$\vec{p} $ on such a way that $\vec{p}$ and $-\vec{p}$ is
counted only once. 
Comparing the Majorana interaction term $H^{(1M)}{ }_L $ of
Eq.\ref{maj4.1} with 
the Dirac interaction term of Eq.\ref{maj3.1}, one sees that in both
cases momentum $\vec{p}$ is conserved as it should be. In Eq.
\ref{maj4.1} the two creation operators appear with opposite momentum,
while in Eq.\ref{maj3.1} the creation and annihilation 
operator appear with the same momentum. Because of that we
could pay attention in the Dirac case to a momentum $\vec{p}$,
without connecting ${\vec{p}}$ with ${-\vec{p}}$, while  
in the Majorana case we have to treat ${\vec{p}}$ and
${-\vec{p}}$ at the same time.
\\

The Majorana  mass 
term of only right handed fields follows from the mass term of
only 
left handed fields of  Eq.\ref{maj4.1} if we
exchange $h = -1$ with $h = 1$ and $h = 1$ with $h = -1$.
We shall treat here the left handed fields only. The
corresponding expressions for the
massive
Majorana right handed fields can be obtained from the left
handed ones by the above mentioned exchange of helicities
of fields.
\\

It is easy to check that the charge conjugation operator ${\cal
C}$ from Eq. \ref{maj2.1} leaves the interaction term of Eq.\ref{maj4.1}
unchanged. Accordingly also the hamiltonian

\be
H^{(M)}_{\vec{p}, L } = H^{(0)}_{\vec{p}, L } + 
H^{(1M)}_{\vec{p}, L }  
\label{maj4.2}
\ee

\noindent
is invariant under charge conjugation: $[H^{(M)}_{\vec{p}, L },
{\cal C} ] = 0. $ \\

As in the Dirac massive case, it is meaningful to use the
operators $ h^{+}_{\vec{p}, h = -1} = b^{(0)+}_{\vec{p}, h = -1}
b^{(0)}_ {\vec{p}, h = -1}  $ and $ h^{-}_{\vec{p}, h = 1}
= d^{(0)+}_{\vec{p}, h = 1} 
d^{(0)}_ {\vec{p}, h = 1},$ to
choose the appropriate basis within which we shall diagonalize
the hamiltonian of Eq.\ref{maj4.1}. One can check that the operators

\be
h^{+}_{\vec{p}} = h^{+}_{\vec{p}, h = -1} - h^{+}_{-\vec{p}, h
= -1},\;\; h^{-}_{\vec{p}} = h^{-}_{\vec{p}, h = 1} -
h^{-}_{-\vec{p}, h = 1}, \label{maj4.3}
\ee

which count the momentum of states, commute with the hamiltonian
\ref{maj4.2} 

\be
[ h^{\pm}_{\vec{p}}, H^{(M)}_{\vec{p},L} ] = 0. \label{maj4.4}
\ee

Since basic states, appropriate to describe the vacuum
state, should have momentum equal zero to guarantee the zero
momentum of the vacuum,  one looks for the basic states with
$< h^{\pm}_{\vec{p}} > = 0. $ One finds four such states

\be
\begin{array}{cc}
|1> = &  b^{(0)+}_{\vec{p}, h = -1} b^{(0)+}_{-\vec{p}, h = -1} |0>, \\
|2> = &  \frac{1}{\sqrt{2}} ( 1 + b^{(0)+}_{\vec{p}, h = -1}
b^{(0)+}_{-\vec{p}, h = -1}  d^{(0)+}_{\vec{p}, h = 1}
d^{(0)+}_{-\vec{p}, h = 1} ) |0>, \\
|3> = & d^{(0)+}_{\vec{p}, h = 1}
d^{(0)+}_{-\vec{p}, h = 1} |0>\\ \hline 
|4> = &  \frac{1}{\sqrt{2}} ( 1 - b^{(0)+}_{\vec{p}, h = -1}
b^{(0)+}_{-\vec{p}, h = -1}  d^{(0)+}_{\vec{p}, h = 1}
d^{(0)+}_{-\vec{p}, h = 1} ) |0>. 
\end{array} \label{maj4.5}
\ee

\noindent
One finds that the state $|4>$ is  the eigenstate of
the hamiltonian of Eq.\ref{maj4.2} with the eigenvalue zero. 
The hamiltonian applied on first three basic states defines a
matrix 

\be
\left( \begin{array}{ccc}
2|\vec{p}| & \sqrt{2} m & 0 \\
\sqrt{2} m & 0 & \sqrt{2} m \\
0 & \sqrt{2} m  & -2|\vec{p}|
\end{array} \right). \label{maj4.6}
\ee

\noindent
Diagonalizing this matrix one finds three vectors and three
eigenvalues. The only 
candidate for the vacuum state is the state 
$\beta_{\vec{p}}{ }^2 |1> + (-) \alpha_{\vec{p}} \beta_{\vec{p}}
|2> + \alpha_{\vec{p}}{ }^2 |3>, $ with $\alpha_{\vec{p}} $ and
$\beta_{\vec{p}} $ defined in Eq.\ref{maj3.3},
 with the 
eigenvalue $-2\sqrt{ |\vec{p}|^2 + m_{ML}{ }^2}$ 
which corresponds to the vacuum state of the $-2\sqrt{
|\vec{p}|^2 + m_{D}{ }^2}$ energy in the 
Dirac massive case $d^{+}_{\vec{p}, h = 1}
d^{+}_ {-\vec{p}, h = 1}  |0>.$ 
The Majorana vacuum state is accordingly

\be
|\phi_{(ML)}> = \prod_{(\vec{p})^{+}} \; |\phi_{M\vec{p}, L}>
, \; |\phi_{M\vec{p}, L} > = ( \beta_{\vec{p}}{ }^2 |1> +
(-) \alpha_{\vec{p}} \beta_{\vec{p}} |2> + \alpha_{\vec{p}}{ }^2
|3> ), \label{maj4.7}
\ee

\noindent
with $ \alpha_{\vec{p}}$ and $ \beta_{\vec{p}}$ defined in Eq.\ref{maj3.3}.
Compared to the Dirac particle case it should be noted that 
we for Majorana had to combine both $vec{p}$ and $-vec{p}$
under the diagonalization and construction of ground state
\ref{maj4.7}.
The energy of the vacuum of Majorana left handed
particles  is
$ <\phi_{(ML)}| H^{(M)}|\phi_{(ML)} > = \sum_{(\vec{p})^+, L} \;
E_{\vec{p}, L}^{(ML)}, $ with $ 
E_{\vec{p}, L}^{(ML)} = 
-2 \sqrt{\vec{p}^2 + m_{ML}{ }^2 } $, which is the energy of two
majorana particles of momentum $p^a = ( -|p^0|, \vec{p}) $ and $p^a
= ( -|p^0|, -\vec{p}) $,
respectively. The energy of the Majorana sea is again infinite.
In the limit $ m_{ML} \longrightarrow 0 $ the Majorana sea
becomes a sea of massless Weyl particles of only left handedness. \\

Concerning charge conjugation we see that with the somewhat
complicated transformation of the ``totally empty vacuum'' 
(Eq.\ref{maj2.2}) the Majorana physical vacuum $|\phi_{ML}>$ is charge
conjugation invariant
\be
{\cal C} |\phi_{ML}> = |\phi_{ml}>. \label{maj4.7a}
\ee
\noindent
The one particle Majorana states can be constructed as  superpositions
of states with $< h^{(+)}{ }_{\vec{p}} + h^{(-)}{ }_{\vec{p}} > =
\pm 1 $. One finds four times two states which fulfil this
condition 

\be
\begin{array}{cc}
|5>\; = &  b^{(0)+}_{\vec{p}, h = -1} |0>,\\
|6>\; = & b^{(0)+}_{\vec{p}, h = -1} d^{(0)+}_{\vec{p}, h =
1} d^{(0)+}_{-\vec{p}, h = 1} |0>, \\
\hline
|7>\; = &  d^{(0)+}_{\vec{p}, h = 1} |0>,\\
|8>\; = &  d^{(0)+}_{\vec{p}, h = 1} b^{(0)+}_{\vec{p}, h =
-1} b^{(0)+}_{-\vec{p}, h = -1} |0>, \\
\hline\hline
|9>\; = &  b^{(0)+}_{-\vec{p}, h = -1} |0>,\\
|10> = & b^{(0)+}_{-\vec{p}, h = -1} d^{(0)+}_{\vec{p}, h =
1} d^{(0)+}_{-\vec{p}, h = 1} |0>, \\
\hline
|5> = &  d^{(0)+}_{-\vec{p}, h = 1} |0>,\\
|6> = &  d^{(0)+}_{-\vec{p}, h = -1} b^{(0)+}_{\vec{p}, h =
-1} b^{(0)+}_{-\vec{p}, h = -1} |0>, 
\end{array} \label{maj4.8}
\ee

\noindent
The first four states have a momentum $\vec{p}$ and the last four 
states  a momentum $-\vec{p}$. The hamiltonian $ H^{(M)}_{
}{\vec{p}, L} $ defines on these states the block diagonal four
two by two matrices. The candidates for the states describing a
one particle state of momentum $\vec{p}$ on a vacuum 
states $|0> $ are states with energy which is for $ p^0 =
\sqrt{\vec{p}^2 + m_{ML}^2 } $ higher than the vacuum state.
One finds the corresponding operators

\be
\begin{array}{cc}
b^{+}_{\vec{p}, h = -1}  = & -\beta_{\vec{p}}
\; b^{(0)+}_{\vec{p}, h = -1} + \alpha_{\vec{p}} \; b^{(0)+}_{\vec{p}, h =
-1} d^{(0)+}_{\vec{p}, h = 1}  d^{(0)+}_{-\vec{p}, h = 1}, \\
d^{+}_{\vec{p}, h = 1}  = & -\alpha_{\vec{p}}
\; d^{(0)+}_{\vec{p}, h = 1} + \beta_{\vec{p}} \; d^{(0)+}_{\vec{p}, h =
1} b^{(0)+}_{\vec{p}, h = -1}  b^{(0)+}_{-\vec{p}, h = -1}, \\ \hline
b^{+}_{-\vec{p}, h = -1}  = & -\beta_{\vec{p}}
\; b^{(0)+}_{-\vec{p}, h = -1} + \alpha_{\vec{p}} \;
b^{(0)+}_{-\vec{p}, h = 
-1} d^{(0)+}_{\vec{p}, h = 1}  d^{(0)+}_{-\vec{p}, h = 1}, \\
d^{+}_{-\vec{p}, h = 1}  = & -\alpha_{\vec{p}}
\; d^{(0)+}_{-\vec{p}, h = 1} + \beta_{\vec{p}} \; d^{(0)+}_{-\vec{p}, h =
1} b^{(0)+}_{\vec{p}, h = -1}  b^{(0)+}_{-\vec{p}, h = -1}, 
\end{array} \label{maj4.9}
\ee

\noindent
which when applied on a true vacuum $|0> $ generates the one
particle states of momentum $\vec{p}$ (the first two operators)
and $-\vec{p}$ (the second two operators), respectively.
\\

We would prefer to know, as in the Dirac massive case, the one
particle operators which when being applied on a Majorana vacuum
state $|\phi_{(ML)}> $ generates a one particle Majorana state with
chosen momentum $\vec{p}$ and which commute with the charge
conjugate operator ${\cal C}$ defined in Eq.\ref{maj2.1}.
Requiring 
$ B^+{ }_{\vec{p}, h = -1} |\phi_{\vec{p}, L} > =
b^+_{\vec{p}, h = -1} |0> $ one finds $ B^+{ }_{\vec{p}, h
= -1} = \alpha_{\vec{p}} \;  b^{(0)+ }_{\vec{p}, h = -1} + 
\beta_{\vec{p}} \;  b^{(0)}_{-\vec{p}, h = -1}. $ Accordingly it follows    
from the requirement $ D^+{ }_{\vec{p}, h = 1} |\phi_{\vec{p},
L} > = d^{+}_{\vec{p}, h = 1} |0> $ that $ D^+{ }_{\vec{p}, h
= 1} = \beta_{\vec{p}} \;  d^{(0)+ }_{\vec{p}, h = 1} + 
\alpha_{\vec{p}} \;  d^{(0) }_{-\vec{p}, h = 1}. $
Taking into account that $ {\cal C} \; B^+{ }_{\vec{p}, h = -1}
{\cal C}^{-1} = -D^{+}_{-\vec{p}, h = 1}  $
we may  conclude that the two operators

\be
{\cal B}^+{ }_{ \pm\vec{p}, h = -1} = \alpha_{\vec{p}} \;(
b^{(0)+ }_{\pm\vec{p}, h = -1} - d^{(0)}_{\mp\vec{p}, h = 1} ) -
\beta_{\vec{p}} \; ( d^{(0)+}_{\pm\vec{p}, h = 1} - b^{(0)}{
}_{\mp\vec{p}, h = -1} ). \label{maj4.10}
\ee

\noindent
$ {\cal B}^+{ }_{ \pm\vec{p}, h = -1} $ operating on the
Majorana vacuum state $|\phi_{ML}>$ generates 
the one particle 
Majorana state of momentum $\pm\vec{p}$. It can easily be
checked that Majorana particle is his own antiparticle $ {\cal
C} \; {\cal B}^+{ }_{ \pm\vec{p}, h = -1} \; {\cal C}^{-1} =
{\cal B}^+{ }_{ \pm \vec{p}, h = -1} $.\\

In the limit when $m_{ML} \longrightarrow 0$, the operator ${\cal
B}^+{ }_{ 
\pm\vec{p}, h = -1} $ operating on a vacuum $|\phi_{(ML)}>$, 
which goes to the vacuum state of the massless case of only left
handed paricles gives a state of a Majorana
massless particle: $ ( b^{(0)+
}{ }_{\pm\vec{p}, h = -1} - d^{(0)}_{\mp\vec{p}, h = 1} )$ $ 
  d^{(0)+}{ }_{\vec{p}, h = 1}  d^{(0)+}{
}_{-\vec{p}, h = 1} $ $ \prod_{\vec{\acute{p}}, \acute{p} \neq p
} \; d^{(0)+}{
}_{\vec{\acute{p}}} |0>. $\\

 One can
accordingly find the operators for 
right handed Majorana 
particles.


We have learned that it is indeed possible to define the
Majorana sea in the way the Dirac sea is defined. This put a new
light on the Majorana particle. It stays to
study whether or not this presentation can be used to better
understand the properties of the Majorana particles.




\newpage
\stepcounter{section}
\setcounter{equation}{0}
\section*{\center Can one connect the Dirac-K\" ahler representation of Dirac
spinors and spinor representations in Grassmann space, proposed
by Manko\v c?}
\centerline{\large  Holger and Norma}

\vspace{5mm}
\noindent
The contribution to this question which follows runs out of 
a complaint like this by Holger and others:

\noindent
It is very suspicious that you, Norma, get Dirac spinors out of
a start with only $\theta^a$ variables, which are clearly
vectors under Lorentz transformations: wave functions depending
on 
the $\theta^a$-variables have a priori no way of being 
spinors. 

\noindent
Really this is much like the Dirac-K\" ahler construction.

\subsection{Introduction:}

\noindent
In Norma's contribution to these proceedings the fermions 
come out of the $d$ dimensional Grassmannian set of
coordinates in the sense that the spin degrees of freedom 
of the fermions appear as represented by wave functions
defined on this space parametrized by Grasmannian variables
called there $\theta^a$. For obtaining the {\bf  spinor}
degrees of 
freedom the ordinary  spatio-temporal  coordinates $x^a$ also
assumed in Norma's model are not important and we here totally
ignore those. The variables $\theta^a$ were assumed to 
transform as a vector under Lorentz transformations.  
One would therefore - a priori - observe that all wave functions
depending on these $\theta^a$'s could only transform as having
integer spin. Nevertheless Norma claims to get half integer
spin degrees of freedom out of her model from these variables!

\noindent
This mysterious result comes about after introducing for the 
Grassmann odd variables the
creation  and anihilation like operators  
\begin{equation}
\tilde{a}^a := i(
 p^{\theta a} - i\theta^a ),\;\;\;
\tilde{\tilde{a}}^a := -( p^{\theta a} + i\theta^a ), 
\label{n2.2}
\end{equation}

\noindent
where

\begin{equation}
p^{\theta a} := -i\overrightarrow{\partial^{\theta}}_a.
\label{ptheta}
\end{equation}

\noindent
A crucial - and by Holger ( but not Norma) considered suspicious
- step consists in hoping for that 
a "constraint" (or whatever) could make the $\tilde{\tilde{a}}^a$ 
vanish so as to justify 
putting 
\begin{equation}
\tilde{\tilde{a}}^a \rightarrow 0
\end{equation}
whenever it occurs in the Hamiltonian ( See Norma's contribution 
Eq.(4.8) ).  
The excuse for this replacement is that it is easily found that
\begin{equation}
\{ \tilde{a}^a, \tilde{\tilde{a}}^a \} =0
\end{equation}
and also that therefore
\begin{equation}
[\tilde{S^{ab}}, \tilde{\tilde{S}}^{ }{cd} ] = 0
\end{equation} 
where the Lorentz transformation generator parts
$\tilde{S}^{ab}$ and $\tilde{\tilde{S}}^{ }{cd}$ are defined by
\begin{equation}
\tilde{S}^{ab} :=\frac{i}{4}[ \tilde{a}^a,\tilde{a}^b]\;\;
;\;\; \tilde{\tilde{S}}^{ }{cd} :=
\frac{i}{4}[\tilde{\tilde{a}}^c,\tilde{\tilde{a}}^d ]
\end{equation}
and together make up the total Lorentz generator
\begin{equation}
S^{ab} = \tilde{S}^{ab} + \tilde{\tilde{S}}^{ }{ab}.
\end{equation}

\noindent
In this way it could be consistent if all the
$\tilde{\tilde{a}}^a$  were set to zero because they commute
with the rest of the variables - the $\tilde{a}^a$.
But the best reason for seeking to put the
$\tilde{\tilde{a}}^a$ to zero is that with a reasonable choice
of the kinetic term in the Lagrangian $-i\dot{\theta^a}\theta^a$
we get as the expression for the conjugate variable to $\theta^a$
\begin{equation}
p^{\theta a} = i\theta^a
\end{equation}
which implies that 
\begin{equation}
\tilde{\tilde{a}}^a =0.
\end{equation}

\noindent
(See Norma's contribution Eq.(4.2))

\noindent
Now, however, the way that Norma chooses to quantize the system, 
that is a particle moving in (ordinary and) Grassmanian
coordinate space,
is to let the wave function be allowed to be any function of the
$d$ Grassmann variables $\theta^a$, so that any such function
represents a state of the system. But in this quantization the
$\tilde{\tilde{a}}^a$ turn out not to be zero. 
In other words that quantization turned out not to obey the
equation expected from expression for the canonical coordinate
$p^{\theta a}$ as derived from the Lagrangian.

\noindent
If, however, in the operators such
as the Hamiltonian and the Lorentz transformation operators 
$\tilde{\tilde{a}}^a$ are just put to zero, 
the expressions obtained after having put the
$\tilde{\tilde{a}}^a$ to zero - i.e. we especially only
use $\tilde{S}^{ab}$ as the Lorentz generator - one has
in principle a new Lorentz transformation instead of the 
a priori one in the wave function on Grassmann-space quantization
used. In that case one could a priori expect that the 
argument for only having integer spin could break down.
Indeed the calculations confirm this to happen.

\noindent
We should now attempt to get an understanding of what goes on
here by using a basis inspired from the
Dirac-K\" ahler construction, which is a way often used on
lattices 
to implement fermions on the lattice. The Dirac-K\" ahler
constructions starts from a field theory with a series of 
fields which are 0-form, 1-form, 2-form, ...,d-form.
They can be thought of as being expanded on a basis of all the
wedge product combinations of the basis $dx^1$, $dx^2$,
...,$dx^d$ for the one-forms, including wedge products from zero
factors to d factors. In the Dirac-K\" ahler construction one
succeeds in constructing out of these``all types of forms'' for
$d = 2n$, with n an integer,
$2^{d/2}$ Dirac spinor fields. This construction should without
cheat be impossible in much the same way as Norma's ought to be.

\noindent
It is the major point of the below calculation  to use the
ideas of the Dirac-K\" ahler construction to make such a basis 
choice for Norma's wavefunctions that the connection between the
two seemingly impossible achievements - Dirac-K\" ahlers and Norma's
- are brought to more light.

\subsection{Dirac-K\" ahler approach in our own way}

\subsubsection{The basis with the spinor indices}

\noindent
With the Dirac-K\" ahler construction in mind we propose to expand
the set of wave functions that was assumed to be the set of all
functions of the $d$ Grassmann variables $\theta^a$ on the
following system of basis wave functions, that are marked by two
spinor indices and for the case of an odd $d$ in addition by an
index $\Gamma$ which can take two values $+1$ and $-1$ and which
reminds of the handedness.

\noindent
For the even d case one has
\begin{equation}
\psi_{\alpha\beta}(\{ \theta^a \}) := \sum_{i=0}^d 
(\gamma_{a_1}\gamma_{a_2} \cdots \gamma_{a_i})_{\alpha\beta}
\theta^{a_1}\theta^{a_2} \cdots \theta^{a_i},
\end{equation}

\noindent
while for the odd d case :
\begin{equation}
\psi_{\alpha\beta\Gamma}(\{ \theta^a \}) := \sum_{i=0}^d
(\gamma_{(\Gamma)a_1}\gamma_{(\Gamma)a_2} \cdots
\gamma_{(\Gamma)a_i})_{\alpha\beta} 
\theta^{a_1}\theta^{a_2} \cdots \theta^{a_i},
\end{equation}
with the convention
$ a_1 < a_2 < a_3<...< a_i.$ 
Here the sums run over the number $i$ of factors in the products
of $\theta^a$ coordinates, a number which is the same as the
number of gamma-matrix factors and it should be remarked that we
include the possibility $i=0$ which means no factors and is
taken to mean that the product of the $\theta^a$-factors is
unity and the product of zero gamma matrices is the unit matrix
(as is natural). The indices $\alpha\beta$ are the spinor
indices and taking the product of gamma matrices as matrices
the symbol $(...)_{\alpha\beta}$ means taking the $\alpha$th
row and $\beta$th column element in the matrix. There is an
understood Einstein convention summation over the contracted 
indices $a_1$,$ a_2$, ...,$a_i$, which are vector indices.
The gamma-matrices are in the even case $2^{d/2}$ by $2^{d/2}$
matrices and in the odd case $2^{(d-1)/2}$ by $2^{(d-1)/2}$
matrices, 
but in the latter case we can choose the sign on say the last 
one of these gamma-matrices depending on the last wavefunction
basis index $\Gamma$ so as to arrange that
\begin{equation}
\gamma_1\gamma_2\cdot\cdots\cdot\gamma_d =\Gamma. 
\end{equation}
The $ \gamma^a$ matrices could be constructed  as follows 
- a construction also showing that they really do exist - by
an easy check of the Clifford algebra
\begin{equation}
\{ \gamma_a,\gamma_b\}= 2 \eta_{ab}
\end{equation}
using a $2^{d/2}$ or $2^{(d-1)/2}$ dimensional spinor space
conceived of as the cartesian product $d/2$ or $(d-1)/2$ spin
one half two dimensional Hilbert spaces:

$$\begin{array}{ccccccccccc}
\gamma_1& :=& i\sigma_2^1 &\times & \sigma_3^2 &\times &
\sigma_3^3& \times &\cdots& 
\times &\sigma_3^n\\ 
\gamma_2 &:=& -i\sigma_1^1&\times &\sigma_3^2&\times & \sigma_3^3
&\times &\cdots& 
\times& \sigma_3^n\\ 
\gamma_3 &:=& iI^1&\times & \sigma_2^2&\times & \sigma_3^3&\times &
\cdots& 
\times &\sigma_3^n\\ 
\gamma_4 &:=& iI^1&\times & (-)\sigma_1^2&\times & \sigma_3^3&\times
& \cdots& 
\times& \sigma_3^n\\ 
\gamma_5 &:=& iI^1&\times & I^2&\times &\sigma_2^3&\times
 & \cdots & 
\times& \sigma_3^n\\ 
\vdots & & \vdots & \vdots & \vdots& \ddots & \vdots \\  
\gamma_{2n-1} &:=&iI^1&\times & I^2&\times & I^3& \times & \cdots &
\times &\sigma_2^n\\ 
\gamma_{2n} &:=&iI^1&\times  &I^2&\times & I^3&\times & \cdots& 
\times &(-)\sigma_1^n\\ 
\end{array}$$

\noindent
for an even $d=2n$, while for an odd $d=2n+1$ the term 
$\gamma_{2n+1}$ has to be added as follows

$$\begin{array}{ccccccccccc}
\gamma_{2n+1}& := &i\Gamma
\sigma_3^1&\times & \sigma_3^2&\times &\sigma_3^3&\times &\cdots&\times&
\sigma_3^n,\\ 
\end{array}$$
with
$\Gamma = \prod_{i}^{2n+1} \gamma_i$.

\noindent
The above metric is supposed to be Euclidean. For the Minkowski
metric $ \gamma_1 \rightarrow -i\gamma_1$ has to be taken, if the
index $1$ is recognized as the "time" index.
We shall make use of the Minkowski metric, counting the
$\gamma^a$ from $0,1,2,3,5,..d$, and assuming the metric
$g^{ab} = diag(1,-1,-1,...,-1)$.

\noindent
It is our main point to show that the action by the operators 
$\tilde{a}^a$ and $\tilde{\tilde{a}}^a$ in the representation
based on the basis $\psi_{\alpha \beta}(\{ \theta^a \})$ 
with $\alpha, \beta = 1, 2, ...,\left\{ 
\begin{array}{c}
2^{(d-1)/2} \;\; {\rm for} \;\; d\;\; {\rm odd} \\ 
2^{d/2}\;\; {\rm for} \;\; d\;\;
{\rm even} \end{array}
\right\}$ transforms the index $\alpha$
and $\beta$, respectively, of the basis $\psi_{\alpha\beta}(\{
\theta^a \})$ as follows:

\begin{equation}
\tilde{a}^a \psi_{\alpha\beta(\Gamma)}(\{ \theta^a \}) \propto
\gamma^a_{\alpha \gamma} \psi_{\gamma\beta(\Gamma)}(\{ \theta^a
\}), 
\nonumber
\end{equation}

\begin{equation}
\tilde{\tilde{a}}^a \psi_{\alpha\beta(\Gamma)}(\{ \theta^a \})
\propto 
\psi_{\alpha\gamma(-\Gamma)}(\{ \theta^a \}) \gamma^a_{\gamma
\beta}, 
\label{tta}
\end{equation}
	
\noindent
which demonstrates the similarities between the spinors of the
Normas approach and the Dirac-K\" ahler approach: The operators
$\tilde{a}^a$ transform the left index of the basis
$\psi_{\alpha\beta(\Gamma)}(\{ \theta^a \}) $, while keeping the right
index fixed and the
operators $\tilde{\tilde{a}}^a$ transform the right index of the basis
$\psi_{\alpha\beta(\Gamma)}(\{ \theta^a \}) $ and keep the left
index fixed. Under the action of either $\tilde{a}^a$ or
$\tilde{\tilde{a}}^a$ the basic functions 
transform  as spinors. The index in parentheses $(\Gamma)$ is defined 
for only odd d.
We can count that the number of spinors is $2^d$ either in the
Norma's approach ( the d dimensional Grassmann space has $2^d$
basic functions ) or in the Dirac-K\" ahler approach (  for $d=2n$
 the number of spinors is $2^{d/2} \cdot
2^{d/2}$, while for $d=2n+1$ the number of spinors is
$2^{(d-1)/2} \cdot 2^{(d-1)/2} \cdot 2$, $n$ is an integer). 

\noindent 
We shall prove the above formulas for action of the
$\tilde{a}^a$ and $\tilde{\tilde{a}}^a$ on basic functions after 
presenting  special cases $ d=1, d=2$
.

\subsubsection{d=1,2
special cases checking basis properties}

\noindent
We shall present the Dirac-K\" ahler basis for 
two 
cases $d =
1,2
,3
$ to see what it means. For comparison with the basis of
the Manko\v c approach see ref.\cite{1man1}.

\noindent
{\bf i) The one dimensional ($d=1$) space.}\\
The Dirac-K\" ahler basis has two ($2^1$)  vectors of the mixed
Grassmann character.
\begin{equation}
\gamma^0 = {\rm I}\; \Gamma,\;\;\;\;
\psi_{11,\Gamma = 1} = 1 + \theta,\;\;\;\;
\psi_{11,\Gamma = -1} = 1 - \theta.
\end{equation}

\noindent
The operator $\tilde{a}^0$ is in this basis a diagonal and 
$\tilde{\tilde{a}}^0$ a 
non-diagonal matrix. 
The superposition of the above basis leads to the new
basis ($1, \theta$) with well defined Grassmann character.

\noindent
{\bf ii) The two dimensional ($d=2$) space.}\\
The Dirac-K\" ahler basis has four ($2^2$) vectors of either
even or odd Grassmann character. According to the definition of
$\gamma$ - matrices we have
\begin{equation}
\gamma^0 = \sigma ^2, \;\;\gamma^1 = -i \sigma^1\\
\nonumber
\end{equation}
and
\begin{equation}
\psi_{\alpha \beta} = 1_{\alpha \beta} + (\gamma^0){ }_{\alpha
\beta} \theta^0 + (\gamma^1){ }_{\alpha \beta} \theta^1 +
(\gamma^0 \gamma^1)_{\alpha \beta} \theta^0 \theta^1.\\
\end{equation}
One finds accordingly
\begin{equation}
\psi_{11} = 1 - \theta^0 \theta^1, \;\;\;\\
\psi_{12} = -i( \theta^0 + \theta^1 ), \;\;\;\\
\psi_{21} = i( \theta^0 - \theta^1 ), \;\;\;\\
\psi_{22} = 1 + \theta^0 \theta^1.\\
\end{equation}

\noindent
One easily checks that $\tilde{a}^a, \;\; a \in 1,2,$ transform
the first index of $\psi_{\alpha, \beta}$, while $\tilde{\tilde{a}}^a,
\;\; a \in 1,2,$ transform the second index of $\psi_{\alpha,
\beta}$, both transforming a 
Grassmann odd function into a Grassmann even function or
opposite. 


\subsubsection{Proof of our formula for action of 
$\tilde{a}^a$ and $\tilde{\tilde{a}}^a$ }

\noindent
Let us first introduce the notation
\begin{equation}
\gamma^A := \gamma^a\gamma^b \cdots \gamma^c,\;\;\;\\
\gamma^{\overline{A}} := \gamma^c\gamma^b \cdots \gamma^a,
\end{equation}
with $a < b < \cdots < c \in A.$
We  recognize that 
\begin{equation}
{\rm Trace} \; (\gamma_A \gamma^{\overline{B}}) = {\rm Trace (I)}
\; \delta_A{ }^B,\;\;\;\\
\nonumber
\sum_A (\gamma_A)_{\alpha \beta} (\gamma^{\overline{A}})_{\gamma
\delta} = {\rm Trace(I)}\; \delta_{\alpha \gamma}
\; \delta_{\beta \delta} 
\end{equation}
and
\begin{equation}
\sum_i (\gamma_{A_i})_{\alpha \beta} (\gamma^c
\gamma^{\overline{A}_i})_{\gamma \delta} = 
{\rm Trace (I)}\;(\gamma^c )_{\alpha
\delta} \delta_{\beta \gamma},\;\;\;\; \\
\nonumber
\sum_i (\gamma_{A_i})_{\alpha \beta} (
\gamma^{\overline{A}_i}(-1)^i \gamma^c)_{\gamma \delta} =
{\rm Trace (I)}\; (\gamma^c )_{\gamma 
\beta} \delta_{\delta \alpha}.
\end{equation}
Using the first equation we find
\begin{equation}
\theta^{A} = \frac{1}{{\rm Trace(I)}}\;
(\gamma^{\overline{A}})_{\alpha \beta}\; \psi_{\beta \alpha
(\Gamma)}(\{\theta^a \}).
\end{equation}
The index $(\Gamma)$ has the meaning for only odd $d$. That is
why we put it in parenthesis.
We may accordingly write
\begin{equation}
\psi_{\alpha\beta(\Gamma)}(\{ \theta^a \}) :=   \sum_{i=0}
\frac{1}{{\rm Trace(I)}}\; (\gamma_{A_i})_{\alpha \beta}
(\gamma^{\overline{A}_i} )_{\gamma \delta} \psi_{\delta
\gamma (\Gamma)}(\{\theta^a \}),\\
\end{equation}
 with $ A_i \in a_1< a_2,\cdots,< a_i$  in ascending order and with
$\overline{A}_i $ in descending order.

\noindent
Then we find, taking into account that $\tilde{a}^a |0> =
\theta^a$,  $\tilde{\tilde{a}}{ }^a |0> = -i \theta^a$, where
$|0>$ is a vacuum state and Eq.(4)
\begin{equation}
\tilde{a}^c \psi_{\alpha\beta(\Gamma)}(\{ \theta^a \}) := 
\sum_{i}(\gamma_{A_i})_{\alpha \beta} \tilde{a}^c \theta^{A_i} =
\sum_{i}(\gamma_{A_i})_{\alpha \beta} \tilde{a}^c
\tilde{a}^{A_i} |0> = 
\sum_{i}
\frac{1}{{\rm Trace(I)}}\; (\gamma_{A_i})_{\alpha \beta}
(\gamma^c \gamma^{\overline{A}_i} )_{\gamma \delta} \psi_{\delta
\gamma (\Gamma)} (\{\theta^a \}).
\end{equation}
Using  the above relations we further find
\begin{equation}
\tilde{a}^c \psi_{\alpha \beta (\Gamma)}(\{ \theta^a \}) :=
(-1)^{\tilde{f} (d,c) } (\gamma^c)_{\alpha \gamma}
 \psi_{\gamma 
\beta (\Gamma)} (\{ \theta^a \}),
\label{t}
\end{equation}
where $ (-1)^{\tilde{f} (d,c) }$ is $\pm 1$, which depends on
the operator $\tilde{a}^c$ and the dimension of the space.

\noindent
We find in a similar way
\begin{equation}
\tilde{\tilde{a}}^c \psi_{\alpha\beta(\Gamma)}(\{ \theta^a \}) := 
\sum_{i}(\gamma_{A_i})_{\alpha \beta} \tilde{\tilde{a}}^c \tilde{a}^{A_i}
|0> = 
\sum_{i} (-1)^i (\gamma_{A_i})_{\alpha \beta} \tilde{a}^{A_i}
\tilde{\tilde{a}}^c |0> = \\
\sum_{i} (-1)^i (\gamma_{A_i})_{\alpha \beta} \tilde{a}^{A_i}
\tilde{a}^c |0>,  
\nonumber
\end{equation}
which gives
\begin{equation}
\sum_{i} \frac{(-1)^i}
{{\rm Trace(I)}};\ (\gamma_{A_i})_{\alpha \beta}
( \gamma^{\overline{A}_i \gamma^c } )_{\gamma \delta}\; \psi_{\delta
\gamma (\Gamma)} (\{\theta^a \}),
\nonumber
\end{equation}
and finally
%
%
\begin{equation}
\tilde{\tilde{a}}^c \psi_{\alpha\beta(\Gamma)}(\{ \theta^a \}) := 
(-1)^{\tilde{\tilde{f}}(d,c)}
\psi_{\alpha
\gamma(-\Gamma)} (\{\theta^a \}) (\gamma^c)_{\gamma \beta}, 
\label{tt}
\end{equation}
with the sign $ (-1)^{\tilde{\tilde{f}}(d,c)}$ depending on
the dimension of the space and the operator
$\tilde{\tilde{a}}^c$. 

\noindent
We have therefore proven the two equations which determines the
action of the operators $\tilde{a}^a$ and $\tilde{\tilde{a}}^a$
on the basic function $\psi_{\alpha
\gamma(-\Gamma)} (\{\theta^a \}) $.

\subsection{Getting an even gamma matrix }

\noindent
According to the Eqs.(\cite[t,tt]) 
it is obvious that the $\gamma^a$ {\bf matrices, entering into
the Dirac-K\" ahler approach for spinors, have an odd
Grassmann character} since both, $\tilde{a}^a$ and
$\tilde{\tilde{a}}^a$, 
have  an odd Grassmann character. They therefore transform
a Grassmann odd basic function into a Grassmann even basic
function changing fermions into bosons. It is clear that
such $\gamma^a$ matrices are not appropriate to enter into the
equations of motion and Lagrangeans for spinors. 

\noindent
Can one find an appropriate definition of the $\gamma^a $
matrices? Yes. Here is an suggestion for the way out!

\noindent
If working  with $\tilde{a}^a$ only, putting
$\tilde{\tilde{a}}^a$ in the Hamiltonian, Lagrangean and all the
operators equal to zero, it is meaningful to define  the
$\gamma^a$ matrices of  an even Grassmann character as follows
\begin{equation}
\tilde{\gamma}^a:= i \tilde{a}^a \tilde{\tilde{a}}^0.
\end{equation}
One can immediately check that 
\begin{equation}
\{\tilde{\gamma}^a,
\tilde{\gamma}^b \} = \{ \tilde{a}^a, \tilde{a}^b\} =
2\eta^{ab}, \;\;\;\; \tilde{S}^{ab} = 
\frac{i}{4} [\tilde{a}^a,
\tilde{a}^b] =  \frac{i}{4} [\tilde{\gamma}^a,
\tilde{\gamma}^b].
\end{equation} 
We have
\begin{equation}
\tilde{\gamma}^a \psi_{\alpha\beta(\Gamma)}(\{ \theta^a \}) = 
\gamma^a_{\alpha \gamma} \psi_{\gamma\delta(-\Gamma)}(\{
\theta^a \}) \gamma^0_{\delta \beta}. 
\end{equation} 
One can  check that $\tilde{\gamma}^a$ have all the
properties of the Dirac $\gamma^a$ matrices.

\noindent
(If working only with $\tilde{\tilde{a}}^a$ the $\gamma^a$
matrices defined as $\tilde{\tilde{\gamma}}^a:= i \tilde{\tilde{a}}^a
\tilde{a}^0$ have again all the properties of the Dirac
$\gamma^a$ matrices.)

\subsection{Conclusion and what we learn}

\noindent
We have shown that the answer to the question:
"{\bf  Can one connect the Dirac-K\" ahler representation of Dirac
spinors and spinor representations in Grassmann space, proposed
by Manko\v c?", is yes.} The action of the operators $\tilde{a}^a$
on the basic functions $\psi_{\alpha\beta}(\{ \theta^a \})$
transforms the index $\alpha$, keeping the index $\beta$ fixed.
The action of the operators 
$\tilde{\tilde{a}}^a$ 
on the basic functions $\psi_{\alpha\beta}(\{ \theta^a \})$
transforms the index $\beta$, keeping fixed the index $\alpha$.
In both cases the basic functions 
transform as spinors and accordingly fulfill the Dirac equation.

\noindent
The Lorentz transformations are in the Manko\v c approach
determined either  only with $\tilde{a}^a $ ($ \tilde{S}^{ab} = 
\frac{i}{4} [ \tilde{a}^a, \tilde{a}^b ] $) or  only with
$\tilde{\tilde{a}}^a $ ( $\tilde{\tilde{S}}^{ab} = 
\frac{i}{4} [ \tilde{\tilde{a}}^a, \tilde{\tilde{a}}^b ] $).
In the Dirac-K\" ahler case the Lorentz transformations
transform either the index $\alpha $ or the index $\beta$. 
If one shifts what is meant by a Lorentz transformation, then
of course it is not so surprising, if it turns out that there
can appear particles/states with an a priori unexpected spin.

\noindent
Not only we have establish the connection between these two
approaches, {\bf we have also shown that the Dirac matrices as appear
in the Dirac-K\" ahler approach have a Grassmann odd character. To
make them having Grassmann even character the transformation of
the index $\alpha$ should be accompanied by the simultaneous
transformation of the index $\beta=0$.} We have also
learned that the Dirac-K\" ahler approach to spinors have for
odd dimensional spaces mixed Grassmann character. Accordingly it
is the appropriate superposition of the Dirac-K\" ahler basis
which can be used to describe spinors.

\noindent
Since in both cases, that is in the Manko\v c approach and the
Dirac-K\" ahler approach, the dimension of the space is $2^d$, it
{\bf means that in the four dimensional space-time there are four
times four spinors, which may be responsible for four families
of quarks and leptons.} This four flavour prediction follows
{\bf both} in Norma's and in other uses of the Dirac-K\" ahler -
like approach. Both  suffer from the empirical evidence
for only three generations.

\setcounter{page}{73}
\setcounter{section}{5}
\newpage
\stepcounter{section}
\setcounter{equation}{0}


\section*{\center Comments on the Hierarchy Problem}
\centerline{\large  Berthold}

\vspace{5mm}

The standard model treated with a momentum space 
cut-off 
has quadratic and logarithmic divergencies.
In perturbation theory fine-tuned large 
subtractions depending
precisely on the standard model parameters 
have to be performed.
This is quite unnatural and known as the 
hierarchy problem. How the
scale of weak interaction and, in particular, 
how scalar masses
are protected, is not understood.
Quadratic divergencies make severe problems 
in higher-order
calculations. They can formally be avoided by 
using dimensional
regularization. They do not occur in supersymmetric 
models
where the supersymmetric partners of each 
particle provide
for a cancellation of the quadratic divergence. 
In spite of
this elegant solution, crude attempts have been 
made and are still
made to obtain a full or near cancellation of 
divergencies by
special choices of the standard model couplings
which could make the
introduction of super partners unnecessary 
\cite{s1},\cite{s2}.

The divergencies can best be studied by looking 
at the logarithm
of the partition function, the ``free energy'' 
$W(J,j)$ \cite{s3}. Here
$J$ denotes a parameter in the Higgs Lagrangian 
of dimension
$m^2$ which multiplies the square of the Higgs field. 
$j$ describes
a field coupled linearly to the neutral Higgs 
component. The
vacuum expectation value of the unrenormalized 
neutral Higgs field is
obtained from
\be\label{1}
<\Phi_0>\sim\frac{\partial W(J,j)}{\partial j}
\Big|_{J=J_0,j=0}~~.
\ee
We denoted the unrenormalized mass term in the 
action by $J_0$.
The vacuum expectation value of the square of
the unrenormalized Higgs field can be found from 
\be\label{2}
<\Phi^\dagger\Phi>\sim\frac{\partial W(J,j)}{\partial J}
\Big|_{J=J_0,j=0}~~.\ee
The calculation of $W(J,j)$ to one-loop order can be
done by the saddle-point method. Taking a universal 
cut-off $\Lambda$
for all propagators, the quadratic divergence arising from
(\ref{1}) is proportional to
\be\label{3}
S_{\Lambda^2}=(3m^2_H+3m^2_Z+6m^2_W-12\sum_fm^2_f)/<\Phi_0>^2\ee
For simplicity the couplings contained in (\ref{3})
are given in terms of the masses.
 The sum in (\ref{3}) extends over all fermions
which obtain their masses through their couplings 
to the Higgs
field. The same expression for $S_{\Lambda^2}$ is
obtained from (\ref{2})
after subtracting the free field divergency as discussed in
reference  \cite{s3}.

The vanishing or smallness
$(\sim1/\Lambda^2)$ of $S_{\Lambda^2}$ has been 
proposed to diminish
the hierarchy problem. The corresponding mass relation
is usually referred to as the Veltman condition. 
It gave a lower
limit on the top quark mass before the top was 
discovered. One of the
problems with the Veltman condition is the question 
of the scale at
which $S_{\Lambda^2}\simeq 0$ should hold. The vacuum 
expectation value
of the unrenormalized Higgs field is scale-independent, but
higher-order calculations and the knowledge of the scale 
dependence of $\Lambda$ are needed ($\Lambda $ may be related 
to high mass states) to take advantage of this
fact. So far it remains open  which scale $\mu$ should 
be taken
such that higher order terms are reasonably small and the 
effect of
the cut-off is suppressed. Because of the strong 
scale dependence
of the top quark Yukawa coupling according to the 
renormalisation 
group equations the predictions for the Higgs mass
from the requirement $S_{\Lambda^2}=0$ ranges from 
$m_H(m_Z)\approx$ 320 GeV
for $\mu\approx m_Z$ down to $m_H(m_Z)\approx$ 140 GeV for
$\mu\approx m_{Planck}$.

Let us now have a look at the logarithmic dependence on the
cut-off $\Lambda$. There is a difference between the results
obtained from the vacuum
expectation value of the linear Higgs field (\ref{1}) 
and the one from
the square of the Higgs field (\ref{2}). In the first case 
the log $\Lambda$
term is governed by the combination \cite{s2}
\be\label{4}
S^L_{Log \Lambda}=(\frac{3}{2}m^4_H+3m^4_Z+6m^4_W-12
\sum_fm^4_f)/<\Phi_0>^4   ~~.\ee
We note that $W(J_0,j)$ for $j\not=0$ is
gauge-dependent, and so is (\ref{4}) as well as the Higgs 
potential $V(J,<\phi_0>)$ obtained by the Legendre transformation 
of $W(J,j)$ with respect to the variable $j$.
Here, the Landau gauge seems to be the most 
appropriate one \cite{s2}.
For the square of the Higgs field, on the other hand, the
corresponding expression takes the same form as in (\ref{4}) 
but with a different coefficient multiplying $m^4_H$ \cite{s3}
\be\label{5}
S^{SQ}_{Log\ \Lambda}=(\zeta m^4_H+
3m^4_Z+6m^4_W-12\sum_fm^4_f)/<\Phi_0>^4 ~.\ee 
The parameter $\zeta$ distinguishes two cases due to a  
subtraction term $\sim J^2$ in $W(J,j)$ which contributes here but 
does not contribute to $<\Phi_0>$ obtaineable from (\ref{1}). 
$\zeta = 1$ follows if no subtraction is performed. 
Presumably, however,
one should take 
$\zeta = 0$ which is obtained by 
fixing the subtraction term such that  
the logarithmic divergence occuring in (\ref{2}) vanishes in the 
unbroken phase.

It is not obvious
whether (\ref{1}) or (\ref{2}) is the better choice for
discussing logarithmic divergencies. The vacuum expectation 
value of the square of the Higgs field
is gauge-invariant and may be prefered. 

To require $S_{Log\ \Lambda}\simeq0$ is of course again a
very speculative assumption. The one-loop results (\ref{4}) and 
(\ref{5}) are strongly
scale-dependent. We set 
$m_t(m_Z)= 173~GeV$ and use the two loop renormalisation 
group equations for the calculation of coupling constants 
at other scales. By taking
the scale $\mu$ from $\mu\approx m_Z$ up to $\mu\approx m_{Planck}$
one obtains $m_H(m_Z)$ values ranging from $\approx$ 290 GeV
down to $\approx$ 150 GeV if using $S^L_{Log\ \Lambda}=0$
(eq. (\ref{4})), and $\approx$ 320 GeV down to $\approx$
120 GeV by using $S^{SQ}_{Log\ \Lambda}=0$
(eq. (\ref{5})) with $\zeta=1$.  $\zeta = 0$ requires a very large
scale of the order of the Planck scale (together with a 
top quark mass
$ m_t(m_Z) \approx 170~GeV$) (see below).

We will now consider the simultaneous suppression of quadratic
and logarithmic divergencies. We then obtain -- at a given 
scale -- two equations containing the Higgs and top coupling 
constants. There are two solutions, one with large values 
for the Higgs and top masses and one with much smaller values. 
We first consider the one with the larger values and choose 
the scale $\mu$ such that $m_t(m_Z) = 173~GeV $. The result
is now different whether $S^L_{Log\ \Lambda}$ or
$S^{SQ}_{Log\ \Lambda}$ is used. In the case of $S^L_{Log\ \Lambda}=0$
suggested in ref. \cite{s2} one gets $m_H(m_Z)\approx$ 178 GeV.
The corresponding scale $\mu$, which can be interpreted as 
the cut-off scale, is found
to be $\approx 10^4 ~TeV$.

If we require the vanishing of $S^{SQ}_{Log\ \Lambda}$ 
together with 
$S_{\Lambda^2}$ and take $\zeta=1$
one obtains $m_H(m_Z) \approx
330~ GeV$ and $m_t(m_Z)\approx 200~GeV$ when 
choosing for the scale 
the very low value $\mu\approx250~GeV$. 
This case is interesting
since $m_H\approx 2m_t$ suggests a bound state picture for 
the Higgs meson. 
The low scale could mean
that the influence of higher states on the standard model
would be noticeable already in the low TeV region. 
Above this scale the
renormalization-group equations of the standard model would
no more be valid. Of course, it could be entirely fortuitous
that $m_H\approx 2m_t$ taken at a low scale leads
to small values for $S_{\Lambda^2}$ and 
$S^{SQ}_{Log\ \Lambda}$, simultaneously. We also note that 
$\zeta=0$ does not lead to admissable solution at this low scale.

For the second type of solutions with the smaller values for the 
Higgs and top couplings the relevant scale $\mu$ must be very 
high, as high as the Planck mass, in order to obtain a value of 
$m_t(m_Z)$ near $170~ GeV$. Let us then take the scale to be equal 
to the Planck scale $ \mu = 10^{19} ~GeV$. Then we can "predict" 
both, the top and the Higgs mass.
Because of the smallness of the Higgs coupling it turns out, that 
the result does not much differ whether we use (\ref{3}) together 
with (\ref{4}) or (\ref{3}) with (\ref{5}) and $\zeta=1$ or 
with (\ref{5}) and $\zeta=0$. One obtains
\be\label{6}
       m_t(m_Z) \approx 170~GeV ~,  ~m_H(m_Z) \approx 140~GeV ~   .\ee
This result is very close to the one obtained by Bennett, Nielsen 
and Froggatt \cite{s4} using a different reasoning. 
It is close since 
these authors are also led to $S^{SQ}_{Log\ \Lambda} = 0 $ at 
the Planck scale. There is no precise agreement 
since the zero value 
of the Higgs coupling at the Planck mass they argue for does not 
occur in the approach presented here.
 
The solution obtained by combining  $S_{\Lambda^2} = 0$ with 
$S_{Log\ \Lambda} = 0$ and taking $\zeta = 0$ is of particular interest. 
First, it requires a very high scale which can be identified with 
the cut-off $\Lambda$. This scale is large enough that even the 
logarithm of 
it can provide for an order of magnitude suppression of new physics.
Secondly, the divergencies are eliminated in the unbroken phase 
as well. Thirdly, (\ref{5}) can be written in terms of 
$\beta$ - functions for the renormalized Higgs 
coupling $\lambda(\mu)$ 
and the renormalized mass parameter in the Higgs 
Lagrangian $J_0(\mu)$ 
\be\label{7}
4\pi^2~S^{SQ}_{Log\ \Lambda} = - \frac{\lambda^2(\mu)}{J^2_0(\mu)} 
\beta(\frac{J^2_0(\mu)}{\lambda(\mu)})   ~~ .     \ee
Thus, the requirement that (\ref{7}) vanishes can formally 
be viewed as a condition for a "fix point" at 
$\mu \approx m_{Planck}$ even though the full vacuum expectation 
value of the square of the unrenormalized Higgs field is 
scale independent. Taking $\mu = 10^{19}~GeV$ and $\alpha_s = 0.12$
one gets
\be\label{8}
 m_t(m_Z) = 168~GeV ~,~ m_H(m_Z) = 137~GeV ~   .\ee

As a last point I like to comment on the question of the
possible participation of a 4th generation. If the particles of 
this generation
obtain their masses through the coupling to the standard model
Higgs particle, one cannot have $S_{\Lambda^2}$ 
and $S_{Log\ \Lambda}$
simultaneously sufficiently small or zero. The reason is
that fermion masses for the $b'$ and $t'$ would enter with masses
which are larger or roughly equal to the mass of the top.
One obtains imaginary solutions of the equations 
or -- for $\zeta = 0$ -- too small values for the fermion masses.

We have seen that within the standard model the possibility
exists that the couplings are arranged such that the influence
from physics beyond the standard model is suppressed. 
This is not a trivial statement. Furthermore, it can only occur for
the known three generations\footnote{Of course, nothing
can be said about particles with a different origin of their
masses.}. When we minimize simultaneously the quadratic and 
the logarithmic dependence of the vacuum expectation value of 
the Higgs field on the cut-off we find interesting 
relations between the Higgs and the top mass. For a very low 
value of the cut-off scale we found $m_H \approx 2 m_t$. 
On the other hand, when using a properly subtracted form of $W(J,j)$ 
such that there is no divergence in the parameter region of no 
symmetry breaking, the required absence of quadratic and logarithmic 
divergencies in the physical region leads to $m_H \approx 140~GeV$~. 

I like to thank Norma Mankoc Borstnic and Holger Bech Nielsen
for inviting me to this workshop. I profited from  
the many discussions, especially also with Colin Froggatt.

\newpage
\stepcounter{section}
\setcounter{equation}{0}
\section*{\center Parity conservation in broken $SO(10)$}
\centerline{\large  Hanns}
\vspace{5mm}


\noindent
Nowadays, it is obvious that the suitable framework of discussing particle
physics is given by spontaneously broken gauge theories. Therefore, also
discrete symmetries like P, CP and T have to be implemented into this
framework [1]. Generally, parity may be broken if left- and righthanded
fermions belong to different representations of the gauge Group $G$. In
vector like gauge theories, like QCD and QED, parity is trivially conserved
since in both cases the left- and righthanded fermions belong to the same
fundamental representation of $SU(3)_C$, or $U(1)_{EM}$, respectively.
Some gauge theories, like $SO(10)$, however allow a more subtle definition 
of parity (called internal parity [1], or D-parity [2,3]) as a special
automorphism of the Lie algebra even if the representation of left
chiral fermions is different from the right chiral one. In such a theory,
parity violation is linked to the spontaneous breakdown of the
corresponding Grand Unified Theory.

For simplicity, we write all fermions of one family as lefthanded fields
$$
f_L = (\psi_{1L},(\psi_2^c)_L = C \bar \psi_{2R}^T).
$$
In $SO(10)$, all 16 quarks and leptons (including a righthanded neutrino
$\nu_R$) of one family, written as lefthanded fields, belong to one
irred. 16 representation:
$$
f_L \sim 16, \qquad f_R \sim \overline{16}.
$$
For convenience, we use the following subgroup of $SO(10)$ for the
spectrum of states:
$$
SO(10) \supset SO(4) \otimes SO(6) \simeq SU(2)_L \otimes SU(2)_R 
\otimes SU(4)_C.
$$
Here, $SU(4)_C$ is the Pati--Salam color group [4] with lepton number as 
fourth color, i.e. $SU(4)_C \supset SU(3)_C \otimes U(1)_{B-L}$. With
this, we find
$$
16 = (2,1,4) + (1,2, \bar 4).
$$
One can define a parity transformation via
$$
P : (f_i)_L \rightarrow U_{ij}(P)(f_j^c)_L.
$$
Since $f_{iL}$ and $f^c_{iL}$ are both members of the 16-representation,
the matrix $U(P)$ can be expressed in terms of the $SO(10)$-generators 
$M_{ij}$ as follows [1]:
$$
U(P) = \exp (i \pi M_{46} \cdot M_{810}).
$$
In most models of $SO(10)$ [5,6], the breaking pattern to the standard
model follows the chain
\begin{eqnarray*}
SO(10) &\rightarrow& SU(2)_L \otimes SU(2)_R \otimes SU(4)_C \otimes P  \\
       &\rightarrow& SU(2)_L \otimes SU(2)_R \otimes SU(3)_C \otimes U(1)_{B-L}
             \otimes P \\
       &\rightarrow& SU(2)_L \otimes U(1)_Y \otimes SU(3)_C
\end{eqnarray*}
or alternatively
\begin{eqnarray*}
SO(10) &\rightarrow& SU(2)_L \otimes SU(2)_R \otimes SU(4)_C \otimes P  \\
       &\rightarrow& SU(2)_L \otimes U(1)_R \otimes SU(4)_C \\
       &\rightarrow& SU(2)_L \otimes U(1)_R \otimes U(1)_{B-L} 
\otimes SU(3)_C \\
       &\rightarrow& SU(2)_L \otimes U(1)_Y \otimes SU(3)_C.
\end{eqnarray*}
In both cases, the breaking of parity (P) is associated with the breaking
of $SU(2)_R$ and is therefore characterized by the same scale $\Lambda_R$,
$\Lambda_R > 350$~TeV. There exists, however, an alternative scenario,
given by Chang et al. [3] where parity breaks separately from $SU(2)_R$
at a much higher scale:
\begin{eqnarray*}
SO(10) &\rightarrow& SU(2)_L \otimes SU(2)_R \otimes SU(4)_C \otimes P  \\
       &\rightarrow& SU(2)_L \otimes SU(2)_R \otimes SU(4)_C \\
       &\rightarrow& SU(2)_L \otimes U(1)_R \otimes U(1)_{B-L} 
\otimes SU(3)_C \\
       &\rightarrow& SU(2)_L \otimes U(1)_Y \otimes SU(3)_C.
\end{eqnarray*}
The reason for the last breaking chain is the assumption of the
existence of a real P-odd $SU(2)_R$
singlet Higgs field which acquires a vacuum expectation value at a high 
scale, independent of $\Lambda_R$. It can be incorporated in a
210-dim. Higgs representation, whereas the usual breaking pattern 
proceeds via 120- and 126-dimensional representations only.

\subsection{Discussion on Majorana particles}
What is the Majorana propagator? A Majorana spinor field $\psi_M(x)$ is a 
four component spinor satisfying the additional constraint
$$
\psi_M = \psi_M^c \equiv C \bar \psi_M^T,
$$
and
$$
\bar \psi_M = \overline{\psi_M^c} \equiv - \psi_M^T C^{-1}
$$
(for anticommuting fields). Written in terms of two-component chiral Weyl
spinors [1], a Majorana spinor is given by
$$
\psi_M = \left( \begin{array}{c} \chi_\alpha \\ \bar \chi^{\dot \alpha} 
\end{array} \right)
$$
whereas a Dirac spinor is written as
$$
\psi_D = \left( \begin{array}{c} \chi_\alpha \\ \bar \lambda^{\dot \alpha} 
\end{array} 
\right).
$$
Correspondingly, the usual Dirac invariant and Majorana invariant are
identical:
$$
\bar \psi \psi = \bar \psi^c \psi = - \psi^T C^{-1} \psi.
$$
Furthermore, the Dirac equation for a Majorana particle reads, in the
Weyl representation where $\gamma_5$ is diagonal,
$$
\left( \begin{array}{cc} 0 & i \sigma^\mu \partial_\mu \\ i \bar 
\sigma_\mu 
\partial^\mu & 0 \end{array} \right)
\left( \begin{array}{c} \chi \\ \bar \chi \end{array} \right) =
m \left( \begin{array}{c} \chi \\ \bar \chi \end{array} \right)
$$
and the Lagrangian is given by
$$
{\cal L} = \frac{i}{4} \bar \psi \stackrel{\leftrightarrow}{/\!\!\!\partial} 
\psi - \frac{m}{2} \bar \psi \psi.
$$
Note the additional factor 1/2, which is due to the fact that $\bar \psi$
is proportional to $\psi$. Decomposing the field operator into creation
and annihilation operators $a_s$, $a_s^\dagger$ and $b_s$, $b_s^\dagger$
 and using
$\psi^c = C \bar \psi^T$, one finds that $a = b$ and therefore
$$
\psi_M(x) = \int \frac{d^3k}{(2\pi)^3 2k_0} \left\{ \sum_s a_s(k) u_s(k)
e^{-ikx} + \sum_s  a_s^\dagger(k) v_s(k) e^{ikx} \right\}
$$
where
$$
\{ a_s(k), a_{s'}^\dagger(k')\}_+ = (2\pi)^3 2k_0 \delta^3(\vec k - \vec k').
$$
>From this, it is straightforward to define the propagator functions for
Majorana particles [2]
$$
\langle 0|T(\psi_M(x),\bar \psi_M(x')|0\rangle = i S_F(x - x')
$$
which is identical to
$$
\langle 0|T(\psi_M(x),\psi_M(x') |0\rangle = i S_F(x - x') C^T
$$
$$
\langle 0|T(\bar \psi_M(x), \bar \psi_M(x')|0\rangle = i C^{-1} S_F(x - x').
$$
With this, it is straightforward to derive the Feynman rules. 
The only difference to
Dirac particles one should be aware of is that internal Majorana particles
have no arrow which indicates the flow of fermion number.

\subsection{Representations of $SO(1,13)$}
Here, I will give a short review of the fundamental spinor representations
of $SO(1,13)$ in terms of the Lorentz group $\otimes SO(10)$, as grand
unification group. Details can be found in Ref.~[1].

The two fundamental representations of $SO(1,13)$ are the spinor
representations
$S^+ = 64$ and $S^- = 64'$. According to the subgroup $SO(1,3) \otimes
SO(10)$, and denoting the representations of the Lorentz group by
${\cal D}(j,j')$, we find
$$
64 = {\cal D}(1/2,0) \otimes 16 + {\cal D}(0,1/2) \otimes \overline{16} = 
(16)_L + (\overline{16})_R,
$$
$$
64' = {\cal D}(1/2,0) \otimes \overline{16} + {\cal D}(0,1/2) \otimes 16 = 
(\overline{16})_L + (16)_R.
$$
This means, the 64-representation contains exactly the desired states of
one family of fermions, a 16-representation of left chiral fermions and a 
$\overline{16}$ of their right chiral antiparticles.

The $64'$, however, has the particle content exactly reversed --
$\overline{16}$
for the left chiral fields, 16 for their antiparticles. Such a family
would consist exclusively of mirror-particles, i.e. left chiral quarks
which are singlets and right chiral quarks which are doublets under
$SU(2)_L$. It should be noted, due to the signature of $SO(1,13)$, both
reps are real and inequivalent, i.e.
$$
64 = \overline{64}, \qquad 64' = \overline{64}' \neq 64.
$$
If we are looking for gauge bosons $V^\mu$ of $SO(1,13)$, they should be 
contained in the (adjoint) 91-representation:
$$
91 = {\cal D}(1,0 + 0,1) \otimes 1 + {\cal D}(1/2,1/2) \otimes 10 + 
{\cal D}(0,0) \otimes 45.
$$
The first term denotes the usual affine connections
$\Gamma^\mu{}_{\nu\lambda}$,
the gauge bosons of the Lorentz group. The last term contains the
familiar 45 gauge bosons $V^\mu{}_{ij}$ of $SO(10)$, whereas the second
term mixes Lorentz- and internal degrees of freedom. Its coupling to the 
fermions in 64 looks like the coupling of a spin-2 tensor meson in the
10-dimensional representation of $SO(10)$.

\newpage
\thispagestyle{empty}
$\,$
\vskip 5 cm
\begin{center}
{\bf LIST OF ACTIVE PARTICIPANTS}

\begin{itemize}
\item Astri Kleppe ({\tt astri@vana.physto.se})
\item Larisa Laperashvili ({\tt larisa@vitep5.itep.ru})
\item Matej Pav\v si\v c ({\tt matej.pavsic@ijs.si})
\item Matja\v z Polj\v sak ({\tt matpoljsak@ijs.si})
\item Svend E. Rugh ({\tt rugh@nbi.dk})
\item Berthold Stech ({\tt b.stech@thphys.uni-heidelberg.de})
\item Hanns Stremnitzer ({\tt STREM@Pap.UniVie.AC.AT})
\item Colin Froggatt ({\tt c.froggatt@physics.gla.ac.uk})
\item Holger Bech Nielsen ({\tt hbech@nbivms.nbi.dk})
\item Norma Manko\v c Bor\v stnik ({\tt norma.s.mankoc@ijs.si})

\end{itemize}

\end{center}
\vfill
\hrulefill\\
{\bf WHAT COMES BEYOND THE STANDARD MODEL}\\[5mm]
Proceedings to the International Workshop, Bled, Slovenia, 29.6 --
9.7.1998\\[3mm]
Publisher DMFA - zalo\v zni\v stvo\\[3mm]
Articles written by active participants\\[3mm]
Edited by Simon Vre\v car \\[3mm]
Printed by Migraf in 150 copies \\[3mm]
\baselineskip 3pt
\begin{tiny}
 Po mnenju Ministrstva za znanost in tehnologijo \v st. 415-01-54/99 z dne
 02.04.1999 \v steje publikacija med proizvode iz 13. to\v cke tarifne \v
 stevilke 3 zakona o prometnem davku, za katere se pla\v cuje 5\% davek od
 prometa proizvodov
\end{tiny}

\end{document}